\documentclass[11pt]{report}
\usepackage{amsmath,amsfonts,amssymb,epsfig,color}
\usepackage{rotating}
\begin{document}
\title{Geometric Complexity Theory VI: the flip via saturated  and 
positive integer programming in representation theory and algebraic 
geometry}
\author{
Dedicated to Sri Ramakrishna \\ \\
Ketan D. Mulmuley
 \\
The University of Chicago\\
http://ramakrishnadas.cs.uchicago.edu
\\  \\
(Technical report TR 2007-04, Comp. Sci. Dept., \\ 
 The University of Chicago, May, 2007) \\ Revised version}

%\includeonly{}

\maketitle

\newtheorem{prop}{Proposition}[section]
\newtheorem{claim}[prop]{Claim}
\newtheorem{goal}[prop]{Goal}
\newtheorem{theorem}[prop]{Theorem}
\newtheorem{hypo}[prop]{Hypothesis}
\newtheorem{guess}[prop]{Guess}
\newtheorem{problem}[prop]{Problem}
\newtheorem{axiom}[prop]{Axiom}
\newtheorem{question}[prop]{Question}
\newtheorem{remark}[prop]{Remark}
\newtheorem{lemma}[prop]{Lemma}
\newtheorem{claimedlemma}[prop]{Claimed Lemma}
\newtheorem{claimedtheorem}[prop]{Claimed Theorem}
\newtheorem{cor}[prop]{Corollary}
\newtheorem{defn}[prop]{Definition}
\newtheorem{ex}[prop]{Example}
\newtheorem{conj}[prop]{Conjecture}
\newtheorem{obs}[prop]{Observation}
\newtheorem{phyp}[prop]{Positivity Hypothesis}
\newcommand{\bitlength}[1]{\langle #1 \rangle}
\newcommand{\ca}[1]{{\cal #1}}
\newcommand{\floor}[1]{{\lfloor #1 \rfloor}}
\newcommand{\ceil}[1]{{\lceil #1 \rceil}}
\newcommand{\gt}[1]{{\langle  #1 |}}
\newcommand{\C}{\mathbb{C}}
\newcommand{\N}{\mathbb{N}}
\newcommand{\R}{\mathbb{R}}
\newcommand{\Z}{\mathbb{Z}}
\newcommand{\frcgc}[5]{\left(\begin{array}{ll} #1 &  \\ #2 & | #4 \\ #3 & | #5
\end{array}\right)}

\newcommand{\cgc}[6]{\left(\begin{array}{ll} #1 ;& \quad #3\\ #2 ; & \quad #4
\end{array}\right| \left. \begin{array}{l} #5 \\ #6 \end{array} \right)}

\newcommand{\wigner}[6]
{\left(\begin{array}{ll} #1 ;& \quad #3\\ #2 ; & \quad #4
\end{array}\right| \left. \begin{array}{l} #5 \\ #6 \end{array} \right)}

\newcommand{\rcgc}[9]{\left(\begin{array}{ll} #1 & \quad #4\\ #2  & \quad #5
\\ #3 &\quad #6
\end{array}\right| \left. \begin{array}{l} #7 \\ #8 \\#9 \end{array} \right)}

\newcommand{\srcgc}[4]{\left(\begin{array}{ll} #1 & \\ #2 & | #4  \\ #3 & |
\end{array}\right)}

\newcommand{\arr}[2]{\left(\begin{array}{l} #1 \\ #2   \end{array} \right)}
\newcommand{\unshuffle}[1]{\langle #1 \rangle}
\newcommand{\ignore}[1]{}
\newcommand{\f}[2]{{\frac {#1} {#2}}}
\newcommand{\tableau}[5]{
\begin{array}{ccc} 
#1 & #2  &#3 \\
#4 & #5 
\end{array}}
\newcommand{\embed}[1]{{#1}^\phi}
\newcommand{\comb}[1]{{\| {#1} \|}}
\newcommand{\stab}{{\mbox {stab}}}
\newcommand{\perm}{{\mbox {perm}}}
\newcommand{\trace}{{\mbox {trace}}}
\newcommand{\polylog}{{\mbox {polylog}}}
\newcommand{\sign}{{\mbox {sign}}}
\newcommand{\proj}{{\mbox {Proj}}}
\newcommand{\height}{{\mbox {ht}}}
\newcommand{\poly}{{\mbox {poly}}}
\newcommand{\std}{{\mbox {std}}}
\newcommand{\m}{\mbox}
\newcommand{\formula}{{\mbox {Formula}}}
\newcommand{\circuit}{{\mbox {Circuit}}}
\newcommand{\core}{{\mbox {core}}}
\newcommand{\orbit}{{\mbox {orbit}}}
\newcommand{\sie}{{\mbox {sie}}}
\newcommand{\pie}{{\mbox {pie}}}
\newcommand{\cycle}{{\mbox {cycle}}}
\newcommand{\ideal}{{\mbox {ideal}}}
\newcommand{\qed}{{\mbox {Q.E.D.}}}
\newcommand{\proof}{\noindent {\em Proof: }}
\newcommand{\weight}{{\mbox {wt}}}
\newcommand{\tab}{{\mbox {Tab}}}
\newcommand{\level}{{\mbox {level}}}
\newcommand{\vol}{{\mbox {vol}}}
\newcommand{\vect}{{\mbox {Vect}}}
\newcommand{\val}{{\mbox {wt}}}
\newcommand{\sym}{{\mbox {Sym}}}
\newcommand{\convex}{{\mbox {convex}}}
\newcommand{\spec}{{\mbox {spec}}}
\newcommand{\strong}{{\mbox {strong}}}
\newcommand{\adm}{{\mbox {Adm}}}
\newcommand{\eval}{{\mbox {eval}}}
\newcommand{\for}{{\quad \mbox {for}\ }}
\newcommand{\Q}{Q}
\newcommand{\mand}{{\quad \mbox {and}\ }}
\newcommand{\invlim}{{\mbox {lim}_\leftarrow}}
\newcommand{\directlim}{{\mbox {lim}_\rightarrow}}
\newcommand{\sformal}{{\cal S}^{\mbox f}}
\newcommand{\vformal}{{\cal V}^{\mbox f}}
\newcommand{\crystal}{\mbox{crystal}}
\newcommand{\conje}{\mbox{\bf Conj}}
\newcommand{\graph}{\mbox{graph}}
\newcommand{\ind}{\mbox{index}}

\newcommand{\rank}{\mbox{rank}}
\newcommand{\id}{\mbox{id}}
\newcommand{\str}{\mbox{string}}
\newcommand{\RSK}{\mbox{RSK}}
\newcommand{\wt}{\mbox{wt}}
\setlength{\unitlength}{.75in}

\begin{abstract} 
This article belongs to a series on geometric complexity theory (GCT),
an approach to the $P$ vs. $NP$ and related  problems through
algebraic geometry and representation theory. 
The basic principle behind this approach is called the {\em flip}. 
In essence, 
it reduces the negative hypothesis  in complexity theory (the lower bound problems),
such as the $P$ vs. $NP$ problem in characteristic zero,
to  the positive hypothesis in complexity theory (the upper bound problems):
specifically, to showing
that the problems of deciding 
nonvanishing of the  fundamental structural
constants in representation theory and algebraic geometry,
such as the well known plethysm  constants \cite{macdonald,fultonrepr}, belong to 
the complexity class $P$.
In this article,  we   suggest a plan for implementing the {flip}, i.e., for 
showing that these decision problems belong to $P$.
This is based on the reduction of the  preceding complexity-theoretic 
positive hypotheses 
to  mathematical positivity hypotheses: specifically, to showing that there exist 
positive formulae--i.e. formulae with nonnegative coefficients--for the structural
constants under consideration and certain functions associated with them. 
These  turn out be intimately related to 
the similar positivity properties of the Kazhdan-Lusztig polynomials \cite{kazhdan,kazhdan1}  and 
the multiplicative structural constants of the canonical (global crystal) bases 
\cite{kashiwara2,lusztigcanonical} 
in the theory of Drinfeld-Jimbo quantum groups. The known proofs of these positivity 
properties 
depend on the Riemann hypothesis over finite fields  (Weil conjectures proved in \cite{weil2}) 
and the related results \cite{beilinson}.
Thus the reduction here, in conjunction with the flip, 
in essence, says that the validity of the $P\not = NP$ conjecture 
in characteristic zero is intimately linked to the Riemann hypothesis over finite 
fields and related problems.

The main ingradients of this  reduction are as follows. 

First, we formulate a general paradigm of saturated, and more strongly, positive
integer programming, and show that it has a polynomial time algorithm,
extending and building on  the techniques in 
\cite{loera,GCT3,GCT5,lovasz,kannan,king,kirillov,knutson}.

Second, building on the work of Boutot \cite{boutot} and  Brion (cf. \cite{dehy}),
we show  that the stretching functions associated with the structural constants
under consideration are quasipolynomials, generalizing the  known result that the 
stretching function associated with the Littlewood-Richardson coefficient 
is a polynomial for
type $A$  \cite{derkesen,kirillov} and a quasi-polynomial for general types \cite{berenstein,dehy}.
In particular, this proves Kirillov's conjecture \cite{kirillov} for the plethysm constants.

Third, using these stretching quasi-polynomials, 
we formulate the  mathematical saturation and positivity hypotheses for the plethysm and
other structural
constants under consideration, which 
generalize the known saturation and  conjectural positivity properties of the
Littlewood-Richardson coefficients   \cite{knutson,loera,king}.
Assuming these hypotheses, it follows that the problem of deciding nonvanishing of any of these
structural constants, modulo a small relaxation,
can be transformed in polynomial time into a saturated,
and more strongly, positive integer programming problem, and hence, can be solved
in polynomial time.

Fourth, we give theoretical and experimental results in support of these hypotheses.

Finally, we suggest an approach to prove these positivity hypotheses motivated by the 
works on Kazhdan-Lusztig bases for Hecke algebras \cite{kazhdan,kazhdan1} and 
the canonical (global crystal) bases of Kashiwara and Lusztig 
\cite{lusztigcanonical,lusztigbook, kashiwara2} 
for representations of Drinfeld-Jimbo quantum groups \cite{drinfeld,jimbo}.
Steps in this direction are taken \cite{GCT4,plethysm,canonical}.

Specifically,  in \cite{GCT4,plethysm} are constructed
{\em nonstandard quantum groups}, with compact real forms, which are
generalizations of the Drinfeld-Jimbo quantum group, 
and also associated {\em nonstandard algebras},
whose relationship with the nonstandard 
quantum groups is conjecturally similar to the relationship of the Hecke 
algebra with the Drinfeld-Jimbo quantum group.
The article  \cite{canonical} gives conjecturally 
correct algorithms to construct canonical bases of 
the matrix coordinate rings of the nonstandard   quantum groups
and of nonstandard algebras that have conjectural positivity 
properties  analogous to those of the 
canonical (global crystal)  bases, as per Kashiwara and Lusztig,
of  the coordinate ring of the Drinfeld-Jimbo quantum group,
and the Kazhdan-Lusztig basis of
the Hecke algebra. These positivity conjectures (hypotheses)
 lie at the heart of this approach. 
In view of \cite{kazhdan1,lusztigcanonical}, their validity 
is intimately linked  to the
Riemann hypothesis over finite fields and the related works  mentioned above.

\end{abstract}

\tableofcontents

\chapter{Introduction}
This article belongs to a series of papers,
\cite{GCT1} to \cite{GCT11},
 on geometric complexity theory (GCT),
which is an approach to the $P$ vs. $NP$ and related problems 
in complexity theory
through algebraic geometry and representation theory. 
We assume here that the underlying field of computation
is of characteristic zero.  %For the problems that arise when
%the field is algebraically closed of   positive characteristic or is 
%finite, see \cite{GCT11}. 
The usual $P$ vs. $NP$ problem is over a finite
field. The characteristic zero version is its weaker, formal implication,
and philosophically, the crux.

The basic 
principle underlying GCT  is called the {\em flip}
\cite{GCTflip}. 
The flip, in essence,  reduces 
the negative hypotheses (lower bound problems) in 
complexity theory,
such as the $P\not =?NP$  problem in characteristic zero, 
to  positive hypotheses 
in complexity theory
(upper bound problems): specifically, to the problem of 
showing  that  a series of decision problems
in representation theory and
algebraic geometry belong to the complexity class $P$. 
Each  of these decision problem
is of the form: Given 
a nonnegative structural constant in representation theory or
geometric invariant theory, such as the well known plethysm constant, 
decide if it is nonzero (nonvanishing), or rather, if is nonzero after 
a small relaxation.
This flip from 
the negative to the positive  may be considered to be a nonrelativizable
form of the flip--from the undecidable to the decidable--that
underlies the proof of G\"odel's incompleteness theorem. But the classical
diagonalization technique in G\"odel's result is relativizable \cite{solovay},
and hence, not applicable to the $P$ vs. $NP$ problem.
The flip, in contrast, is nonrelativizable. It is furthermore nonnaturalizable
\cite{GCT10}); i.e., it crosses  the natural proof barrier 
\cite{rudich} that 
any approach to the $P$ vs. $NP$ problem must  cross.

We  suggest  here  a plan for  implementating the flip; i.e., for showing that the 
decision problems above belong to $P$. 
This is based on the reduction in this paper of the complexity-theoretic positivity
hypotheses mentioned above to mathematical positivity hypotheses:
specifically, to showing that there exist 
positive formulae for the structural
constants under consideration and certain functions associated with them.
We also give  theoretical and experimental evidence in support of the
latter  hypotheses.

\begin{figure}
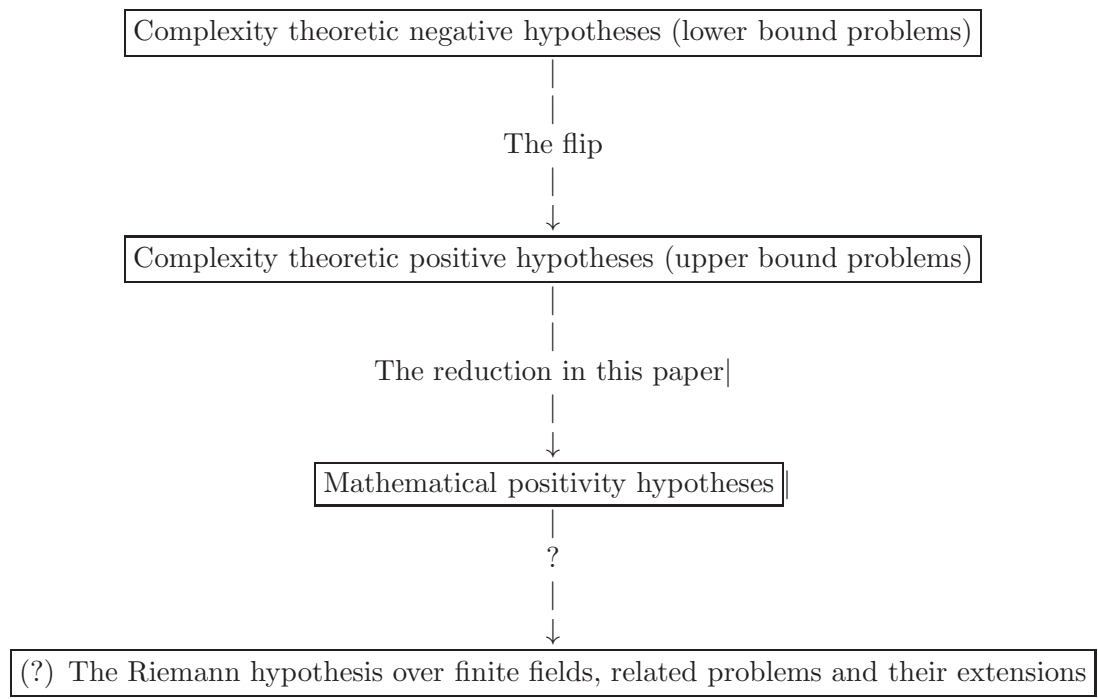
 
\[ \begin{array}{c}
\fbox{Complexity theoretic negative hypotheses (lower bound problems)} \\
 | \\
 | \\
\mbox{The flip} \\
 | \\
\downarrow \\
\fbox{Complexity theoretic positive hypotheses (upper  bound problems)} \\
 | \\
 | \\
\mbox{The reduction in this paper}
 | \\
 | \\
\downarrow \\
\fbox{Mathematical positivity hypotheses}
 | \\
 | \\
\mbox{?}\\
 | \\
\downarrow \\
\fbox{(?) The Riemann hypothesis over finite fields, related problems and their extensions}
\end{array} \]
\caption{Pictorial depiction of the basic plan for implementing the flip}
\label{fbasicintro}
\end{figure}

Here we say that a formula is positive if its coefficients are nonegative.
The problem finding the positive formulae  as above turns  out be intimately related to 
the analogous problem for 
the Kazhdan-Lusztig polynomials \cite{kazhdan}  and 
the multiplicative structural constants of the canonical (global crystal) bases 
\cite{kashiwara2,lusztigcanonical} 
in the theory of Drinfeld-Jimbo quantum groups. The known solution to the latter problem
 \cite{kazhdan1,lusztigcanonical} depends on 
the Riemann hypothesis over finite fields, proved in  \cite{weil2}, 
and the related results in \cite{beilinson}.
Thus the flip and the reduction here together roughly say 
that the validity of the $P\not = NP$ conjecture 
in characteristic zero is intimately linked to the Riemann hypothesis over finite 
fields and related problems. This is illustrated in  Figure~\ref{fbasicintro};
 the question marks there
indicate unsolved problems.
It seems that substantial extension of the techniques related to the Riemann hypothesis
over finite fields may be needed to prove the required mathematical positivity hypotheses here.
We do not have the necessary mathematical expertize for this task. But it
is our hope that the experts in algebraic geometry and representation theory will have
something to say on this matter.

It  may be 
conjectured that the flip paradigm would also work in the context of
the usual $P$ vs. $NP$ problem over $F_2$ (the boolean field) or the
finite field $F_p$. But implementation of the flip over a finite field
is expected to be much harder than in characteristic zero. That 
is why we focus on characteristic zero here, deferring discussion of the
problems that arise over finite field to \cite{GCT11}.

Now we turn to  a more detailed exposition of the main results in this paper and 
of Figure~\ref{fbasicintro}.

\subsection*{Acknoledgements}
We are  grateful to the authors of \cite{rosas2} for
pointing out an error in the saturation hypothesis (SH) in the earlier 
version of this paper. It has been corrected in this version with 
appropriate relaxation without affecting the overall approach of GCT 
(cf. Section~\ref{sintroplethysm} and also \cite{GCT6erratum}). 
We are also grateful to 
Peter Littelmann for bringing the
reference \cite{dehy} to our attention, to
H. Narayanan for suggesting the use of \cite{kannan} in the proof of
Theorem~\ref{tindexquasi} and bringing the positivity conjecture in
\cite{loera}
to our attention, and to Madhav Nori for a helpful discussion.
The experimental results in Chapter~\ref{cevidence} were obtained using  Latte 
\cite{latte}.

\section{The decision problems} \label{sdecision}
We now describe the relevant decision problems 
in representation theory and algebraic geometry.
The actual decision problems that arise  in the flip (cf. the second
box in Figure~\ref{fbasicintro})  are  relaxed versions
of these problems  described later (cf. Hypothesis~\ref{PHfliprevised}).

\begin{problem} (Decision version of the  Kronecker problem) \label{pintrokronecker} 

Given partitions $\lambda,\mu,\pi$, decide nonvanishing of the Kronecker 
coefficient  $k_{\lambda,\mu}^\pi$. This is the multiplicity  
of the irreducible representation (Specht module)
$S_\pi$ of the symmetric group $S_n$ in the
tensor product $S_\lambda \otimes S_\mu$.

Equivalently \cite{fultonrepr}, let $H=GL_n(\C)\times GL_n(\C)$ and 
$\rho: H \rightarrow G=GL(\C^n \otimes \C^n)=GL_{n^2}(\C)$ the natural embedding. Then
 $k_{\lambda,\mu}^\pi$ is the multiplicity of the 
$H$-module $V_\lambda(GL_n(\C)) \otimes V_\mu(GL_n(\C))$ in the $G$-module
$V_\pi(G)$, considered as an $H$-module via the embedding $\rho$.
\end{problem} 

Here $V_\lambda(GL_n(\C))$ denotes the irreducible representation (Weyl module) of
$GL_n(\C)$ corresponding to the partition $\lambda$; $V_\pi(G)$ is the Weyl module
of $G=GL_{n^2}(\C)$.

 Problem~\ref{pintrokronecker} is a special case of the following  generalized plethysm problem.

\begin{problem} (Decision version of the  plethysm problem) \label{pintroplethysm}

Given partitions $\lambda,\mu,\pi$, decide nonvanishing of the plethysm
constant $a_{\lambda,\mu}^\pi$. This is the multiplicity  of 
the irreducible representation 
$V_\pi(H)$ of $H=GL_n(\C)$ 
in the irreducible representation 
$V_\lambda(G)$ of $G=GL(V_\mu)$,
where $V_\mu=V_\mu(H)$ is an irreducible representation $H$.
Here $V_\lambda(G)$ is considered an $H$-module via 
the representation map $\rho:H\rightarrow G=GL(V_\mu)$.

\noindent (Decision version of the generalized plethysm problem) 

The same as above, allowing  $H$ to be any connected reductive group.
\end{problem}

This is
a special case of the following 
fundamental  problem of representation theory (characteristic  zero):

\begin{problem} (Decision version of the subgroup restriction problem) \label{pintrosubgroup}

Let $G$ be connected reductive group, $H$ a reductive group,
possibly disconnected, and 
$\rho:H \rightarrow G$  an explicit, polynomial 
homomorphism (as defined in  Section~\ref{sssubgroup}). 
Here $H$ will generally be a subgroup of $G$, and $\rho$ its embedding.
Let  $V_\pi(H)$ be 
 an irreducible representation
of $H$, and $V_\lambda(G)$ an irreducible representation  of $G$.
Here $\pi$ and $\lambda$ denote the classifying labels 
of the irreducible representations $V_\pi(H)$ and $V_\lambda(G)$,
respectively.  Let 
$m_\lambda^\pi$ be the multiplicity   of $V_\pi(H)$
in $V_\lambda(G)$, considered as an $H$-module via $\rho$.

Given specifications of the embedding $\rho$ and the labels $\lambda,\pi$, as described 
in Section~\ref{sssubgroup}, decide nonvanishing of the multiplicity $m_\lambda^\pi$.
\end{problem} 

All reductive groups in this paper are over $\C$.
The reductive groups that arise in GCT in characteristic zero are: the general and special 
linear groups $GL_n(\C)$ and  $SL_n(\C)$, algebraic tori,
the symmetric group $S_n$, and the groups formed from these by (semidirect)
products.  The reader may wish to focus  on just these 
concrete  cases, since
all  main ideas in this paper  are illustrated therein.

Problem~\ref{pintrosubgroup} is, in turn, a special case of the following
most  general problem.

\begin{problem} (Decision problem in geometric invariant theory) 
\label{pintrogit}

Let $H$ be a reductive group, possibly disconnected, $X$ a projective $H$-variety ($H$-scheme),
i.e., a variety with $H$-action. Let $\rho$ denote this $H$-action.
Let $R=\oplus_d R_d$ be the homogeneous
coordinate ring of $X$. Assume that the singularities of $\spec(R)$ are rational.

We  assume that $X$ and $\rho$ have
special properties (as described in Section~\ref{sspgit}), so that,
in particular, they have  short specifications.
Let $V_\pi(H)$ be an irreducible representation 
of $H$.  Let $s_d^\pi$ be the multiplicity of  $V_\pi(H)$ in $R_d$, considered as an $H$-module
via the action $\rho$.

Given $d,\pi$, the specifications of $X$ and $\rho$,
decide nonvanishing of the multiplicity  $s_d^\pi$.
\end{problem} 

This last problem is hopeless for general $X$. Indeed the usual specification of $X$,
say in terms of the generators of the ideal of its appropriate embedding, is so large as 
to make this problem meaningless for a general $X$.
But the instances of this decision problem that arise in GCT are for  the  following
very special kinds of projective $H$-varieties $X$, which, in particular, have small
specifications (Section~\ref{sspgit}): 

\begin{enumerate} 
\item $G/P$, where $G$ is a connected, reductive group,
 $P\subseteq G$ its parabolic subgroup, and $H\subseteq G$ a reductive subgroup with
an explicit polynomial embedding.
Problem~\ref{pintrosubgroup} reduces to  this special case of 
Problem~\ref{pintrogit};  cf. Section~\ref{sspgit}.
\item {\em Class varieties} \cite{GCT1,GCT2}, which are associated with the
fundamental complexity classes such as $P$ and $NP$. 
They are very special like $G/P$, with 
conjecturally rational singularities \cite{GCT10}.
Each class variety is specified  by 
the  complexity class and the parameters of the lower bound problem under consideration.
Briefly, the $P$ vs. $NP$ problem in characteristic zero is reduced in \cite{GCT1,GCT2} 
to showing that the class variety corresponding to the complexity class $NP$ and the parameters
of the lower bound problem (such as the input size)  cannot
be embedded in the class variety corresponding
to  the complexity class $P$ and the same parameters. 
Efficient criteria for the decision problems stated above are needed to  construct 
{\em explicit obstructions} \cite{GCT2} to such embeddings, thereby proving their nonexistence.
Specifically, Problems~\ref{pintrosubgroup} and \ref{pintrogit}
are the decision problems
associated with  Problems 2.5 and 2.6 in \cite{GCT2}, respectively. 
See Sections~\ref{sdetvsperm}-\ref{spvsnp} for a brief review of this story.
\end{enumerate} 
For these varieties Problem~\ref{pintrogit} turns out to be qualitatively similar  to
Problem~\ref{pintrosubgroup}  (cf. Section~\ref{sspgit} and \cite{GCT2,GCT10}). 
For this reason,
the Kronecker and the plethysm problems, which lie at the heart of the subgroup 
restriction problem,  can be taken as the main prototypes of the
decision problems that arise here.

One can now ask: 

\begin{question} \label{qdecision}
Do the decision problems above 
(Problems~\ref{pintrokronecker}-\ref{pintrosubgroup} and 
Problem\ref{pintrogit}, when $X$ therein is $G/P$ or a class variety)
belong to $P$? 
That is, can the nonvanishing of any of structural constants in these
problems be decided in 
$\poly(\bitlength{x})$ time, where $x$ denotes the input-specification of the 
structural constant and $\bitlength{x}$ its bitlength?
\end{question} 

For Problem~\ref{pintroplethysm}, the input specification for the 
plethysm constant $a_{\lambda,\mu}^\pi$ is given in the form of a triple 
$x=(\lambda,\mu,\pi)$. Here 
the partition $\lambda$ is specified 
as a sequence of positive integers 
$\lambda_1 \ge \lambda_2 \ge \cdots \lambda_k>0$ (the zero parts of 
the partition are suppressed); 
$k$ is called the height or length of $\lambda$, and is denoted by
$\height(\lambda)$. 
The bitlength $\bitlength{\lambda}$ is defined to be 
the total bitlength of the integers $\lambda_r$'s. The bitlength 
$\bitlength{x}$ is defined to be $\bitlength{\lambda}+\bitlength{\mu}+\bitlength{\pi}$.
A detailed specification of the input specification  $x$ and its 
bitlength $\bitlength{x}$ 
for the other problems 
is given in Section~\ref{sphypo}.

For the reasons described in 
Section~\ref{sintroplethysm},  Question~\ref{qdecision}  may not have an
affirmative answer  in general; i.e., these problems may not be in $P$ in their
strict form stated above. 
The following main conjectural complexity-theoretic positivity hypothesis 
governing the flip says that  the  relaxed forms of these decision problems
described in Section~\ref{sphypo}
belong to $P$. As we shall see in Chapter~\ref{cobstruction}, these 
relaxed forms suffice for  the purposes of the flip.

\begin{hypo} \label{PHfliprevised} {\bf (PHflip)}
The relaxed forms (cf. Section~\ref{sphypo}) of
Problems~\ref{pintrokronecker}, \ref{pintroplethysm}, \ref{pintrosubgroup}, and
the special cases of Problem~\ref{pintrogit}, when $X$ therein is $G/P$ or  a class variety--which
together include  all  decision problems  that arise in the flip--belong
to the complexity class $P$. 

This means nonvanishing of any of these structural constants,
modulo a small  relaxation (as described in Section~\ref{sphypo}),
can be decided in
$\poly(\bitlength{x})$ time, where $x$ denotes the input-specification of the 
structural constant and $\bitlength{x}$ its bitlength.
\end{hypo} 

The phrase ``modulo a small relaxation'' in the
relaxed  form of the plethysm 
problem means the following:

\noindent (a) Let $h=\dim{G}+\height{\lambda}+\height{\pi}$, where 
$\dim(G)$ is the dimension of the group $G$ in Problem~\ref{pintroplethysm}.
Then there exist absolute nonnegative
constants $c$ and $c'$, independent of $\lambda$, $\mu$ and $\pi$,
such that nonvanishing of the relaxed (stretched)  plethysm constant 
$a_{b \lambda, b \mu}^{b \pi}$, 
for any positive integral relaxation parameter $b> c h^{c'}$,
can be decided in $O(\poly(\bitlength{\lambda},\bitlength{\mu},\bitlength{\pi},
\bitlength{b}))$ time, where $\bitlength{b}$ denotes the bitlength $b$.
The notation
$\poly(\bitlength{\lambda},\bitlength{\mu},\bitlength{\pi},\bitlength{b})$ here
means bounded by a polynomial of constant degree in 
$\bitlength{\lambda},\bitlength{\mu},\bitlength{\pi}$ and $\bitlength{b}$.
In particular, the time is 
$O(\poly(\bitlength{\lambda},\bitlength{\mu},\bitlength{\pi})$
if the relaxation parameter $b$  is small;
 i.e. if its bitlength $\bitlength{b}$ is 
$O(\poly(\bitlength{\lambda},\bitlength{\mu},\bitlength{\pi}))$. (Observe
that the bit length of $h$ is $O(\poly(\bitlength{\lambda},\bitlength{\mu},\bitlength{\pi}))$.)

\noindent (b) There exists a polynomial time algorithm for deciding 
nonvanishing of $a_{\lambda,\mu}^\pi$, which works correctly on
almost all $\lambda,\mu$ and $\pi$. Here polynomial time means
$O(\poly(\bitlength{\lambda},\bitlength{\mu},\bitlength{\pi})$ time.
The meaning of ``correctly on almost all'' is specified in 
Hypothesis~\ref{hshplethysmintro} below.

A detailed specification 
of the  relaxation, i.e., the meaning of
the phrase ``modulo a small relaxation'' for the other problems 
is given in Section~\ref{sphypo}.

The structural constants in Problems~\ref{pintrokronecker}-\ref{pintrosubgroup} are 
of fundamental importance in representation theory. The kronecker and the plethysm 
constants in Problems~\ref{pintrokronecker} and \ref{pintroplethysm}, in particular,
have been studied intensively; see \cite{fultonrepr,macdonald,stanleypos} for their
significance. There are many known formulae for these structural constants based on 
on the character formulae in representation theory.
Several formulae for the characters of connected, reductive groups are
known by now \cite{fultonrepr}, starting with the Weyl character formula. For the
symmetric group, there is the 
Frobenius character formula \cite{fultonrepr},
for the general linear group over a finite 
field, Green's formula \cite{macdonald}, and for finite simple groups of Lie type, 
the character formula of Deligne-Lusztig \cite{deligne}, and Lusztig \cite{lusztig}.
(Finite simple groups of Lie type, other than $GL_n(F_q)$,
are not needed in GCT.) 

One obvious method for deciding nonvanishing of the structural constants in 
Problems~\ref{pintrokronecker}-\ref{pintrogit} is to compute them exactly. 
But all known algorithms for exact computation of the structural constants in
Problems~\ref{pintrokronecker}-\ref{pintrosubgroup}
take exponential time. This is expected,
since  this problem is  $\#P$-complete. In fact, even the problem of 
exact computation of  a Kostka number, which is a very special
case of these structural constants, is $\#P$-complete \cite{hari}.
This means there
is no polynomial time algorithm for computing any of them,
assuming $P \not = NP$. 

Of course, there are $\#P$-complete quantities--e.g. the permanent of a 
nonnegative matrix \cite{valiant}--whose
nonvanishing can still   be decided in polynomial time \cite{schrijver}. But 
the decision problems above are of a totally different kind and, at the surface, appear to have
inherently exponential complexity.
This is  because the dimensions of the irreducible representations that occur in
their statements can be exponential in the ranks 
of the groups involved and the bit lengths of the classifying labels of
these representations. For example, the dimension of the Weyl module
$V_\lambda(GL_n(\C))$
can be exponential in $n$ and   the bit length of the partition $\lambda$.
Furthermore, the number of terms in any of  the preceding character formulae
is also exponential. 
All these decisions problems ask if one exponential dimensional
representation can occur within another exponential dimensional 
representation. To solve them, it may seem necessary 
to take a detailed look into  these representations and/or the
character  formulae of exponential complexity. 
Hence,  it seemed 
hard to believe  that  nonvanishing of 
these structural constants can, nevertheless,
be decided in polynomial time (modulo a small relaxation). This 
constituted the main philosophical obstacle in the course of GCT.

\section{Deciding nonvanishing of Littlewood-Richardson coefficients} \label{sintroLR}
The first result, which indicated that this obstacle may be  removable, came in
the wake of the saturation theorem of Knutson and Tao \cite{knutson}.
This concerns the following special case  of Problem~\ref{pintrosubgroup},
with $G=H \times H$, the embedding $\rho:H \rightarrow G$ being diagonal.

\begin{problem} \label{pintrolittle} (Littlewood-Richardson problem)

Given a complex semisimple, simply connected  Lie group $H$,
and its dominant weights $\alpha,\beta,\lambda$, 
decide nonvanishing of a
generalized Littelwood-Richardson coefficient $c_{\alpha,\beta}^{\lambda}$.
This is the multiplicity of the irreducible representation
$V_\lambda(H)$ of $H$ in the tensor product 
$V_{\alpha}(H) \otimes V_{\beta}(H)$.
\end{problem}

It was shown in \cite{GCT3,knutson2,loera} independently 
that nonvanishing of the Littlewood-Richardson 
coefficient of type $A$ 
can be decided in polynomial time; i.e., polynomial in the bit lengths of $\alpha,\beta,\lambda$.
Furthermore, the algorithm in \cite{GCT3} works  in strongly 
polynomial time in the terminology of  \cite{lovasz}; 
cf. Section~\ref{sstandard}. The three main ingradients in this result are:
\begin{enumerate}
\item {\bf PH1}: 
The Littlewood-Richarson rule, which goes back to 1940's, and whose
most important feature is that it is {\em positive}--i.e., it involves no alternating signs
as in character-based formulae--and
 its strengthening in \cite{berenstein}, which gives a positive, polyhedral 
formula for the Littlewood-Richardson coefficient as the number of integer points in
a polytope; this can be the BZ-polytope \cite{berenstein} or the hive polytope \cite{knutson}.
We shall refer to this positivity property as the first positivity hypothesis (PH1).

\item  The 
polynomial and strongly polynomial time algorithms for linear programming
\cite{khachian,tardos}, and
\item  {\bf SH}: The saturation theorem of Knutson and Tao \cite{knutson}. This says that 
$c_{\alpha,\beta}^\lambda$ is nonzero if $c_{n \alpha,n \beta}^{n \lambda}$ is nonzero 
for any $n\ge 1$. We shall refer to this saturation property as the saturation hypothesis (SH).
\end{enumerate} 

Brion \cite{zelevinsky} 
observed that the verbatim translation of the saturation property in \cite{knutson} 
 fails to hold
for the  the generalized  Littlewood-Richardson coefficients of types 
$B$, $C$, $D$ (it  also fails 
for the Kronecker coefficients, as well as
the plethysm  constants \cite{kirillov}). Hence,  the algorithms in
\cite{GCT3,knutson2,loera}  do not work
in types $B$, $C$ and $D$. Fortunately, this situation can be remedied.
It is shown in \cite{GCT5} that nonvanishing
of the generalized Littewood-Richardson coefficient 
$c_{\alpha,\beta}^\lambda$  of arbitrary type can be decided in
(strongly) polynomial time, 
assuming the positivity conjecture of De Loera and McAllister \cite{loera}.
This conjectural hypothesis, based on 
considerable experimental evidence, is as follows.
Let 
\begin{equation} \label{eqintrostretch1}
\tilde c_{\alpha,\beta}^\lambda(n)=c_{n \alpha, n \beta}^{n \lambda}
\end{equation}
be the stretching function associated with the Littlewood-Richardson
coefficient $c_{\alpha,\beta}^\lambda$. It is known to be a polynomial in type $A$
\cite{derkesen,kirillov},
and a quasi-polynomial, in general  \cite{berenstein,dehy,loera}. 
Recall that a fuction $f(n)$ is called a quasi-polynomial if there
exist $l$ polynomials $f_j(n)$, $1\le j \le l$, 
such that $f(n)=f_j(n)$ if $n=j$ mod $l$. Here $l$ is supposed to be
the smallest such integer, and  is called the period of $f(n)$. 
The period of $\tilde c_{\alpha,\beta}^\lambda(n)$ for  types $B,C,D$ is either $1$ or $2$
\cite{loera}. In general, it is bounded by 
a fixed constant depending on the types of the simple
factors the Lie algebra. 

\begin{defn} \label{dintropos1}
We say that the quasi-polynomial 
$f(n)$ is {\em  strictly positive}, 
if all coefficients of $f_j(n)$, for all $j$, are
nonnegative; i.e., the nonzero coefficients are positive.
In general, we define the {\em positivity index} $p(f)$ of $f$ to be 
the smallest nonnegative integer such that $f(n+p(f))$ is strictly  positive. 
We also say that $f(n)$ is positive with index $p(f)$. 
\end{defn}
Thus  $f(n)$ is strictly positive, iff its  positivity index is 
zero.

With this terminology, the hypothesis mentioned above is the following.
We say a connected reductive group  $H$ is {\em classical},
if each  simple factor of its Lie algebra ${\cal H}$ is of type 
$A,B,C$ or $D$. We also say that the type of $H$ or ${\cal H}$ is classical.

\begin{hypo} \label{hph2little} {\bf (PH2)}:  \cite{king,loera}
Assume that $H$ in Problem~\ref{pintrolittle} is classical. Then
the Littlewood-Richardson stretching quasi-polynomial $\tilde c_{\alpha,\beta}^\lambda(n)$ 
is strictly positive.
\end{hypo} 

We shall refer to this as the second positivity hypothesis (PH2).
This was conjectured by King, Tollu and Toumazet \cite{king} 
for type $A$, and 
De Loera and McAllister for types $B,C,D$.
Since the stretching function above is a polynomial in type $A$, 
the positivity conjecture of King et al clearly implies 
the  saturation theorem of Knutson and Tao. That is, PH2 implies SH for  type $A$.

We can formulate an analogue of SH for a Lie algerbra of arbitrary classical
type so that
PH2 implies SH for an arbitrary type. 
For this, we need to  formulate the
notion of a   saturated 
quasi-polynomial,  which is not contradicted by the 
counterexamples, mentioned above,    to verbatim translation of the saturation 
property in \cite{knutson,kirillov} to the  setting of quasi-polynomials. 
Specifically,
the notion of saturation in \cite{knutson,kirillov} works well
if the stretching function is a polynomial, but not so if 
it is a  quasipolynomial.
Let $f(n)$ be a quasi-polynomial with period $l$. Let 
 $f_j(n)$, $1\le j \le l$, be the polynomials
such that $f(n)=f_j(n)$ if $n=j$ mod $l$. 
The index of $f$, $\ind(f)$, is defined to be the smallest $j$ 
such that the polynomial $f_j(n)$ is not identically zero. If $f(n)$ is identically
zero, we let $\ind(f)=0$. 
If $f(1)\not =0$, then clearly $\ind(f)=1$. 

\begin{defn} \label{dintrosat}
We say that $f(n)$ is {\em  strictly saturated} if for any $i$:
$f_i(n) > 0$ for every $n\ge 1$ whenever $f_i(n)$ is not identically zero.
The {\em saturation index} $s(f)$ of $f$ is defined to be the smallest 
nonnegative integer such that $f(n+s(f))$ is strictly saturated. 
We also say that $f(n)$ is saturated with index $s(f)$.

\end{defn} 

Thus $f(n)$ is strictly saturated iff its saturation index is 
zero.
Clearly the saturation idex is bounded above by the positivity index.
Thus if $f(n)$ is strictly positive, it is strictly saturated.
Hence, PH2 (Hypothesis~\ref{hph2little}) implies:

\begin{hypo} \label{shlittle}
{\bf (SH)}: The 
Littlewood-Richardson stretching quasi-polynomial $c_{\alpha,\beta}^\lambda(n)$ of arbitary classical
type is strictly saturated.
\end{hypo} 

The polynomial time algorithm in \cite{GCT5} works assuming  SH as well.
For the Littlewood-Richardson coefficient 
of type $A$, the notion of strict saturation here 
coincides with the notion of saturation in  \cite{knutson} since $c_{\alpha,\beta}^\lambda(n)$
is a polynomial in that case.
Knutson and Tao \cite{knutson} also conjectured a generalized saturation property 
for arbitrary types. But that property,
unlike the one defined above,
is only  conjectured to be  sufficient, but
not claimed to be, or expected to be necessary. For this reason,
it cannot be used in the complexity-theoretic applications in this paper.

There is another   positivity conjecture for Littlewood-Richardson coefficients
that also implies  the saturation theorem 
of Knutson and Tao. For this consider
the generating function 
\begin{equation} \label{eqlittlerational}
C_{\alpha,\beta}^\lambda(t)=\sum_{n\ge 0} \tilde c_{\alpha,\beta}^\lambda(n) t^n.
\end{equation}
It is a rational function since $\tilde c_{\alpha,\beta}^\lambda(n)$ is a quasi-polynomial
\cite{stanleyenu}. 
For type $A$, if  $\tilde c_{\alpha,\beta}^\lambda(n)$ is not identically zero, then
$C_{\alpha,\beta}^\lambda(t)$  is a rational function of form
\begin{equation}\label{eqintro1} 
\f{h_d t^d + \cdots + h_0}{(1-t)^{d+1}},
\end{equation}
since $\tilde c_{\alpha,\beta}^\lambda(n)$ is a polynomial \cite{stanleyenu}.
It is  conjectured in \cite{king} that:

\begin{hypo} \label{hintrolittleph3} ({\bf PH3}:)
The coefficients $h_i$'s in eq.(\ref{eqintro1}) 
are nonnegative (and $h_0=1$).
\end{hypo}
We shall call this the third positivity hypothesis (PH3). 
It clearly implies SH for Littlewood-Richardson coefficients of type $A$.
To describe its analogue for arbitrary classical type we need a definition.

Let
$F(t)=\sum_n f(n) t^n$ be the generating function associated with the quasi-polynomial
$f(n)$. It is a rational function \cite{stanleyenu}. 

\begin{defn} \label{dintroposform}
We say that $F(t)$  has a
{\em   positive form}, 
if, when $f(n)$ is not identically zero,
it can be expressed in the form
\begin{equation} \label{eqpos2intro}
F(t)=\f{h_d t^d +\cdots + h_0}{\prod_{i=0}^k (1-t^{a_i})^{d_i}},
\end{equation} 
where (1) $h_0=1$, and $h_i$'s are nonnegative integers, (2) 
$a_i$'s and $d_i$'s are positive integers, 
(3) $\sum_i d_i=d+1$, where $d=\max{\deg(f_j(n))}$
 is the degree of $f(n)$.

We define the modular index of this  positive form to be $\max\{a_i\}$.
\end{defn}

If $F(t)$ has a   positive form with $a_0=1$,  then  
$f(n)$ is strictly saturated (Definition~\ref{dintrosat}); this easily
follows from the power series expansion of the right hand side of eq.(\ref{eqpos2intro}).

The analogue of Hypothesis~\ref{hintrolittleph3} for arbitrary classical 
type is:

\begin{hypo} \label{hintrolittleph3gen} ({\bf PH3}:)
The rational function $C_{\alpha,\beta}^\lambda(t)$ has a  positive form,
with $a^0=1$, 
of modular index bounded by a constant  depending only on the types of the simple factors 
of the Lie algebra of $H$. 
\end{hypo} 

This too implies SH for arbitrary classical type. For types $B,C,D$, 
the constant above is $2$.
Experimental evidence 
for this hypothesis is given in Section~\ref{sevilittle}.

The analogue of the PH3, even in the more general $q$-setting,
is known to hold for the generating 
function of the Kostant partition function of type $A$, and more generally, for
a parabolic Kostant partition function; cf. Kirillov  \cite{kirillov}. 
This also gives a support for the PH3 above,
given a close relationship between Littlewood-Richardson
coefficients and Kostant partition functions \cite{fultonrepr}.

\section{Back to the general decision problems}
It may be
remarked that the Littlewood-Richardson problem  actually never arises
in the flip. It is only used as a  simplest  proptotype of the actual (much harder)
problems that arise--namely relaxed forms of 
Problems~\ref{pintrokronecker}-\ref{pintrogit}.

Now we turn to these  problems.
The goal is to generalize the preceding results and
hypotheses for the Littlewood-Richardson coefficients to the structural constants that
arise in these problems. 
The problem of finding  a positive, combinatorial formula
for the plethysm constant (Problem~\ref{pintroplethysm}),
akin to the positive Littlewood-Richardson rule, has already been recognized 
as an  outstanding, classical  problem 
in  representation theory \cite{stanleypos}--the known formulae 
based  on  character theory  mentioned in Section~\ref{sdecision} 
are not positive, because they involve alternating
signs.
Indeed, existence of such a formula is  a part of the first positivity hypothesis (PH1)
below for the plethysm constant, and this problem is the main focus of the 
work  in 
\cite{GCT4,plethysm,canonical,algcomb}. 
In view of the intensive work on the plethym constant in the literature, it has now become 
clear that  the complexity of the plethysm problem (Problem~\ref{pintroplethysm}) 
is far higher than that of the Littlewood-Richardson problem 
(Problem~\ref{pintrolittle}).
This gap in the complexity is the main source of difficulties that has to be addressed.
We now state the main ingradients in the plan in this paper to show that 
the relaxed forms of
Problems~\ref{pintrokronecker}, \ref{pintroplethysm}, \ref{pintrosubgroup}, and
\ref{pintrogit}, with $X=G/P$ or a class variety, belong to $P$.

\section{Saturated and positive integer programming} \label{ssatpospgm}
First, we   formulate  a general algorithmic paradigm of 
saturated and 
positive integer programming that can be applied in the context of 
these problems.

Let $A$ be an $m\times n$ integer matrix, and $b$ an integral $m$-vector.
An integer programming problem asks if the polytope $P: A x \le b$ 
contains an integer point. In general, it is NP-complete.  
We want to define its relaxed version, which will turn out to have a polynomial
time algorithm.

We allow $m$, the number of constraints, to be exponential in $n$. Hence, 
we cannot assume that $A$ and $b$ are explicitly specified. Rather, 
it is assumed that the polytope $P$ is specified  in the form of a (polynomial-time) 
separation oracle in the spirit of Gr\"otschel, Lov\'asz
and Schrijver
\cite{lovasz}; cf. Section~\ref{sseporacle}.
Given a point $x \in \R^n$, the separation oracle tells if
$x \in P$, and if not, gives a hyperplane that separates $x$ from $P$.

Let $f_P(n)$ be the 
Ehrhart quasi-polynomial of $P$ \cite{stanleyenu}. 
By definition, $f_P(n)$ is the number of integer points in the
dilated polytope $n P$.

An integer programming problem 
 is called {\em saturated}, if 
\begin{enumerate} 
\item The specification of $P$ also contains a
number  $\sie(P)$, called the {\em saturation index estimate},
with the guarantee 
that the saturation index $s(f_P) \le \sie(P)$; cf. Definition~\ref{dintrosat}.
In particular, this
means $f_P(n+\sie(P))$ is strictly saturated. 
\item the goal of the problem is to give an efficient algorithm 
to decide if, given an integral
relaxation parameter $c > sie(P)$, if $c P$ contains an integer point.
\end{enumerate} 
The algorithm has to work only for  relaxation parameters $c > \sie(P)$.
In particular, if $\sie(P) \ge  1$, the algorithm 
problem does not have to determine if $P$ contains an integer point.

An integer programming problem 
 is called {\em positive}, if 
\begin{enumerate} 
\item  the specification of $P$ also contains 
a number  $\pie(P)$, called the {\em positivity index estimate},
with the guarantee 
that the positivity index $p(f_P) \le \pie(P)$; 
cf. Definition~\ref{dintropos1}.
 In particular, this
means $f_P(n+\pie(P))$ is strictly positive. 
\item the goal of the problem is to give an efficient algorithm 
to decide if, given an integral
relaxation parameter $c > \pie(P)$, if $c P$ contains an integer point.
\end{enumerate} 
Again, the algorithm has to work only for  relaxation parameters $c > \pie(P)$.
Since $s(f_P) \le p(f_P)$, a positive integer programming problem is also
saturated.

The following is the main complexity-theoretic result in this paper.

\begin{theorem} \label{tintrosat} (cf. Section~\ref{ssaturated}) 
\begin{enumerate} 
\item Index of the Ehrhart quasi-polynomial $f_P(n)$  of 
a  polytope $P$ presented by a separation oracle can be computed
in oracle-polynomial time, and hence, in polynomial time, assuming that the 
oracle works in polynomial time.
\item A saturated, and hence positive, 
 integer programming problem  has a polynomial
time algorithm.
\item Suppose  the polytopes $P$'s  that arise in a specific 
decision problem have the following property: whenever $P$ is nonempty,
the Ehrhart quasi-polynomial $f_P(n)$ is ``almost always'' strictly saturated.
Then there exists a polynomial time algorithm for deciding if
$P$ contains an integer point that works correctly ``almost always''.
\end{enumerate} 
\end{theorem} 
The meaning of the phrase ``almost always'' in the context of the
decision problems in this paper will be specified later 
(cf. Theorem~\ref{tindexquasi}).

It may be remarked that the index as well as the 
period of the Ehrhart quasi-polynomial can be exponential in the bit length
of the specification of $P$. In contrast to the polynomial time algorithm
above to compute the index, the known  algorithms   to compute the 
period (e.g. \cite{woods})
take time that is  exponential in the dimension of $P$. It may be conjectured 
that one cannot do much better: i.e., 
the period, unlike the index here,  cannot be computed 
in polynomial time, in fact, even 
in $2^{o(\dim(P))}$ time.

The algorithm  in Theorem~\ref{tintrosat} 
is based on the separation-oracle-based
linear programming algorithm of Gr\"otschel, Lov\'asz and Schrijver \cite{lovasz}, and
a polynomial time algorithm for computing the Smith normal form \cite{kannan}.

The paradigm of saturated  integer programming is useful 
when one knows, a priori, a good estimate for the saturation index of the
polytope under consideration, or when the saturation index is almost always 
zero. For example,
if $P$ is the hive polypolype for  the Littlewood-Richardson 
coefficient  (type A), then $\sie(P)=0$, 
by the saturation theorem \cite{knutson},
and $\pie(P)=0$, by PH2 (Hypothesis~\ref{hph2little}). 
For the polytopes $P$ that would arise in this paper, 
$\sie(P)$ and $\pie(P)$ would in general be nonzero, but conjecturally always
small, and $\sie(P)$ would conjecturally be almost always zero.

\section{Quasi-polynomiality, positivity hypotheses, and the canonical models} \label{sintroquasi}
The basic goal now is to use Theorem~\ref{tintrosat} to get polynomial time algorithms to decide
nonvanishing, modulo small relaxation, of the structural constants
 in Problems~\ref{pintrokronecker}, \ref{pintroplethysm},
\ref{pintrosubgroup} and ~\ref{pintrogit}, with
$X=G/P$ or a class variety. The main results in this paper which go towards this goal are
as follows.

\subsubsection*{Quasi-polynomiality}
We associate stretching functions with the structural constants in 
Problems~\ref{pintrokronecker}-\ref{pintrogit},
akin to the stretching function $\tilde c_{\alpha,\beta}^\lambda(n)$
in eq.(\ref{eqintrostretch1}) associated with the Littlewwod-Richardson coefficient,
and show that they are quasipolynomials; cf. Chapter~\ref{cquasipoly}.
(But their periods need not be constants, as in
the case of Littlewood-Richardson coefficients; in fact, they may be exponential in general.)
In particular, this proves Kirillov's conjecture \cite{kirillov}  for the plethysm constants.
The proof is an extension of Brion's remarkable proof (cf. \cite{dehy}) 
of quasi-polynomiality of the stretching
function  associated with the Littlewood-Richardson
coefficient. The main ingradient in the proof is Boutot's result \cite{boutot} 
that singularities of the  quotient of an affine variety with rational
singularities with respect to the action of a reductive group
are also rational. This is a generalization of an earlier result of Hochster and Roberts 
\cite{hochster}
in the theory of Cohen-Macauley rings.

\subsubsection*{Saturation and positivity hypotheses}
Using the stretching quasipolynomials above, 
we formulate (cf. Section~\ref{sphypo}) 
analogues of the saturation and positivity hypotheses SH, PH1,PH2,PH3 in 
Section~\ref{sintroLR} 
for the structural constants in Problems~\ref{pintrokronecker}-\ref{pintrosubgroup} and
Problem~\ref{pintrogit}, with $X=G/P$ or a class variety.
As for Littlewood-Richardson coefficients, it turns out that  PH2  implies SH.
The hypotheses PH1 and SH (more strongly, PH2) together imply
that the problem of deciding nonvanishing of the structural constant
in any of these
problems, modulo a small relaxation,   can be transformed in polynomial time into a saturated (more strongly, positive)
integer programming problem, and hence, can be solved
in polynomial time by Theorem~\ref{tintrosat}.
In particular, this shows that all the relaxed 
decision problems that arise in flip 
(cf. Hypothesis~\ref{PHfliprevised}) have polynomial time algorithms, assuming these positivity 
hypotheses.
Though these  algorithms are elementary,
the  positivity hypotheses on which their correctness depends turn out to be nonelementary. They
are intimately linked  to  the fundamental  phenomena in algebraic geometry  and
the theory of quantum groups, as we shall see.

We also give 
theoretical  and experimental results in support of these hypotheses; cf. 
Chapter~\ref{cquasipoly}-\ref{cevidence}.

\subsubsection*{Canonical models}
The proofs of quasi-polynomiality mentioned above
also associate with each structural constant under consideration a projective scheme, called the 
{\em canonical model}, whose Hilbert function coicides with the stretching quasi-polynomial
associated with that structural constant, akin to the model associated by Brion \cite{dehy} 
with the Littlewood-Richardson coefficient. These canonical models play a crucial role in
the approach to the posivity hypotheses suggested in Section~\ref{sapproach}.

\section{The plethysm problem} \label{sintroplethysm}
We now give precise statements of these results and hypotheses 
for the plethysm problem (Problem~\ref{pintroplethysm}). It
is the main prototype in this paper, which  illustrates  the basic ideas. 
Precise statements for the more general Problems~\ref{pintrosubgroup} and 
\ref{pintrogit} appear in Section~\ref{sphypo}.

As for the Littlewood-Richardson coefficients (cf.(\ref{eqintrostretch1})),
Kirillov \cite{kirillov} associates
with a plethysm constant $a_{\lambda,\mu}^\pi$ a 
stretching function 
\begin{equation} 
\tilde a_{\lambda,\mu}^\pi (n)=a_{n \lambda,\mu}^{n \pi},
\end{equation} 
and a generating function
\[ 
A_{\lambda,\mu}^\pi(t)=\sum_{n\ge 0} a_{n \lambda,\mu}^{n \pi}  t^n.
\]
(Note that $\mu$ is not stretched in these definitions.)

He  conjectured that  $A_{\lambda,\mu}^\pi(t)$ 
is a rational function. This is verified here in a 
stronger form:

\begin{theorem} \label{tquasiplethysm}
\noindent (a) (Rationality) The generating function 
$A_{\lambda,\mu}^\pi(t)$ is rational.

\noindent (b) (Quasi-polynomiality) The stretching function
$\tilde a_{\lambda,\mu}^\pi(n)$ is a quasi-polynomial function of
$n$. This is equivalent to saying that  all poles of $A_{\lambda,\mu}^\pi(t)$ are roots of
unity, and the degree of the numerator of $A_{\lambda,\mu}^\pi(t)$
is strictly smaller than that of
the denominator.

\noindent (c) 
There exist  graded, normal 
$\C$-algebras $S=S(a_{\lambda,\mu}^\pi)=\oplus_n S_n$, and   
$T=T(a_{\lambda,\mu}^\pi)=\oplus_n T_n$ such that:
\begin{enumerate} 
\item The schemes $\spec(S)$ and  $\spec(T)$
are normal and have rational singularities.
\item $T=S^H$, the subring of $H$-invariants in $S$, where $H=GL_n(\C)$ as in 
Problem~\ref{pintroplethysm},
\item The quasi-polynomial 
$\tilde a_{\lambda,\mu}^\pi(n)$ is the Hilbert function of $T$.
In other words, it is the Hilbert function of the homogeneous coordinate
ring of the projective scheme $\mbox{Proj}(T)$.
\end{enumerate}

\noindent (d) (Positivity) The rational function $A_{\lambda,\mu}^\pi(t)$
can be expressed in a positive form: 

\begin{equation} \label{eqtquasip}
A_{\lambda,\mu}^\pi(t)=\f{h_0+h_1 t+ \cdots+h_d t^d}{\prod_j(1-t^{a(j)})^{d(j)}},
\end{equation} 
where $a(j)$'s and $d(j)$'s are positive integers, 
 $\sum_j d(j)=d+1$, where $d$ is the degree of the quasi-polynomial 
$\tilde a_{\lambda,\mu}^\pi(n)$, 
$h_0=1$, and $h_i$'s  are nonnegative integers.
\end{theorem}

The specific rings $S(a_{\lambda,\mu}^\pi)$ and  $T(a_{\lambda,\mu}^\pi)$
 constructed in the proof of 
Theorem~\ref{tquasiplethysm} are very special. We call them {\em canonical rings} 
associated with the plethysm constant $a_{\lambda,\mu}^\pi$. We call 
$Y(a_{\lambda,\mu}^\pi)=\proj(S(a_{\lambda,\mu}^\pi))$,
and $Z(a_{\lambda,\mu}^\pi)=\proj(T(a_{\lambda,\mu}^\pi))$ the {\em canonical models} associated 
with  $a_{\lambda,\mu}^\pi$. The canonical rings are their homogenous coordinate rings.

It may be remarked that the analogue of Theorem~\ref{tquasiplethysm} (b)
for Littlewood-Richardson
coefficients has an elementary polyhedral proof. 
Specifically,  the 
 Littlewood-Richardson stretching function $\tilde c_{\alpha,\beta}^\lambda(n)$ of any type
is a quasi-polynomial since 
it coincides with the Ehrhart quasi-polynomial of the 
BZ-polytope \cite{berenstein}. 
Similarly, the analogue of Theorem~\ref{tquasiplethysm} (d) 
for Littlewood-Richardson coefficients
follows from Stanley's positivity theorem for the Ehrhart series of a 
rational polytope (which  is implicit in \cite{stanleytoric}).
These polyhderal proofs cannot be extended to the  plethysm constant at this
point, since no polyhedral expression for them is known so far--in fact,
this is a part of the conjectural positivity hypothesis PH1 below. 
In contrast, Brion's proof in \cite{dehy} of quasi-polynomiality of 
$\tilde c_{\alpha,\beta}^\lambda(n)$ can be extended to prove Theorem~\ref{tquasiplethysm}
since it does not need a polyhedral interpretation for $a_{\lambda,\mu}^\pi$. But
Boutot's result \cite{boutot} that it relies on is nonelementary (because it needs 
resolution of singularities in characteristic zero \cite{hironaka},
among other things). 
We also give an elementary (nonpolyhedral proof) for Theorem~\ref{tquasiplethysm} (a) 
(rationality). But this does not extend to a proof of quasipolynomiality for all $n$, which
turns out to be a far delicate problem. It is crucial in the context of saturated 
integer programming.

\begin{theorem} \label{tconeplethysm} (Finitely generated cone)

For a fixed partition $\mu$, 
let $T_\mu$ be the set of pairs $(\pi,\lambda)$  such that the irreducible
representation 
$V_\pi(H)$ of $H=GL_n(\C)$ occurs in the irreducible 
representation $V_\lambda(G)$ of $G=GL(V_\mu(H))$
with nonzero multiplicity. Then
$T_\mu$ is a finitely generated semigroup with respect to addition.
\end{theorem} 

This is proved by an extension of 
Brion and Knop's proof  of the analogous  result for Littlewood-Richardson
coefficients based on  invariant theory.
In the case of Littlewood-Richardson coefficients,
this again has an elementary polyhedral proof \cite{zelevinsky}.

\begin{theorem} \label{tpspaceplethysm} (PSPACE) 

Given partitions $\lambda,\mu,\pi$, the plethysm constant 
$a_{\lambda,\mu}^\pi$ can be computed in 
$\poly(\bitlength{\lambda},\bitlength{\mu},\bitlength{\pi})$ 
space.
\end{theorem}

The main observation in the proof of Theorem~\ref{tpspaceplethysm} is that 
the oldest algorithm for computing the plethysm constant,
based on the Weyl character formula, can be
efficiently parallelized so as to work in polynomial parallel time
using exponentially many processors. After this, the result follows
from the relationship between parallel and space complexity classes.
It may be remarked that the known algorithms for computing 
$a_{\lambda,\mu}^\pi$ in the literature--e.g., the one based on Klimyk's 
formula \cite{fultonrepr}--take exponential time as well as space.

Theorems~\ref{tquasiplethysm}, \ref{tconeplethysm} and \ref{tpspaceplethysm} lead to the following 
conjectural saturation and  positivity hypotheses for the plethysm constant.
These are analogues of PH1,PH2,PH3, SH in Section~\ref{sintroLR} for Littlewood-Richardson 
coefficients. 

\begin{hypo} {\bf (PH1)} \label{hph1plethysmintro}

For every $(\lambda,\mu,\pi)$ there exists a polytope 
$P=P_{\lambda,\mu}^\pi \subseteq \R^m$ such that:

\noindent (1)  The Ehrhart 
quasi-polynomial of $P$
coincides with the stretching quasi-polynomial $\tilde a_{\lambda,\mu}^\pi(n)$
in Theorem~\ref{tquasiplethysm}. (This  means  $P$ is given by a linear system of
the form 
\begin{equation} \label{eqph1hypoplethysm}
A x \le b,
\end{equation}
 where $A$ does not depend on $\lambda$ and $\pi$ and 
$b$ depends  only  on $\lambda$ and $\pi$ in  a homogeneous, linear
fashion.) In particular,
\begin{equation} \label{eqph1int}
a_{\lambda,\mu}^\pi=\phi(P),
\end{equation}
where $\phi(P)$   is equal to the number of integer points in $P$.

\noindent (2)  The dimension 
$m$ of the ambient space, and hence the dimension of $P$ as well, and 
the bitlength of every entry in $A$ 
are   polynomial 
in the bitlength of $\mu$ and the heights of $\lambda$ and $\pi$.

\noindent (3)  Whether a point $x \in R^m$ lies in $P$ can 
be decided in
$\poly(\bitlength{\lambda},\bitlength{\mu},\bitlength{\pi},\bitlength{x})$
time.
That is, the membership problem belongs to the complexity class $P$.
If $x$ does not lie in $P$, then this membership algorithm 
also outputs, in the spirit of \cite{lovasz}, the 
specification of a hyperplane separating
$x$ from $P$.
\end{hypo}

The first statement here,
in particular, would imply a {\em positive}, polyhedral  formula 
for $a_{\lambda,\alpha}^\mu$,  in the spirit of the known positive 
polyhedral formulae   for the Littlewood-Richardson coefficients
in terms of the BZ- \cite{berenstein},
 hive  \cite{knutson} or other types of polytopes \cite{dehy}.
It would also imply polyhedral proofs for Theorem~\ref{tquasiplethysm} (a), (b), (d),
and Theorem~\ref{tconeplethysm}.
Conversely, Theorem~\ref{tquasiplethysm} (a), (b), (d),
and Theorem~\ref{tconeplethysm} constitute a theoretical evidence for existence of 
such a positive polyhedral formula.

The second statement in PH1 is justified by Theorem~\ref{tpspaceplethysm}. Specifically,
it should be possible to compute the number of integer points in 
$P$ in PSPACE in view of Theorem~\ref{tpspaceplethysm}.
If $\dim(P)$ and $m$ were exponential, then the usual algorithms 
for this problem, e.g. Barvinok \cite{barvinok},   cannot be made to  work
in PSPACE.
Indeed, it may be conjectured that the number of integer points
in a general polytope $P\subseteq \R^m$  can not be computed in
 $o(m)$ space.

The number of constraints in the hive \cite{knutson} or the  BZ-polytope  \cite{berenstein} 
for the  Littlewood-Richardson coefficient $c_{\alpha,\beta}^\lambda$
is polynomial  in the number of parts of $\alpha,\beta,\lambda$. 
In contrast, the number of constraints defining $P_{\lambda,\mu}^\pi$ 
may  be exponential in the $\bitlength{\mu}$ and the number of
parts of $\lambda$ and $\pi$. But this is not a serious problem.
As long as the faces of the polytope $P$ have a nice description,
the third statement in PH1  is a reasonable 
assumption. This has been demonstrated in \cite{lovasz} for the 
well-behaved polytopes in combinatorial optimization with exponentially many
constraints. The situation in representation theory
should be similar, or even better. For example, 
the facets of  the hive polytope \cite{knutson} 
are far nicer than the facets of a typical polytope in
combinatorial optimization.

It is known that 
membership in a polytope is a ``very easy'' problem.
Formally, if a polytope has polynomially many constraints, 
this problem belongs to the complexity class  $NC \subseteq P$ \cite{karp}, 
the subclass of problems with efficient parallel algorithms, which is very low 
in the usual complexity hierarchy. Even if the number of
constraints of $P_{\lambda,\mu}^\pi$ in PH1 is exponential, the membership problem may still
be conjectured to  be in $NC$ (cf. Remark{rnc})--which would 
be ``very easy'' compared to the decision problem we began with (Problem~\ref{pintroplethysm}). 
For this reason, PH1 is primarily a mathematical positivity 
hypothesis as against  PHflip (Hypothesis~\ref{PHfliprevised}),
 and the positive, polyhedral   formula for
$a_{\lambda,\mu}^\pi$ in (\ref{eqph1int}) is its main content.

The remaining positivity hypotheses are purely mathematical. 
They generalize SH,PH2 and PH3 for the Littlewood-Richardson
coefficients to the  plethysm constants. We turn their specification next.
We can begin by asking  if the stretching quasipolynomial 
$\tilde a_{\lambda,\mu}^\pi(n)$ is strictly saturated or positive. 
This need not be so. The  recent article \cite{rosas} shows that 
strict saturation need not hold for the Kronecker coefficients,
as was conjectured in the earlier version of this paper.
A similar phenomenon was also reported in \cite{plethysm,canonical},
where it was observed that the structural constants of the 
nonstandard quantum groups associated with the  plethysm 
problem (of which the Kronecker problem is a special case) need not satisfy
an analogue of PH2. But it was observed there that
the positivity (and hence 
saturation) indices of these structural constants  are  small, though
not always zero; eg. see Figures 30,33,35 in \cite{canonical}. 
The same can be expected here. 
This is also supported by 
the experimental evidence in \cite{rosas2}
where too it may be observed that the positivity index is  small.
Furthermore, in the special case ($n=2$) of the Kronecker problem 
analysed in \cite{rosas2}, the saturation index is zero for almost
all Kronecker coefficients. 

These considerations suggest:

\begin{hypo} {\bf (SH)} \label{hshplethysmintro}

\noindent (a): The  saturation index  (Definition~\ref{dintrosat}) of 
$\tilde a_{\lambda,\mu}^\pi(n)$ is 
bounded by a polynomial in
the dimension of $G$ in Problem~\ref{pintroplethysm}
and the heights of $\lambda$ and $\pi$.
This means there exist absolute nonnegative constants $c$ and $c'$,
independent of $n,\lambda,\mu$ and $\pi$, such that the 
saturation 
index is bounded above by $c h^{c'}$, where 
$h=\dim{G}+\height{\lambda}+\height{\pi}$.

\noindent (b): The quasi-polynomial
$\tilde a_{\lambda,\mu}^\pi(n)$ is strictly saturated, i.e. the
saturation index is zero,  for almost all $\lambda, \mu$, $\pi$. 
Specifically, the density of the triples $(\lambda,\mu,\pi)$ 
of total bit length $N$ with nonzero $a_{\lambda,\mu}^\pi$
for which the saturation index is not zero
is less than $1/N^{c''}$, for any positive constant $c''$, as 
$N\rightarrow \infty$. 
\end{hypo}

A stronger form of (a) is:

\begin{hypo} \label{phph2nonstdplethysmrevised}
{\bf (PH2)}
The  positivity index (Definition~\ref{dintropos1}) of 
the stretching  quasi-polynomial 
$\tilde a_{\lambda,\mu}^\pi(n)$ is bounded by a polynomial in
the dimension  of $G$ and the heights of $\lambda$ and $\pi$.
\end{hypo}

The following is another stronger form of SH (a).
For this, we observe that 
the positive rational form in Theorem~\ref{tquasiplethysm} (d) is not unique.
Indeed, there is one such form  for every  h.s.o.p. (homogeneous sequence of parameters)
of the homogenenous coordinate ring $S$; the $a(j)$'s in eq.(\ref{eqtquasip}) 
are the degrees of these parameters. 

Kirillov    asked if the only possible 
pole of $A_{\lambda,\mu}^\pi$ is at $t=1$--i.e. if 
$a_{\lambda,\alpha}^\mu(n)$   is a polynomial.
This  is not so (cf. Section~\ref{sevikron}). But it may be  conjectured that the 
structural constants $a(j)$'s are small. Specifcally, 
consider an h.s.o.p. of $S$  with a (lexicographically) minimum degree sequence, and
call the (unique) positive rational form in Theorem~\ref{tquasiplethysm} (d)
associated with such an h.s.o.p. {\em minimal}. 
The modular index $\chi(a_{\lambda,\mu}^\pi)$ of the plethysm constant
is defined to be the modular index (Definition~\ref{dintroposform})
of this minimal
positive form.
Then:

\begin{hypo} {\bf (PH3)} 
\label{hPH3plethysmminimal}

The function $A_{\lambda,\mu}^\pi(t)$ associated with $a_{\lambda,\mu}^\pi$
has a positive rational form with modular index bounded 
by a polynomial 
in the dimension of $G$ and the heights of  $\lambda$ and $\pi$.

More specifically, this is so for the minimial positive rational form
of $A_{\lambda,\mu}^\pi(t)$ as above; i.e., 
the  modular index $\chi(a_{\lambda,\mu}^\pi)$
is itself bounded by a polynomial 
in the dimension of $G$ and the heights of  $\lambda$ and $\pi$.
\end{hypo}

This is a conjectural analogue of  a stronger form
of  PH3 for Littlewood-Richardson
coefficients (Hypothesis~\ref{hintrolittleph3}),  which says that
the  modular index of a Littlewood-Richardson coefficient, defined similarly,
is one. 
PH3 here would imply that the period of 
$A_{\lambda,\mu}^\pi(t)$ is smooth--i.e. has small prime factors--though it may be
exponential in the heights of $\lambda,\mu,\pi$.
It can be shown that PH3 implies SH (a)  (Section~\ref{sphypo}). 

The following result addresses the second arrow in Figure~\ref{fbasicintro}
in the context of the relaxed decision problem for the plethysm constant:

\begin{theorem} 
The complexity theoretic positivity hypothesis PHflip 
(Hypothesis~\ref{PHfliprevised}) 
for the plethysm constant is implied by 
the mathematical positivity hypotheses PH1 and SH above. Specifically,
assuming PH1 and SH:

\noindent (a) 
Nonvanishing of $a_{b \lambda, b \mu}^{b \pi}$
for any $b> c h^{c'}$, with $c,c',h$ as in SH, 
can be decided in 
$O(\poly(\bitlength{\lambda},\bitlength{\mu}, \bitlength{\pi},\bitlength{b}))$
 time. 

\noindent (b) There is 
an $O(\poly(\bitlength{\lambda},\bitlength{\mu}, \bitlength{\pi}))$ time
algorithm for deciding if $a_{\lambda,\mu}^\pi$ is nonvanishing,
which works correctly on almost all $\lambda,\mu$ and $\pi$; almost all
means the same as in SH.
\end{theorem} 

Here (a)  follows by applying  Theorem~\ref{tintrosat} (2) 
to the polytope $P_{\lambda,\mu}^\pi$ in PH1, and 
letting the positivity index estimate for this polytope
be $c h^{c'}$;
(b) follows from Theorem~\ref{tintrosat} (3).

\subsection*{Evidence for the positivity hypotheses in special cases}
Littlewood-Richardson coefficients are special cases of (generalized) plethym constants.
We have already seen that PH1 holds in this case, and that there is considerable 
experimental evidence for PH2 and PH3 (Section~\ref{sintroLR}).
Another crucial special case of the plethym problem is the Kronecker problem 
(Problem~\ref{pintrokronecker})--in fact, this may be considered to be the crux of the plethysm
problem.
It follows from the results  in \cite{algcomb} that  PH1 
holds for the Kronecker problem when $n=2$;
the earlier known formulae \cite{remmel,rosas} for the Kronecker coefficient 
in this case are not positive. It can also be seen from the experimental 
evidence  in \cite{rosas2} that the saturation and positivity indices
of the Kronecker coefficient, for $n=2$, are very small, and
almost always zero.
%Experimental evidence for PH2 and PH3 in this case is given in Section~\ref{sevikron}.
We also give in Chapter~\ref{cevidence} additional experimental evidence for PH2 for 
another basic special case of Problem~\ref{pintrosubgroup}, with $H$ therein
being the symmetric group.

\section{Towards PH1, SH, PH2,PH3 via
canonial  bases and canonical models} \label{sapproach}
In this section,
we suggest an approach to prove PH1, SH, PH2 and PH3  for the plethysm constant
and the analogous hypotheses for the other structural constants in Problems~\ref{pintrosubgroup},
and \ref{pintrogit}, with $X=G/P$ or a class variety.
In the case of Littlewood-Richardson coefficients of type A, PH1 and SH have 
purely combinatorial proofs.
But it seems  unrealistic to expect such proofs of the saturation and
positivity  hypotheses for the plethysm and other  structural
constants under consideration here given their  substantially higher complexity.

The approach that we suggest is motivated by the proof of PH1 for Littlewood-Richardson
coefficients of arbitrary types based on  the  canonical (local/global crystal) bases 
of Kashiwara and Lusztig 
for representations of  Drinfeld-Jimbo quantum groups
\cite{dehy,kashiwara2,littelmann,lusztigcanonical,lusztigbook}. 
By a Drinfeld-Jimbo quantum group we shall mean in this paper quantization
$G_q$ of a complex, semisimple group $G$
as in \cite{rtf} that is dual to the Drinfeld-Jimbo quantized enveloping algebra \cite{drinfeld}. 
Canonical bases for representions of a Drinfeld-Jimbo quantum group in type $A$ 
are intimately linked \cite{gro}  to the Kazhdan-Lusztig basis
for Hecke algebras \cite{kazhdan,kazhdan1}. A starting point for the approach suggested here is:

\begin{obs} {\bf (PH0)} \label{obslusztig}
The homogeneous coordinate rings of the canonical models
associated 
by Brion with the Littlewood-Richardson coefficients
have quantizations endowed with canonical bases as per Kashiwara and Lusztig.
\end{obs}
This is  a consequence of the work of Kashiwara \cite{kashiwaraglobal} and 
Lusztig \cite{lusztigpnas,lusztigbook}; see  Proposition~\ref{plusztigkashi} for its precise
statment.
This is why we call the models here  canonical models.

We shall refer to the  property above as the zeroeth positivity hypothesis PH0.
Positivity here refers to the deep characteristic
positivity property of the  canonical basis  proved by Lusztig:
namely its multiplicative and comultiplicative  structure constants are nonnegative.
For this
reason, we say  that a canonical basis is  positive.
Similar positivity property  is also known for the Kazhdan-Lusztig basis
\cite{kazhdan1}. 
The proofs of these positivity properties are 
based on  the Riemann hypothesis over finite fields (Weil conjectures) \cite{weil2} 
and the related 
work of Beilinson, Bernstein, Deligne \cite{beilinson}. 

The property above is called PH0 because it implies PH1 for Littlewood-Richardson 
coefficients of arbitrary types. Specifically, the latter 
 is a formal consequence of 
the abstract  properties of these canonical bases and is intimately related to their
positivity; cf. Section~\ref{ssignilittle}, 
and \cite{dehy,kashiwara2,littelmann,lusztigbook}. The saturation hypothesis 
SH in type A \cite{knutson} is a refined property of the  polyhedral formulae in PH1. In Section~\ref{scanonicalmodel}
 we suggest an approach to prove SH, PH2 and PH3 
for arbitrary types based on the properties of these canonical bases.
All this  indicates that  for the Littlewood-Richardson problem
PH1,  SH, PH2 and PH3  are intimately linked to PH0.

\begin{figure}
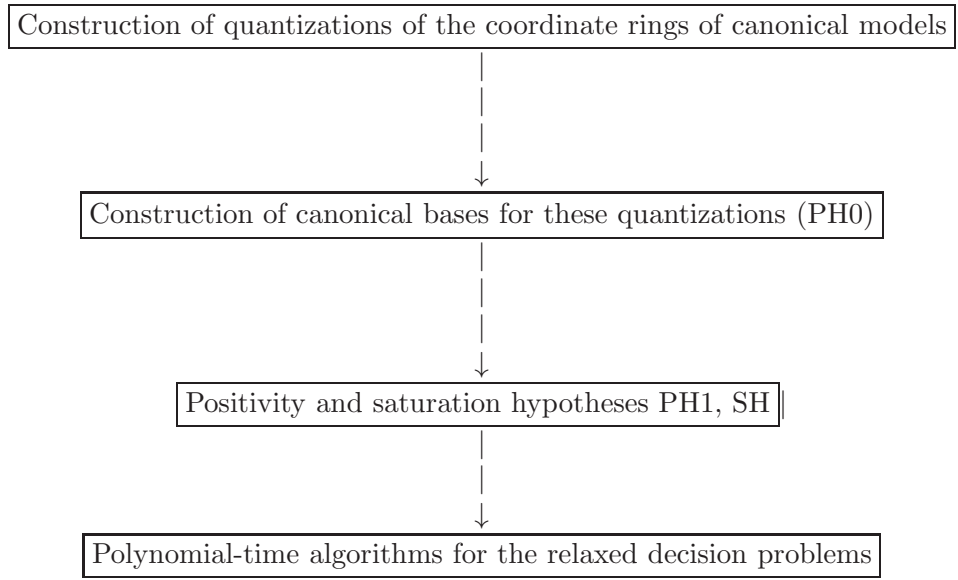
 
\[ \begin{array}{c}
\fbox{Construction of quantizations of the coordinate rings of canonical models}\\
 | \\
 | \\
 | \\
\downarrow \\
\fbox{Construction  of  canonical bases for these quantizations (PH0)} \\
 | \\
 | \\
 | \\
\downarrow \\
\fbox{Positivity and saturation hypotheses PH1, SH} 
 | \\
 | \\
 | \\
\downarrow \\
\fbox{Polynomial-time algorithms for the  relaxed decision problems}
\end{array} \]
\caption{Pictorial depiction of the approach}
\label{fbasicapp}
\end{figure}

This suggests the following approach for proving PH1, SH, PH2 and PH3
for the
plethysm and other structural constants under consideration in this paper
(cf. Section~\ref{sph0general}):
\begin{enumerate} 
\item Construct quantizations of the  homogeneous coordinate rings of the
canonical models associated with
these structural constants,
\item Show that they have  canonical bases in some appropriate sense 
thereby extending 
 PH0 to this general setting.  
\item Prove PH1, SH, PH2, and PH3
  by a detailed analysis and study 
of these  canonical   bases as per this extended PH0, just as in the
case of Littlewood-Richardson coefficients.
\end{enumerate} 
Pictorially, this is depicted in Figure~\ref{fbasicapp}. 

Quantizations  of the homogeneous coordinate rings of the
canonical models associated with Littlewood-Richardson coefficients and their positive 
canonical bases are constructed using 
the theory Drinfeld-Jimbo quantum group. In  type $A$, it is intimately related to the 
theory of Hecke algebras.
But, as expected, the  theories of Drinfeld-Jimbo quantum groups and Hecke algebras 
do not work for the plethysm problem. 
What is needed is a quantum group and a quantized algebra 
 that can play the same role in the plethysm problem
that the Drinfeld-Jimbo quantum group  and the Hecke algebra 
play in the Littlewood-Richardson problem.
These   have been 
constructed in \cite{GCT4} for the Kronecker problem (Problem~\ref{pintrokronecker})
and in \cite{plethysm} for the generalized plethysm problem (Problem~\ref{pintroplethysm}). We shall call them {\em nonstandard quantum groups} and
{\em nonstandard quantized algebras};
cf. Section~\ref{squantumgroup} for their  brief overview.
In the special case of the Littlewood-Richardson problem, these specialize to  the 
Drinfeld-Jimbo quantum group and the Hecke algebra, respectively.
The article  \cite{canonical} gives conjecturally 
correct algorithms to construct canonical bases of 
the matrix coordinate rings of the nonstandard   quantum groups
and of nonstandard algebras that have conjectural positivity 
properties  analogous to those of the 
canonical (global crystal)  bases, as per Kashiwara and Lusztig,
of  the coordinate ring of the Drinfeld-Jimbo quantum group,
and the Kazhdan-Lusztig basis of
the Hecke algebra. 
These conjectures lie at
the heart of the approach suggested here, since they are crucial for the
extension of PH0 (cf. Figure~\ref{fbasicapp}) to the general setting here.
Their verification
seems to need substantial extension  of the work surrounding the Riemann 
hypothesis over finite fields mentioned above.

\section{Basic plan for implementing  the flip} 

The main application of the results and hypotheses in this paper
in the context of 
the flip is the following result. As mentioned in Section~\ref{sdecision},
and described in more detail in Sections~\ref{sdetvsperm}-\ref{spvsnp}, 
each    lower bound problem, such as the  $P$ vs. $NP$ problem over $\C$,
is reduced in \cite{GCT1,GCT2} to the problem
of proving {\em obstructions} to
embeddings among the class varieties that arise in the problem.
In Chapter~\ref{cobstruction}
 we define  a {\em robust obstruction}, which is an
obstruction that is well behaved with respect 
to  relaxation, and whose validity (correctness)
depends only on an appropriate PH1 but not SH. It is conjectured that
in each of the lower bound problems under consideration, robust obstructions
exist (Section~\ref{srobustex}).
In the lower bound problems under consideration, ultimately
one is only interested in proving
existence of obstructions. So one may as well search for only robust 
obstructions.

\begin{theorem} (cf. Chapter~\ref{cobstruction}) \label{tflipapplication} 
Consider the $P$ vs. $NP$ or the $NC$ vs. $P^{\#P}$ problem over 
$\C$ \cite{GCT1}. Assume that the homogeneous coordinate rings
of the  relevant class varieties  \cite{GCT1,GCT2} 
in this  context have rational singularities. Also assume that the
structural constants associated with these class varieties
satisfy analogous PH1 as specified in
Chapter~\ref{cobstruction}. Then:

\noindent (a) The problem of verifying a robust  obstruction 
in each of these problems  belongs to $P$, so also the relaxed form
of the problem of verifying any obstruction (not necessarily robust).

\noindent (b) There exists an explicit family of robust obstructions 
 in each of these problems assuming an
additional hypothesis OH specified 
in Chapter~\ref{cobstruction}; the meaning of the term explicit is 
also given there.

\noindent (b) The problem of deciding existence of a  geometric obstruction 
also belongs to $P$, assuming a stronger form of PH1 specified 
in Chapter~\ref{cobstruction}.
Here geometric obstruction is a simpler  type of  robust
obstruction,
defined in Chapter~\ref{cobstruction}, which is conjectured to exist
in the lower bound problems under consideration.
\end{theorem}
For a precise statement of this theorem, see Chapter~\ref{cobstruction}.

This theorem needs only PH1, but not SH, which 
 is only needed to argue why
robust obstructions should exist (Section~\ref{srobustex}), and 
furthermore, it is only needed  for Problems
\ref{pintrokronecker}-\ref{pintrosubgroup} and not for
the GIT Problem \ref{pintrogit}. 
Thus PH1 is the main positivity hypothesis of
GCT in the context proving existence of (robust) obstructions for the 
lower bound problems under consideration.

\begin{figure}[!p]
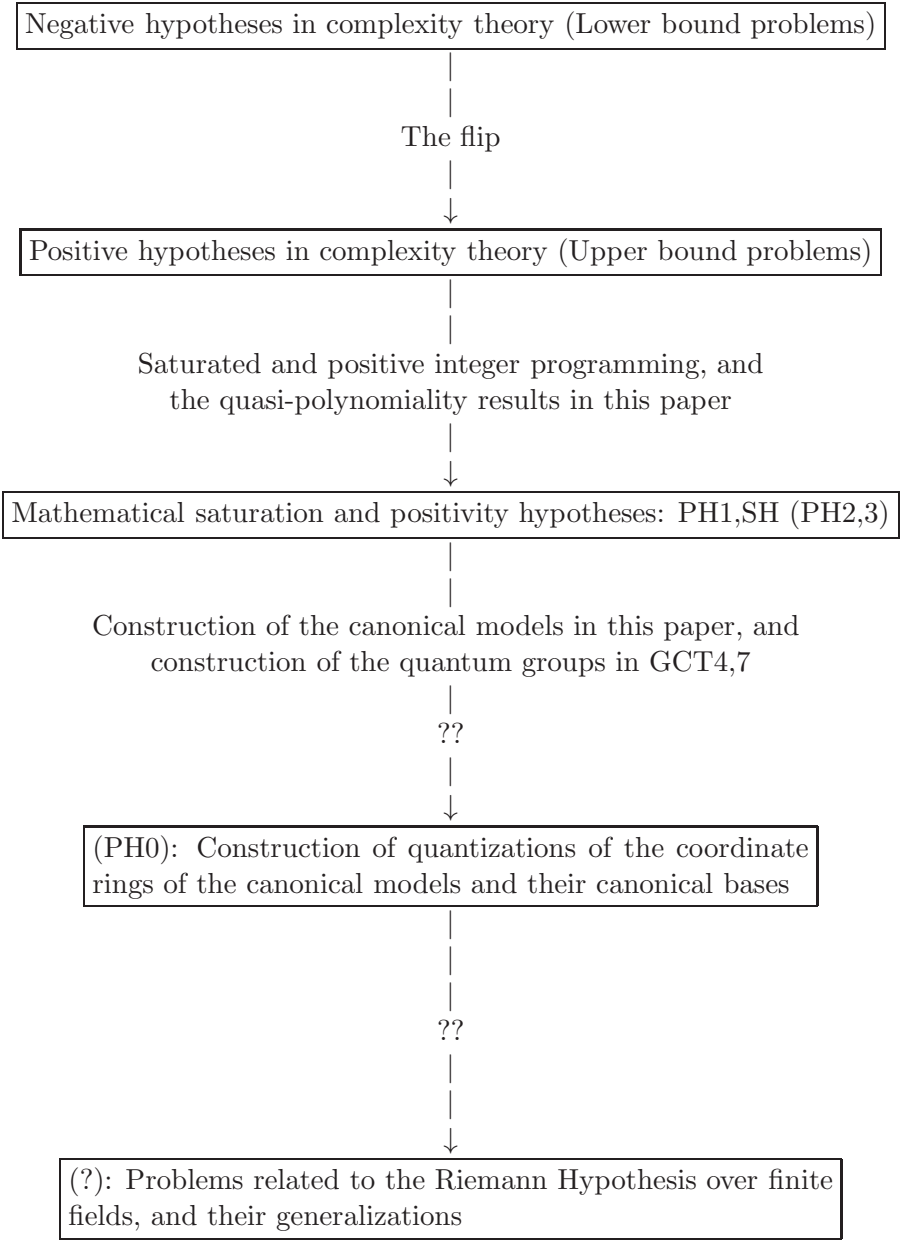
 
\centering
\[\begin{array} {c} 
\fbox{Negative hypotheses in complexity theory  (Lower bound problems)} \\
|\\
| \\
\mbox{The flip}\\
|\\
\downarrow \\
\fbox{Positive hypotheses in complexity theory (Upper bound problems)} \\
|\\
| \\
\mbox{Saturated and positive integer programming, and} \\
\mbox{the quasi-polynomiality results in this paper}\\
|\\
\downarrow \\
\fbox{Mathematical saturation and positivity  hypotheses: PH1,SH (PH2,3)} \\
|\\
|\\
\mbox{Construction of the canonical models in this paper, and }\\
\mbox{construction of the quantum groups in GCT4,7}\\
|\\
?? \\
|\\
\downarrow \\
\fbox{\parbox{3.75in}{(PH0): Construction  of
quantizations of the 
coordinate rings of the canonical models and their canonical bases}} \\
|\\
|\\
|\\
?? \\
|\\
|\\
\downarrow \\
\fbox{\parbox{4in}{(?): Problems related to 
 the Riemann Hypothesis over finite fields, and their generalizations}}
\end{array}\]
\caption{A  basic plan  for implementing  the flip} 
\label{fflip}
\end{figure}

A basic plan for implementing the flip suggested by the considerations above is 
summarized in Figure~\ref{fflip}. It is an elaboration of Figure~\ref{fbasicintro}.
Question marks  in the figure indicate open problems.

\section{Organization of the paper}
The rest of this paper is organized as follows.

In Chapter~\ref{cprelim} we describe the
basic complexity theoretic notions that we need in this paper and 
describe their significance in the context of representation theory.

In Chapter~\ref{csatpos}, we give a polynomial time algorithm for saturated integer programming
(Theorem~\ref{tintrosat}), and give precise 
statements of the  results 
and positivity hypotheses for
Problems~\ref{pintrosubgroup} and \ref{pintrogit} (with $X=G/P$ or a class variety) 
mentioned in Section~\ref{sintroquasi}. These 
generalize the ones given
in Section~\ref{sintroplethysm} for the plethysm constant.
The framework of saturated integer programming in this paper may be applicable to many
other structural constants in representation theory and algebraic geometry, 
such as the   Kazhdan-Lusztig polynomials (cf. Sections~\ref{sother}).

In Chapter~\ref{cquasipoly}, we prove 
the basic quasi-polynomiality results--Theorem~\ref{tquasiplethysm} and its generalizations for
Problems~\ref{pintrosubgroup} and \ref{pintrogit}. We also define 
canonical models for the structural constants under consideration,
and briefly describe the relevance of
the nonstandard quantum groups and the related
results in \cite{GCT4,plethysm,canonical} 
in the context of quantizing the coordinate rings of these canonical models
and  extending PH0 to them (Figure~\ref{fbasicapp}).

In Chapter~\ref{cpspace}, we prove the basic PSPACE results--Theorem~\ref{tpspaceplethysm}
 and its extensions for
the various cases of Problem~\ref{pintrosubgroup}.

In Chapter~\ref{cevidence}, we give experimental evidence for the positivity hypotheses PH2 and 
PH3 in some  special cases of the Problems~\ref{pintrokronecker}-\ref{pintrogit}.

In Chapter~\ref{cobstruction}, 
we   describe an 
application (Theorem~\ref{tflipapplication}) of the results and
positivity hypotheses in this paper to  the problem of verifying 
or discovering a  {\em robust  obstruction}, i.e., a ``proof of
hardness'' \cite{GCT2}
in the context of the $P$ vs. $NP$ and the permanent vs. determinant 
 problems in characteristic zero.

\section{Notation}
We let  $\bitlength{X}$ denote the total bitlength of the specification of $X$. Here $X$
can be an integer, a partition, a classifying 
label of an irreducible representation of a reductive group,
a polytope, and so on. The exact meaning of $\bitlength{X}$ will be clear from  the context.
The notation $\poly(n)$ means $O(n^a)$, for some constant $a$.
The notation $\poly(n_1,n_2,\ldots)$ similarly means bounded by 
a polynomial of a constant degree in $n_1,n_2,\ldots$.
Given a reductive group $H$, $V_\lambda(H)$ denotes the irreducible representation of 
$H$ with the classifying label $\lambda$. The meaning  depends on $H$. 
Thus if $H=GL_n(\C)$, $\lambda$ is a partition and $V_\lambda(H)$ the Weyl module
indexed by $\lambda$, if $H=S_m$, then $\lambda$ is a partition of size $|\lambda|=m$,
and $V_\lambda(H)$ the Specht module indexed by $\lambda$, and so on.

\chapter{Preliminaries in complexity theory}  \label{cprelim}
In this chapter, we recall  basic definitions in complexity theory,
introduce  additional ones, and illustrate 
their significance in the context of representation theory.

\section{Standard complexity classes} \label{sstandard}
As usual, $P$, $NP$ and $PSPACE$  are the classes of problems that 
can be solved
in polnomial time,  nondeterministic polynomial time, and polynomial space,
respectively. The class of functions that can be computed in polynomial 
time (space)  is sometimes denoted by $FP$ (resp. $FPSPACE$).
But, to keep the notation simple,
we shall denote these  classes by $P$ and $PSPACE$ again.

Let $SPACE(s(N))$ denote the class of problems that can 
be solved in $O(s(N))$ space on inputs of bit length $N$; by convention $s(N)$
counts only the size of the work space. In other words, 
the size of the input, which
is on the read-only input tape, and the output, which is on the write-only
output tape is not counted. Hence $s(N)$ can be less than the size of the input
or the output, even
logarithmic compared to these sizes. The class $space(\log(N))$ is denoted by LOGSPACE.

An algorithm is called strongly polynomial \cite{lovasz}, if given an
input $x=(x_1,\ldots,x_k)$, 
\begin{enumerate} 
\item the total number of arithmetic steps ($+,*,-$ and comparisones) in
the algorithm is polynomial in $k$, the total number of input parameters,
but does not depend $\bitlength{x}$, where  $\bitlength{x}=\sum_i \bitlength{x_i}$ denotes the
bitlength  of $x$.
\item
the bit length of every intermediate operand in the computation is polynomial
in $\bitlength{x}$.
\end{enumerate} 
Clearly, a strongly polynomial algorithm is also polynomial.
let strong $P \subseteq P$ denote the subclass of problems with strongly polynomial time
algorithms.

\ignore{Similarly,  a strongly polynomial space algorithm means 
\begin{enumerate} 
\item the total number of intermediate operands (in the work space) 
is polynomial in $k$, the total number of input parameters,
but does not depend on the bit lengths of these parameters, and
\item
the bit length of every intermediate operand in the computation is polynomial
in the total bit length $||a||$ of the input.
\end{enumerate} 
Let $strongPSPACE \subseteq PSPACE$ be the subclass of problems having 
strongly polynomial space algorithms.
}

The counting
class associated with $NP$ is denoted by $\#P$. Specifically, a function 
$f:\N^k\rightarrow \N$, where $\N$ is the set of nonnegative integers,
is in $\#P$ if it has a formula of the 
form:
\begin{equation} \label{eqsharpp1}
f(x)=f(x_1,\cdots,x_k)=\sum_{y \in \N^l} \chi(x,y),
\end{equation}
where $\chi$ is a polynomial-time computable function that takes values
$0$ or $1$, and $y$ runs over all tuples such that 
$\bitlength{y}=\poly(\bitlength{x})$.  
The formula (\ref{eqsharpp1}) is called a $\#P$-formula. 
An important feature of a $\#P$-formula in the context of representation theory is 
that it is {\em positive}; i.e., it does not contain any alternating signs. 

The formula (\ref{eqsharpp1}) is called a  strong $\#P$-formula, if, in addition,
$l$ is polynomial in $k$ and $\chi$ is a strongly polynomial-time 
computable function. 
Let strong $\#P$ be the class of functions 
with strong $\#P$-fomulae.

It is known and easy to see that
\begin{equation} 
\#P \subseteq PSPACE.
\end{equation}

\subsection{Example: Littlewood-Richardson coefficients} 
By  the Littlewood-Richardson rule  \cite{fultonrepr}, the coefficient 
$c_{\alpha,\beta}^\lambda$ (cf. Problem~\ref{pintrolittle}) in type $A$  is given by: 
\begin{equation} \label{eqdefnlittle1} 
c_{\alpha,\beta}^\lambda=\sum_T \chi(T),
\end{equation} 
where $T$ runs over all numbering of 
the  skew   shape $\lambda / \alpha$, and
$\chi(T)$ is $1$ if $T$ is a Littlewood-Richardson skew tableau of  content 
$\beta$, and 
zero, otherwise.
The total number
of entries in $T$ is quadratic  in the total number of nonzero parts in
$\alpha,\beta,\lambda$, and the number of arithmetic steps 
needed to compute  $\chi(T)$ is  linear in this total number. 
Hence (\ref{eqdefnlittle1})  is a strong $\#P$-formula, and Littlewood-Richardson
function $c(\alpha,\beta,\lambda)=c_{\alpha,\beta}^\lambda$
belongs to strong $\#P$. 
It may be remarked that the character-based formulae for the Littlewood-Richardson
coefficients   are not $\#P$-formulae, since they involve 
alternating signs. But the algorithms 
based on the these  formulae for computing Littlewood-Richardson
coefficients  run in polynomial space.
Thus, from the perspective of  complexity theory,
the main significance of the Littlewood-Richardson rule is  that it puts
the problem, which at the surface is only in $PSPACE$, in its  smaller subclass
(strong) $\#P$.

Though the Littlewood-Richardson rule is  often called efficient 
in the representation
theory literature, it is not really so from the perspective of
complexity theory.
Because  computation of $c_{\alpha,\beta}^\lambda$ using this  formula 
takes  time that is exponential in both the total
number of parts of $\alpha,\beta$ and $\lambda$, and their bit lengths.
This is inevitable, 
since this problem is $\#P$-complete \cite{hari}. Specifically, this means
there is no polynomial time algorithm to compute $c_{\alpha,\beta}^\lambda$,
assuming $P\not = NP$. 

As remarked in earlier,
nonzeroness (nonvanishing) of $c_{\alpha,\beta}^\lambda$ can be decided 
in $\poly(\bitlength{\alpha},\bitlength{\beta},\bitlength{\lambda})$ time;
\cite{loera,GCT3,knutson}.
Furthermore, the algorithm in \cite{GCT3} is strongly polynomial; i.e., the 
number of arithemtic steps in this algorithm is 
a polynomial in the total number of parts
of $\alpha,\beta,\lambda$, and does not depend on the bit lengths of
$\alpha,\beta,\lambda$. Hence the problem of deciding nonvanishing of
$c_{\alpha,\beta}^\lambda$ (type $A$) belongs to strong $P$.

The discussion above shows that the Littlewood-Richardson problem 
is akin to the problem of computing the permanent of an integer matrix with 
nonnegative coefficients. The latter  is known to be $\#P$-complete \cite{valiant}, but its
nonvanishing can be decided in polynomial time, using the polynomial-time algorithm
for finding a perfect matching in bipartite graphs \cite{schrijver}.
If the  positivity hypotheses in this paper hold, the
situation would be  similar for many fundamental structural constants in
representation theory and algebraic geometry in a relaxed sense.

\section{Convex $\#P$} \label{sconvexsharpp}
Next we want to introduce a subclass of $\#P$ called convex $\#P$.

Given a polytope $P \subseteq R^l$, let $\chi_P$ denote the 
characteristic (membership) function of $P$: i.e., 
 $\chi_P(y)=1$, if $y\in P$, and zero  otherwise.
We say that $f=f(x)=f(x_1,\ldots,x_k)$
has a convex $\#P$-formula if, for every 
$x \in \Z^k$, there exists a convex polytope (or, more generally,
a convex body)
$P_x \subseteq \R^l$, such that 
\begin{enumerate} 
\item The membership  function 
$\chi_{P_x}(y)$ can be computed in $\poly(\bitlength{x},\bitlength{y})$
time,  each integer point in 
$P_x$ has $O(\poly(\bitlength{x}))$ bitlength, and 
\item 
\begin{equation} \label{eqconvex0}
f(x)=\phi(P_x),
\end{equation} 
where $\phi(P_x)$ denotes the number of integer points in
$P_x$. Equivalently, 
\begin{equation} \label{eqconvex1}
f(x)=\sum_{y\in \Z^l} \chi_{P_x}(y),
\end{equation}
where $y$ runs over tuples in $\Z^l$ of $\poly(\bitlength{x})$
bitlength, and 
$\chi_{P_x}$ denotes the membership  function of the polytope 
$P_x$.
\end{enumerate} 
Equation (\ref{eqconvex1}) 
is similar to eq.(\ref{eqsharpp1}).  The main difference 
is that $\chi$ is now the membership  function of a convex polytope.
Clearly, eq.(\ref{eqconvex1}), and hence, eq.(\ref{eqconvex0})  is
a $\#P$-formula, when $\chi_{P_x}$ can be computed in polynomial time.
Let convex $\#P$ be the subclass of $\#P$ consisting of functions with 
convex $\#P$-formulae.

We say that eq.(\ref{eqconvex0}) is a strongly convex $\#P$-formula, 
if the characteristic function of $P_x$ is computable in strongly polynomial time.
Let strongly convex $\#P$ be the subclass of $\#P$ consisting of functions with 
strongly convex $\#P$-formulae.

We do not assume in eq.(\ref{eqconvex0}) that the polytope $P_x$ is 
explicitly  specified by its defining constraints. Rather, we only assume,
following \cite{lovasz}, 
that we are given a computer  program, called a {\em  membership oracle}, which, given
input parameters $x$ and $y$, 
tells  whether $y \in P_x$ in $\poly(\bitlength{x},\bitlength{y})$  time. 

If the number of  constraints defining $P_x$ 
is polynomial in $\bitlength{x}$, 
then it is possible to specify $P_x$ by simply writing down these
constraints.  In this case the membership question can be trivially decided
in polynomial time--in fact, even in LOGSPACE--by verifying each constraint one at a time. 
This would not work if $P_x$ has exponentially many constraints.
In good cases, 
it is possible to answer the membership question in
polynomial time even if $P_x$ has exponentially many facets.
Many such examples in combinatorial optimization are given in \cite{lovasz}.
One such illustrative example in representation theory is given in Section~\ref{slittlecone}.
The polytopes that would arise in the plethysm 
 and other problems of main interest in this paper 
are also expected to be of this kind.

We now illustrate the notion of convex $\#P$ with a few 
examples in representation theory.

\subsection{Littlewood-Richardson coefficients}  \label{sconvexlittle}
A geeneralized Littlewood-Richardson coefficient 
$c_{\alpha,\beta}^\lambda$  for arbitrary semisimple Lie algebra (Problem~\ref{pintrolittle})
 has a strong,
convex $\#P$-formula, because
\[c_{\alpha,\beta}^\lambda=\phi(P_{\alpha,\beta}^\lambda),\]
where $P_{\alpha,\beta}^\lambda$ is the BZ-polytope \cite{berenstein}
 associated with
the triple $(\alpha,\beta,\lambda)$. It is easy to see from
the description in \cite{berenstein} that the number of
 defining constraints of 
$P_{\alpha,\beta}^\lambda$ is polynomial in the total number of 
parts (coordinates)  of $\alpha,\beta,\lambda$. Given $\alpha,\beta,\lambda$,
these constraints can be computed in strongly polynomial time.
Hence, the membership problem for $P_{\alpha,\beta}^\lambda$ belongs to $LOGSPACE \subseteq P$.
It follows that   the Littlewood-Richardson function 
$c(\alpha,\beta,\lambda)=c_{\alpha,\beta}^\lambda$ belongs to strongly convex  $\#P$.

\subsection{Littlewood-Richardson cone} \label{slittlecone}
We now give a natural example of a polytope in representation theory,
the number of whose defining constraints is exponential, but  whose 
membership function
can still be computed in polynomial time.

Given a complex, semisimple, simply connected  group $G$,
let the Littlewood-Richardson semigroup $LR(G)$ be the set of all triples
$(\alpha,\beta,\lambda)$ of dominant weights of $G$ such that 
the irreducible module $V_\lambda(G)$ appears in the tensor product 
$V_\alpha(G) \otimes V_\beta(G)$ with nonzero multiplicity \cite{zelevinsky}. 
Brion and Knop \cite{elashvili} have shown that $LR(G)$ is a finitely generated 
semigroup with respect to addition. This also follows from the
polyhedral expression for Littlewood-Richardson coefficients in terms
of BZ-polytopes \cite{zelevinsky}. Let $LR_\R(G)$ be the polyhedral cone generated by 
$LR(G)$. 

When  $G=GL_n(\C)$, the  facets of $LR_\R(G)$ 
have an explicit description  by the affirmative solution
to Horn's conjecture in \cite{kly,knutson}.
But their number can be quite large (possibly
exponential). 
Nevertheless,  membership of any rational $(\alpha,\beta,\lambda)$ (not necessarily
integral) in $LR_\R(G)$ can be decided in
strongly polynomial time.

This is because $LR_\R(G)$ is the projection of a
polytope $P(G)$, the number of whose constraints is
polynomial in the heights of $\alpha,\beta,\lambda$ \cite{zelevinsky}. 
If $\phi: P(G)\rightarrow LR(G)$ is this projection, we 
can choose $P(G)$ so that 
for any integral $(\alpha,\beta,\lambda)$, 
$\phi^{-1}(\alpha,\beta,\lambda)$ is the BZ-polytope 
associated with the triple $(\alpha,\beta,\lambda)$.
To decide if $(\alpha,\beta,\lambda) \in LR(G)$, we only have
to decide if the polytope $\phi^{-1}(\alpha,\beta,\lambda)$ is
nonempty. This can be done in strongly polynomial time using
Tardos' linear programming algorithm \cite{tardos}.

\subsection{Eigenvalues of Hermitian matrices} 
Here is  another example of a polytope in representation theory 
with exponentially many facets, whose membership problem can still belong to $P$. 

For a Hermitian matrix $A$, let $\lambda(A)$ denote the 
sequence of eigenvalues of $A$ arranged in a weakly decreasing order. 
Let $HE_r$ be the set of triple $(\alpha,\beta,\lambda) \in \R^r$ 
such that $\alpha=\lambda(A+B)$, $\beta=\lambda(A)$,
$\lambda=\lambda(B)$ for some Hermitian matrices $A$ and $B$ of dimension $r$.
It is closely related to the Littlewood-Richardson semisgroup $LR_r=LR(GL_r(\C))$:
 $HE_r \cap P_r^3=LR_r$,
where $P_r$ is the semigroup  of partitions of length  $\le r$.
I. M. Gelfand asked for an explicit description of $HE_r$.
Klyachko \cite{kly} showed that 
$HE_r$ is a convex polyhedral cone. An explicit description of
its facets is now known by the affirmative answer to Horn's conjecture.
But their number  may  be exponential.
Hence, membership in $HE_r$ is still not easy to check using this
explicit description. This leads to the following complexity theoretic variant
of Gelfand's question:

\begin{question} 
Does the memembership problem for $HE_r$ belong to $P$?
\end{question} 

Given that the answer is yes for the closely related $LR_r=LR(GL_r(\C))$
(Section~\ref{slittlecone}),
this may be  so. 
If $HE_r$ were a projection of some polytope with polynomially many facets,
this would follow as in Section~\ref{slittlecone}. But this is not necessary.
For example,  Edmond's perfect matching polytope  for non-bipartite
graphs is not known to be a projection of any polytope with polynomially
many constraints. Still the associated membership problem belongs to
$P$ \cite{schrijver}.

\section{Separation oracle} \label{sseporacle}
Suppose $P \subseteq \R^l$
is a convex polytope whose membership function $\chi_P$ is
polynomial time computable. If $\chi_P(y)=0$ for some $y \in \R^r$,
 it is natural to ask, in the spirit of \cite{lovasz},  for a 
``proof'' of nonmembership 
in the form of a hyperplane that separates $y$ from $P$.

In this paper, we assume that all polytopes are specified by the 
{\em separation oracle}. This is a computer program, which given $y$, tells if $y \in P$, 
and if $y\not \in P$,  returns such a separting hyperplane as a proof of 
nonmembership. 
We assume that the hyperplane  is given in the form $l=0$, where 
a linear function $l$ such that 
$P$ is contained in the half space $l\ge 0$, but $l(y)<0$.
Furthermore. we  assume that $P$ is a well-described 
polyhedron in the sense of \cite{lovasz}. This means $P$ is specified 
in the form of a triple
$(\chi_P,n,\phi)$, where $P\subseteq \R^n$, 
$\chi_P$ is a program for computing 
the  membership function given $y\in \R^n$,  and there exists a 
system of inequalities with rational coefficients having $P$ as its
solution set such that the encoding bit length of each inequality is
at most $\phi$. We define the encoding length $\bitlength{P}$ of
$P$ as $n+\phi$. 
We also assume that the separation oracle works in $O(\poly(\bitlength{P},\bitlength{y})$ time.

For example, the polynomial time algorithm for the membership function
of the Littlewood-Richardson cone (cf. Section~\ref{slittlecone}) can be
easily modified to return a separating hyperplane as a proof of
nonmembership.

In what follows, we shall assume, as a part of the definition of a 
convex $\#P$-formula,  that $P_x$ in (\ref{eqconvex0}) is a well-described polyhedron 
 specified by a separation oracle that
works in polynomial time with $\bitlength{P_x}=\poly(\bitlength{x})$.
These additional requirements 
are  needed for the saturated integer programming algorithm in Chapter~\ref{csatpos}.

\chapter{Saturation and positivity} \label{csatpos}
In this chapter we describe (Section~\ref{ssaturated}) 
a polynomial time algorithm for saturated and 
positive integer programming 
(Theorem~\ref{tintrosat}). In Section~\ref{sphypo} we 
state the main results and positivity hypotheses for the relaxed forms of
 Problem~\ref{pintrosubgroup} and Problem~\ref{pintrogit}, with
 $X=G/P$ or a class variety therein. Together they say that these relaxed
decision problems can be efficiently transformed into 
saturated (more strongly, 
 positive)
integer programming problems, and hence can 
be solved in polynomial time.

\section{Saturated and positive integer programming} \label{ssaturated}
We begin by proving  Theorem~\ref{tintrosat}.

Let $P\subseteq R^n$ be a polytope   given by a separation oracle (Section~\ref{sseporacle}). 
Let $\bitlength{P}$ be the encoding length of $P$  as defined in Section~\ref{sseporacle}. 
An oracle-polynomial time algorithm \cite{lovasz} is an algorithm whose
running time is $O(\poly(\bitlength{P}))$, where each
call to the separation oracle  is computed as one step.
Thus if the separation oracle works in polynomial time, 
then such an  algorithm works in polynomial time in the usual sense.
Let $\phi(P)$ be the number of integer points in $P$. Let $f_P(n)=\phi(n P)$ be
the Ehrhart quasi-polynomial \cite{stanleyenu} of $P$.
Let $l(P)$ be the least period of $f_P(n)$, if $P$ is nonempty. 
Let $f_{i,P}(n)$, $1\le i \le l(P)$, be the polynomials such that $f_P(n)=f_{i,P}(n)$ if $n=i$ 
modulo $l(P)$.  Let $F_P(t)=\sum_{n\ge 0} f_P(n) t^n$ denote the Ehrhart series of
$P$. It is a rational function.

\begin{theorem} \label{tindexquasi}

\noindent (a) The index of  $f_P(n)$, $\ind(f_P)$,   can be computed
in oracle-polynomial time, and hence, in polynomial time, assuming that the 
oracle works in polynomial time.
Furthermore, if $\ind(f_P)\not = 0$ (i.e. if $P$ is nonempty),
then $f_{i,P}(n)$ is not
an identically zero polynomial for every $i$ divisible by $\ind(f_P)$.

\noindent (b) The saturated, and hence,  positive 
integer programming problem, as defined in Section~\ref{ssatpospgm},
can be solved in 
oracle-polynomial time. Here it is assumed that the specification of $P$
also contains the saturation  index estimate $\sie(P)$, or 
the positivity   index estimate $\pie(P)$, 
and that the  bitlength of this estimate  is $O(\poly(\bitlength{P}))$. 
Given a relaxation parameter $c> \sie(P)$ (or $\pie(P)$), the 
problem is to determine if $cP$ contains an integer point in
$O(\poly(\bitlength{P},\bitlength{c}))$ time.

\noindent (c) 
Suppose  $\{P_x\}$ is a family of  polytopes, indexed by some parameter $x$,
with  the following property: wherenver $P_x$ is nonempty,
the Ehrhart quasi-polynomial $f_{P_x}(n)$ is ``almost always''
strictly saturated. Almost always means, the density of $x$'s of
bitlength $\le N$, with nonempty
$P_x$ for which $f_{P_x}(n)$ is not strictly saturated is less than
$1/N^{c''}$, for any positive $c''$, as $N\rightarrow 0$. 
We also assume that $P_x$ is given by a separation oracle that works 
in $O(\poly(\bitlength{x}))$ time, where $\bitlength{x}$ is the bitlength
of $x$, and $\bitlength{P_x}=O(\poly(\bitlength{x}))$. 

Then there exists a $O(\poly(\bitlength{x}))$ time algorithm for deciding if
$P_x$ contains an integer point that works correctly ``almost always'';
i.e., on almost all $x$.
\end{theorem}

\proof 
 
\noindent (a): 

Nonemptyness of $P$ can be decided in oracle-polynomial time using 
the algorithm of Gr\"otschel, Lov\'asz and Schrijver \cite{lovasz} (cf.
Theorem 6.4.1 therein). An extension of this algorithm, furthermore, yields
a  specification of the affine space $\mbox{span}(P)$ 
containing $P$ if $P$ is nonempty (cf. Theorems  6.4.9,
and 6.5.5 in \cite{lovasz}).  Specifically, it outputs 
an integral matrix $C$ and an integral vector $d$ such that $span(P)$ 
is defined by $C x=d$. This final specification is exact, even though
the first part of the algorithm in \cite{lovasz} uses the ellipsoid method. 
Indeed, the use of simultneous diophantine approximation based on
basis reduction in lattices is precisely to ensure this exactness in
the final answer. This is crucial for the next step of our algorithm.

If $P$ is empty, $\ind(f_P)=0$. So assume that it is
nonempty. Let $\bar C$ be the Smith normal form of $C$; i.e.,
$\bar C= A C B$ for some unimodular matrices $A$ and $B$, where 
the leftmost principal submatrix  of $\bar C$ is a diagonal,  integral 
matrix, and all other columns are zero.

The matrices $\bar C, A$ and $B$ can be computed in polynomial time using
the algorithm in \cite{kannan}. After a unimodular change of coordinates, by
letting $z=B^{-1} x$,   $\mbox{span}(P)$ is specified by the linear system
$\bar C z=\bar d=A d$. The equations in this system are of the form:
\begin{equation} \label{eqspan1}
\bar c_i z_i=\bar d_i,
\end{equation}
$i \le \mbox{codim}(P)$, for some integers $\bar c_i$ and $\bar d_i$.
By removing common factors if necessary, we can assume that 
$\bar c_i$ and $\bar d_i$ are relatively prime for each $i$.
Let $\tilde c$ be the l.c.m. of $\bar c_i$'s.

The statement (a)  follows from:

\begin{claim}   $\ind(f_P)=\tilde c$  and $f_{i,P}(n)$ is not
an identically zero polynomial for every $i$ divisible by $\tilde c$.
\end{claim} 
\noindent {\em Proof of the claim:} Indeed, $n P=\{n z \ | \ z \in P\}$
contains no integer point unless $\tilde c$ divides $n$. Hence, it is easy to see that
 $F_P(t)=F_{\bar P}(t^{\tilde c})$, where $F_{\bar P}(x)$ is the Ehrhart series
of the dilated polytope $\bar P=\tilde c P$. By eq.(\ref{eqspan1}), the equations defining
$\bar P$ are:
\begin{equation} \label{eqspan2}
z_i=\bar d_i (\tilde c/\bar c_i),
\end{equation}
Clearly,   $\tilde c$ divides the least period $l(P)$ of $f_P$, and
$l(\bar P)=l(P)/\tilde c$ is the period of the Ehrhart quasipolynomial $f_{\bar P}(n)$.
It suffices to show that the index of $f_{\bar P}(n)$ is one 
and that $f_{j,\bar P}(n)$ is not an identically zero polynomial for every 
$1 \le j \le l(\bar P)$. 
This is equivalent to showing that $\bar P$ contains a point $z$ with 
with  $z_i=a_i/b$, for some integers
$a_i$'s and $b$  such that $b=j$
modulo $l(\bar P)$. Let us call such a point $j$-admissible.
Because of the form of the equations (\ref{eqspan2}) defining
$\mbox{span}(\bar P)$, we can assume, without loss of generality,
that $\bar P$ is full dimensional. This means the system (\ref{eqspan2}) is empty.
Then this follows from denseness of the set of $j$-admissible points. 
This proves the claim, and hence (a).

\noindent  (b): Let $s=\sie(P)$ be the given saturation index estimate.
This means $f_P(n+s)$ is strictly saturated. This in conjunction with (a)
implies that, 
given a relaxation parameter $c > s$,
$c P$ contains an integer point, iff $c$ is divisible by $\ind(f_P)$
(by letting $n=c-s$).
This can be checked in $O(\poly(\bitlength{P},\bitlength{c}))$
time since $\ind(f_P)$ can be
computed in polynomial time by (a). 

\noindent (c) The algorithm computes $\ind(f_{P_x})$ and says ``Probably Yes''if the index is one, and ``No''  otherwise. Since 
the saturation index of $f_{P_x}(n)$ is zero almost
always, by the argument in  (b) with $s=0$ and $c=1$,
``Probably Yes''  really means ``Yes'' almost always. 
\qed

The algorithm in (c) has one drawback. If the answer is ``Probably Yes'',
we have no easy way of checking if $P_x$ really contains an integer point.
Ideally, we would like an algorithm that says  ``Yes'', with an
integer point in $P_x$ as a proof certificate, or ``No'', or ``Unsure'', and
the density of $x$'s on which it says ``Unsure'' should be very small.
This problem can be overcome if the family 
$\{P_x\}$ has the following stronger  property, akin to the family of
hive polytopes \cite{knutson}: 
there is a linear function $l_x$ such that, for almost all $x$,
if $\{P_x\}$ is nonempty, then the $l_x$-optimum of $P_x$ is integral
(this is stronger than saying that $f_{P_x}(n)$ is strictly saturated).
In this case, the algorithm in (c) can be extended to yield 
the integral $l_x$-optimum as a proof certificate. If the $l_x$-optimum
is not integeral, the algorithm says ``Unsure''. 
PH1 and SH (Section~\ref{sintroplethysm})
for the plethysm (and more generally, the subgroup restriction) 
problem may be strengthened by stipulating that  the polytopes therein
have this property. But this is not needed in this paper.

We  note down one corollary of the proof of Theorem~\ref{tindexquasi} 
 (this should be well known):

\begin{prop} \label{pcorsatint}
The rational function $F_P(t)=F_{\bar P}(t^{\tilde  c})$, where $F_{\bar P}(x)$ is the
Ehrhart series of the dilated polytope $\bar P=\tilde c P$, and 
$\tilde c$ is the index of $f_P(n)$.
\end{prop} 

If $P$ is explicitly specified in the form a linear system 
\begin{equation} \label{eqAB}
A x \le b,
\end{equation} 
where $A$ is an $m\times n$ matrix, $b$ an $m$ vector and 
 $m=\poly(n)$, then the following stronger version of Theorem~\ref{tindexquasi}
holds. Let $\bitlength{A}$ and $\bitlength{A,b}$ denote
the bitlength of the specification of $A$ and of the linear system (\ref{eqAB}).

\begin{theorem} \label{tstrongindex}
Suppose  $P$ is specified in terms of an explicit linear system (\ref{eqAB}).
Then the index of the  Erhart quasi-polynomial $f_P(n)$ can be computed
in $\poly(\bitlength{A,b})$ time, using $\poly(\bitlength{A})$ arithmetic
operations.

Thus, saturated, and hence, positive 
integer programming problem 
specified in the form (\ref{eqAB}) can be solved in 
in $\poly(\bitlength{A,b,c})$ time, where $c$ is the relaxation parameter,
 using $\poly(\bitlength{A})$ arithmetic
operations.
\end{theorem} 
\proof This is proved exactly as Theorem~\ref{tindexquasi}, but with Tardos'
strongly polynomial time algorithm for combinatorial linear programming \cite{tardos} 
used in place of the algorithm in \cite{lovasz}. \qed

\subsection{A general estimate for the saturation index} 

Now we give a general estimate for the saturation index of 
any polytope $P$ with a specification of the form 
\begin{equation} \label{eqAB2}
A x \le b,
\end{equation} 
where $A$ is an $m\times n$ matrix, $m$ possibly exponential.
Let $\comb{P}=n+\psi$, where $\psi$ is the maximum bitlength of any entry
of $A$. Trivially, $\comb{P} \le \bitlength{P}$.
We do not assume that we know the specification (\ref{eqAB2}) 
of $P$ explicitly. We only assume that it exists, and that we are 
told $\comb{P}$. Then:

\begin{theorem} \label{tconservative}
The saturation index of $P$ is $O(2^{\poly(\comb{P})})$. Thus the 
bitlength of the saturation index is $O(\poly(\comb{P}))$. 
\end{theorem}

Conjecturally, this also holds for the positivity index. 
This estimate is very conservative, but useful when no better estimate
is available.

\proof 
There exists a triangulation of $P$ into simplices such that 
every vertex of any simplex is also a vertex of $P$. Then
\[ f_P(n)= \sum_{\Delta} f_\Delta(n), \]
where $\Delta$ ranges over all open simplices in this triangulation;
a zero-dimensional open simplex is a vertex. The saturation index of
$f_P(n)$ is clearly bounded by the maximum of the saturation indices of
$f_\Delta(n)$. 

Hence, we can assume, without loss of generality, that
$P$ is an open simplex. Let $v_0,\ldots,v_n$ be its vertices.
Then, 
by Ehrhart's result (cf. Theorem 1.3 in \cite{stanleydecomp}), 
\begin{equation} \label{eqehrhart}
F_P(t) = \f{\sum_i h_i t^i}{\prod_{j=0}^{n} (1-t^{a_j})}, 
\end{equation}
where $h_0=1$, 
$h_i$'s are nonnegative, and $a_j$ is the least positive integer 
such that $a_j v_j$ is integral. By Cramer's rule, the bit length of 
each $a_j$ is $\poly(\comb{P})$. Without loss of generality, we
can also assume that $a_j$'s are relatively prime. Otherwise, the 
estimate on the saturation index below has to be multiplied by 
the g.c.d. of $a_j$'s. Then the result follows by applying 
the following lemma to $F_P(t)$, since $\bitlength{a_j}=O(\poly(\comb{P}))$. 
\qed 

\begin{lemma} \label{lmodularindex}
Let $f(n)$ be a quasipolynomial whose generating function $F(t)$
has a positive form 
\begin{equation} \label{eqehrhart2}
F(t) = \f{\sum_i h_i t^i}{\prod_{j=0}^{n} (1-t^{a_j})}, 
\end{equation}
where $h_0=1$, 
$h_i$'s are nonnegative, and $a_j$'s are positive and relatively prime. Let 
$a=\max \{a_j\}$. 
Then the saturation index $s(f)$ of $f(n)$ is $O(\poly(a,n))$.
\end{lemma} 

\proof 
Let $g(n)$ be the  quasi-polynomial whose generating function
$G(t)=\sum g(n)t^n$ is $1/{\prod_{j=0}^{n} (1-t^{a_j})}$.
It is known that this is the Ehrhart quasipolynomial of the 
polytope $N(a_0,\ldots,a_n)$ defined by the linear system 
\[ \sum a_j x_j =1, x_j >0.\] 
The saturation index $s(g)$ of $g(n)$ is bounded by the Frobenius number 
associated with the set of integers $\{a_j\}$--this is the 
largest positive integer $m$ such that the diophantine equation
\[ \sum_j  a_j x_j = m\] 
has no positive integeral solution $(x_0,\ldots,x_n)$. 
It is known (e.g. \cite{beck}) that the Frobenius number is bounded by 
\[ \sum_j a_j + \sqrt{a_0 a_1 a_2 (a_0+ a_1 +a_2)} = O(\poly(a)),\]
assumming that $a_0 \le a_1 \ldots$. 
Hence, $s(g)=O(\poly(a))$. 

Since $f(n)$ is a quasi-polynomial, the degree of the numerator of
$F(t)$ is less than the degree of the denominator. Thus the maximum
value of $i$ that occurs in (\ref{eqehrhart2}) is $a n$. 

Let $g_i(n)$, $i\le an$,  be the  quasi-polynomial whose generating function
is $t^i/{\prod_{j=0}^{n} (1-t^{a_j})}$. Then 
\[s(g_i)\le i + s(g)= O(\poly(a,n)).\]
Since, $h_i$'s in (\ref{eqehrhart2}) are nonnegative,
$s(f)=\max{s(g_i)}$. The result follows.
\qed

\subsection{Extensions} \label{ssextension}
We now mention a few straightforard extensions of Theorem~\ref{tindexquasi}.

First, it is not necessary that $P$ be a closed polytope. We can allow it to be half-closed.
Specifically, it can be a solution set of 
a system of inequalitites of the form:

\begin{equation} 
A_1 x \le b_1 \quad \mbox{and} \quad A_2 x < b_2,
\end{equation} 
where we have allowed strict inequalities. The function $F_P(n)=\phi(n P)$, the number
of integer points in $n P$, is again a quasi-polynomial. Hence, the notions of 
saturation  and positivity can be generalized to this setting in a natural way.

Second,  the algorithm in Theorem~\ref{tindexquasi} (b) only needs 
a nonnegative number $s(P)$ such that, for any positive integer $c>s(P)$:

\noindent {\bf Saturation guarantee:} If the affine span of $c P$,
 contains an integer point,
then $c P$ is guaranteed to contain an integer point.

If $s(P)=\sie(P)$,
then this guarantee holds, as can be seen
from the proof of Theorem~\ref{tindexquasi}.

\subsection{Is there a simpler algorithm?} \label{sistheresimpler}
Though the algorithm for saturated integer programming in Theorem~\ref{tindexquasi} is 
conceptually very simple, in reality it is quite intricate, because the work of Gr\"otschel,
Lov\'asz and Schrijver \cite{lovasz} needs a 
delicate extension of the ellipsoid algorithm \cite{khachian}  and the polynomial-time algorithm 
for basis reduction in lattices due to Lenstra, Lenstra and Lov\'asz \cite{lenstra}. As has been
emphasized in \cite{lovasz}, such a polynomial-time  algorithm should only be taken as 
a proof of existence 
of an efficient  algorithm for the problem under consideration.
It may be conjectured that for the problems under consideration in this paper
such simple, combinatorial
algorithms exist. But for the design of such algorithms, saturation alone does  not suffice.
The stronger property (PH3), and  more,  is necessary. 
We shall address this issue in  Section~\ref{ssissimpler}.

\section{Littlewood-Richardson coefficients again} \label{slittleagain}
Theorem~\ref{tstrongindex} applied to the BZ-polytope \cite{berenstein},
with saturation index estimate equal to zero, 
 specializes to the following in the setting of the Littlewood-Richardson 
problem (Problem~\ref{pintrolittle}):

\begin{theorem} \label{tsatlit} \cite{GCT5} 
Assuming SH (Hypothesis~\ref{shlittle}),
nonvanishing of $c_{\alpha,\beta}^\lambda$, given $\alpha,\beta,\lambda$, can be decided in
strongly polynomial time (Section~\ref{sstandard}) 
 for any semisimple  classical Lie algebra ${\cal G}$.
\end{theorem} 

It is assumed here that $\alpha,\beta,\lambda$ are specified by their coordinates 
in the basis of fundamental weights. 
For type $A$, this reduces to the result in \cite{GCT3}, which holds
unconditionly.

The saturation conjecture for type $A$ arose \cite{zelevinsky} in 
the context of Horn's conjecture and the related result 
of Klyachko \cite{kly}. We now turn to 
implications of Theorem~\ref{tsatlit} in this context.

Given a complex, semisimple, simply connected, classical
  group $G$, let  $LR(G)$ 
be the Littlewood-Richardson semigroup as in Section~\ref{slittlecone}.
The following is a  natural generalization of the problem raised by Zelevinsky
\cite{zelevinsky} to this general setting:
\begin{problem} 
Give an efficient description of $LR(G)$.
\end{problem} 

Zelevinsky asks for a mathematically explicit description. 
This is a computer scientist's variant of his problem.

Let $LR_\R(G)$ be the polyhedral convex cone generated by $LR(G)$.
For $G=GL_n(\C)$, by the saturation theorem, a triple 
$(\alpha,\beta,\lambda)$ of dominant weights belongs to $LR(G)$ iff
it belongs to $LR_\R(G)$. 
Assuming SH (Hypothesis~\ref{shlittle}), Theorem~\ref{tsatlit} provides the following 
efficient description for $LR(G)$ in general. Recall that the period 
of the  Littlewood-Richardson stretching polynomial $\tilde c_{\alpha,\beta}^\lambda(n)$ 
divides a fixed constant  $d(G)$, which only depends on the types of simple factors of $G$
\cite{loera,GCT5}.
Let $\alpha_i$'s denote the coordinates of $\alpha$ in the basis of fundamental weights.

\begin{cor} \label{csatlit}

\noindent (a) Assuming SH, whether a given $(\alpha,\beta,\lambda)$ belongs to $LR(G)$ can be determined 
in strongly polynomial time.

\noindent (b) There exists a decomposition of $LR_\R(G)$ into a set of polyhedral
cones, which form a cell complex ${\cal C}(G)$,
 and, for each chamber $C$ in this
complex, 
 a set $M(C)$ of $O(\rank(G)^2)$ modular equations, each 
of the form 
\[ \sum_i a_i \alpha_i + \sum_i b_i \beta_i + 
\sum_i c_i \lambda_i = 0 \quad (\mbox{mod}\  d),\] 
for some  $d$ dividing $d(G)$, 
such that 
\begin{enumerate} 
\item SH (Hypothesis~\ref{shlittle})  is equivalent to saying that:
$(\alpha,\beta,\lambda) \in LR(G)$ iff 
$(\alpha,\beta,\lambda) \in LR_\R(G)$ and 
$(\alpha,\beta,\lambda)$ satisfies the modular equations in the 
set $M(C_{\alpha,\beta,\lambda})$ associated with the cone 
$C_{\alpha,\beta,\lambda}$ containing $\alpha,\beta,\lambda$.

\item Given $(\alpha,\beta,\lambda)$, whether $(\alpha,\beta,\lambda) \in LR_\R(G)$ can
be determined in strongly polynomial time (cf. Section~\ref{shlittle}).
\item If so, the cone 
$C_{\alpha,\beta,\lambda}$ and the associated set 
$M(C_{\alpha,\beta}^\lambda)$ of modular equations can also be determined 
in strongly polynomial time. After this, whether $(\alpha,\beta,\lambda)$ 
satisfies the equations in $M(C_{\alpha,\beta}^\lambda)$ can be trivially
determined in strongly polynomial time.
\end{enumerate} 
\end{cor}
\proof (a) is a consequence of Theorem~\ref{tsatlit}.
(b) follows from a careful analysis of the algorithm therein; see the proof of 
a more general result (Theorem~\ref{tconesatform}) later. \qed

We  call the labelled cell complex ${\cal C}(G)$, in which each  cell
$C \in {\cal C}(G)$ is labelled with the set of modular  equations 
$M(C)$, the {\em modular complex}, associated with $LR_\R(G)$. 
When $G=SL_n(\C)$, the modular complex is trivial: it just consists of
the whole cone $LR_\R(G)$ with only one 
obvious  modular equation attached to it.
But, for general $G$, the modular complex and the map $C\rightarrow M(C)$ are
nontrivial. We do not know their explicit description.
Corollary~\ref{csatlit} says that, given
$x=(\alpha,\beta,\lambda)$,
whether $x \in LR_\R(G)$, and whether 
the relevant modular equations are satisfied  can be 
quickely verified on a computer, though the modular equations  cannot 
be easily
determined and verified  by hand, as in type $A$.
This is the main difference between 
type $A$ and general types. 

This naturally leads to:

\begin{question} 
Is there a mathematically
explicit description of the modular complex ${\cal C}(G)$ for
a general $G$? 
\end{question}

\section{The saturation and positivity hypotheses} \label{sphypo}
Now let $f(x)$, $x\in \N^k$,  be a counting function associated with a structural
constant in representation theory or algebraic geometry. Here $x$ denotes
the  sequence of parameters associated with the constant. Let $\bitlength{x}$ denote the
bitlength of $x$. Let $\comb{x}$ and $\rank(x)$ 
 denote its combinatorial size and combinatorial rank--these measure complexity
of the nonstretchable part  in the 
specification of $x$ and  will
be specified later for the $f$'s of interest in this paper.

For example, in the Littlewood-Richardson problem, 
$x$ is  the triple $(\alpha,\beta,\lambda)$, 
$f(x)=f(\alpha,\beta,\lambda)=c_{\alpha,\beta}^\lambda$, $\bitlength{x}$ is the
total bitlength of the coordinates of $\alpha,\beta,\lambda$,
$\comb{x}$ is the
total number of coordinates of $\alpha,\beta$ and $\lambda$, and
 $\rank(x)=\comb{x}$. 
The number of coordinates does not change during stretching, and hence,
constitute the nonstretchable part of the input specification here.

Assume that $f(x)$ is nonnegative for all $x\in \N^k$,
Then we can successively ask the following questions:

\begin{enumerate}
\item Does  $f \in PSPACE$? That is,
can  $f(x)$  be computed in $\poly(\bitlength{x})$ space?
\item Does $f \in \#P$? (cf. Section~\ref{sstandard}) 
\item  Does  $ f\in \convex \#P$? (cf. Section~\ref{sconvexsharpp})
\item Can a stretching function $\tilde f(x,n)$ be associated with $f(x)$
intrinsically so that $\tilde f(x,n)$ is quasi-polynomial? 
\item {\bf (PH1?):} 
Is there a polytope $P_x$, for every $x$, 
with $\bitlength{P_x}=O(\poly(\bitlength{x}))$ and 
$\comb{P_x}=O(\poly(\comb{x}))$,
such that $\tilde f(x,n)=f_{P_x}(n)$? 
\item Are there good analogues of SH and/or PH2, PH3 for $\tilde f(x,n)$?
If so,  nonvanishing of $f(x)$, modulo small relaxation,  can 
be decided in $O(\poly(\bitlength{P_x}))$ time by  Theorem~\ref{tindexquasi}.
\end{enumerate} 

In the rest of this paper, we study these questions when
$f=f(x)$ is a nonnegative function associated with a structural constant in 
any of the decision problems in Section~\ref{sdecision}. 
Exact specifications of $x,\bitlength{x},\comb{x},\rank(x), f(x)$, and
$\tilde f(x,n)$ for these decision problems are 
given in Sections~\ref{sssubgroup}-\ref{sspgit}.
It is shown in Chapter~\ref{cpspace} that  $f(x) \in PSPACE$
for Problem~\ref{pintroplethysm} and
the special cases of Problem~\ref{pintrosubgroup} that arise in the flip.
This may be   conjectured to be so for the $f$'s in Problem~\ref{pintrogit}, with $X$ therein
a class variety; cf \cite{GCT10} for its justification.
Quasipolynomiality of $\tilde f(x,n)$ is 
addressed    in Chapter~\ref{cquasipoly}. 
The hypotheses PH1, SH, PH2, and PH3 in these cases 
have the following unified form.

\begin{hypo} \label{phph1} {\bf (PH1)}
Let $f=f(x)$ be the function associated with a structural constant in 
\begin{enumerate} 
\item Problem~\ref{pintrokronecker}, or
\item \ref{pintroplethysm}, or
\item Problem~\ref{pintrosubgroup},  or
\item Problem~\ref{pintrogit}, with $X$ being a class variety
therein.
\end{enumerate} 
Then the function $f(x)$ has a convex $\#P$-formula 
(cf. (\ref{eqconvex0}))
\[ f(x)=\phi(P_x),\] 
such that:
\begin{enumerate} 
\item for every fixed $x$,
the Ehrhart quasi-polynomial $f_{P_x}(n)$ of $P_x$ coincides with $\tilde f(x,n)$. 
\item $\bitlength{P}=O(\poly(P))$ and $\comb{P}=O(\poly(\comb{x}))$.
\end{enumerate}
\end{hypo}

\begin{hypo} \label{phsh} {\bf (SH)}

\noindent (a) Suppose $f(x)$ is a structural constant as in  PH1 above.
Then for every $x$, the saturation index $s(\tilde f)$ of  $\tilde f(x,n)$ is
$O(\poly(\rank(x)))$. This means there exist absolute nonnegative  constants 
$c, c'$ such that $s(\tilde f) \le c (\rank(x))^{c'}$.

\noindent (b) For $f(x)$ in 
Problems~\ref{pintrokronecker}-\ref{pintrosubgroup},
 the saturation index of $\tilde f(x,n)$ is zero--i.e.,
 $\tilde f(x,n)$ is strictly saturated--for almost all $x$.
This means the density of $x$, with $\bitlength{x} \le N$ and $f(x)$ nonzero,
for which the saturation index $s(\tilde f)$ is nonzero is $\le 1/N^{c''}$,
for any positive costant $c''$, as $N\rightarrow \infty$. 
\end{hypo}

More strongly than (a),

\begin{hypo} \label{phph2} {\bf (PH2)} 
For $f(x)$ as in PH1, 
the positivity index of  $\tilde f(x,n)$ is 
$O(\poly(\rank(x)))$.
\end{hypo}

\begin{hypo} \label{phph3} {\bf (PH3)} 
For $f(x)$ as in PH1,  the generating function $F(x,t)=\sum_n \tilde f(x,n) t^n$
has a positive rational form of modular index $O(\poly(\rank(x)))$.
More specifically, 
the modular index of $\tilde f(x,n)$, as defined in Section~\ref{sminimal} 
for $f$'s that arise in this paper, is 
$O(\poly(\rank(x)))$.
\end{hypo} 

PH3 implies SH (a); this follows from Lemma~\ref{lmodularindex}.

The following conservative bound follows from Theorem~\ref{tconservative}.

\begin{theorem} {\bf (Weak SH)} \label{tweaksh}

Assuming PH1  (Hypothesis~\ref{phph1}), 
the saturation index of $\tilde f(x,n)$ is bounded by $2^{O({\poly(\comb{x})})}$;
hence  its bitlength is bounded by $O({\poly(\comb{x})})$.
\end{theorem}

The following result addresses the relaxed forms of the decision problems 
for the structural constants under consideration 
(cf. Section~\ref{sdecision}). 

\begin{theorem}  \label{tmainphyp}
Suppose   $f(x)$ is a structural constant as in  PH1 above. 
Then  PH1 (Hypothesis~\ref{phph1}) and SH (Hypothesis~\ref{phsh}) imply 
Hypothesis~\ref{PHfliprevised} (PHflip) in this case. Specifically:

\noindent (a) For $f(x)$ in Problems~\ref{pintrokronecker}-\ref{pintrogit},
nonvanishing of $\tilde f(x,a)$, for a given $x$ and a relaxation
parameter $a > c (\rank(x))^{c'}$, with $c,c'$ as in Hypothesis~\ref{phsh}, 
can be decided in  $\poly(\bitlength{x},\bitlength{a})$ time.

\noindent (b)  For $f(x)$ as in 
Problems~\ref{pintrokronecker}-\ref{pintrosubgroup},
there is a $\poly(\bitlength{x})$ time algorithm for deciding
nonvanishing of $f(x)$ that works correctly on almost all $x$. 
\end{theorem}

This follows from  Theorem~\ref{tindexquasi}.

The following sections 
give precise descriptions of $x$, $\bitlength{x}, \comb{x}, \rank(x)$
and $\tilde f(x,n)$ for the structural constants under consideration.

\section{The subgroup restriction problem} \label{sssubgroup}
In this section we  consider the subgroup restriction problem (Problem~\ref{pintrosubgroup}).
The Kronecker and the plethysm problems (Problems~\ref{pintrokronecker}, \ref{pintroplethysm}) 
are its special cases. 

Let $G,H,\rho,\lambda,\pi,m_\lambda^\pi$ be as in Problem~\ref{pintrosubgroup}.
We  shall define below an explicit polynomial homomorphism $\rho:H \rightarrow G$, as needed
in the statement of Problem~\ref{pintrosubgroup}, and also 
the precise specifications 
$[H],[\rho],[\lambda],[\pi]$ of $H,\rho,\lambda,\pi$,
respectively. We shall also define the bitlengths
 $\bitlength{H},\bitlength{\rho},\bitlength{\lambda},\bitlength{\pi}$ and the 
combinatorial bit lengths
 $\comb{\lambda},\comb{\pi}$. We let
$\comb{H}=\bitlength{H}$ and
$\comb{\rho}=\bitlength{\rho}$, since $H$ and $\rho$ belong to
the  nonstretchable part of the input.  
On the other hand, $\lambda$ and $\pi$ will be stretched in the 
definition of $\tilde f(x,n)$, and hence their combinatorial bit lengths
will differ from the usual bit lengths.
The input $x$ in the subgroup restriction  problem 
is  the tuple $([H],[\rho],[\lambda],[\pi])$. 
Its bitlength $\bitlength{x}$ is defined to be the sum of the bitlengths 
$\bitlength{H}, \bitlength{\rho},\bitlength{\lambda},\bitlength{\pi}$, and
 $\comb{x}$ is
defined to be the sum of  $\comb{H},\comb{\rho},\comb{\lambda}$ and
$\comb{\pi}$. Finally $\rank(x)$ is defined to the sum of the ranks of $H$ 
and $G$ and $\comb{\lambda}$ and $\comb{\pi}$. Here that rank of a (reductive)
group is defined in a standard way. For example,
the rank of the symmentric group $S_n$
is $n$, that of $GL_n(\C)$ is $n$. The rank of a general 
finite or connected  simple 
group can be  defined similarly, 
and the rank of a more complex reductive group is
defined to be  the sum of the ranks of its simple components. 
With this terminology, we let $f(x)=m_\lambda^\pi$, with  $x$ as defined here
in   Hypotheses~\ref{phph1}-\ref{phph3} and 
Theorem~\ref{tmainphyp}  for the
subgroup restriction problem. 
Here $H$ and $\rho$ are implicit in the definition
of $m_\lambda^\pi$.

For example, in the plethym problem (Problem~\ref{pintroplethysm}), these specifications are
as follows. The specification 
$[H]$ is just the root system for $H=GL_n(\C)$. Its 
bitlength $\bitlength{H}$ is $n$.
The specification $[\rho]$ of the representation map
$\rho: H \rightarrow G=GL(V_\mu(H))$ consists of just the partition  $\mu$ specified 
in terms of its nonzero parts. 
Its bitlength $\bitlength{\rho}=\bitlength{\mu}$. The ranks of $H$ and $G$ are
as usual.
The partitions $\lambda$ and $\mu$ are specified 
in terms of their nonzero parts.
Their bitlength is the total bitlength of the parts, and the combinatorial
bit length  is the total number of parts (the height).
It is crucial here that only nonzero parts of $\lambda$ are specified,  because
the rank of $G$ can be exponential in the rank of $H$ and the bitlength of $\mu$. Hence,
the bitlength of this compact representation of $\lambda$ can be  polynomial in the rank of 
$H$ and the bitlength of $\mu$, even if the dimension of $G$ is exponential.
The main difference between $\bitlength{x}$ and $\comb{x}$ is that
the stretchable data $\lambda$ and $\pi$ contribute their bitlengths 
to the former, and  their heights to the latter.
The plethysm problem is the main prototype of the subgroup restriction problem.
If the reader wishes, (s)he
can skip the rest of  this subsection and 
jump to  Section~\ref{sstretching} 
in  the first reading.

In general, we 
assume that $H$ in Problem~\ref{pintrosubgroup} is a  finite simple group, or a complex simple,
simply connected  Lie group, or an algebraic  torus $(\C^*)^k$,
or a direct product of such groups.
The results and hypotheses in this paper are also applicable if we 
allow simple types of semidirect products, such as wreath products, which is 
all that we need for the sake of the flip. But these  extensions  are routine, and 
hence, for the sake of simplicity, we shall confine ourselves to direct products.

\subsection{Explicit polynomial homomorphism} 
Now let us define an {\em explicit polynomial homomorphism}. This will be done by 
defining basic explicit homomorphisms, and composing them functorially.

\noindent{\em Basic explicit homomorphisms:}

Let $V$ be an  irreducible polynomial representation of $H$ (characteristic  zero),
or more generally,
an explicit polynomial representation that is constructed functorially
from the irreducible polynomial representations using the operations
 $\oplus$ and $\otimes$. Then the corresponding 
homomorphism  $\rho:H\rightarrow G=GL(V)$ is an explicit polynomial homomorphism. 
The identity map $H\rightarrow H$ is also an  explicit polynomial homomorphism.

The polynomiality restriction here only applies to the 
torus component of $H$. If $H$ is
 a finite simple group, or a complex semisimple group, then any irreducible
representation of $H$ is, by definition,  polynomial. In general, a representation is
polynomial if its restriction to the torus component is polynomial; i.e., a sum of
polynomial (one dimensional) characters.

To see why the polynomiality restriction is essential,
let $H$ be a torus,  $V$  its rational
representation, and $G=GL(V)$. Let  $V_\lambda(G)=\sym^d(V)$, the symmetric representation
of $G$, and let $\pi$ be the label of the trivial character of $H$.
Then the multiplicity $m_\lambda^\pi$ is the number of $H$-invariants in $\sym^d(V)$.
This is easily seen to be the number of nonnegative solutions of a system of
linear diophontine equations. 
But the problem of deciding whether a given system of linear diophontine equations has 
a nonnegative solution is, in general, $NP$-complete. 
Though the system that arises above is of a special form,
it is not expected to be in $P$ if $V$ is allowed to be any rational representation; 
the associated decision problem may
be $NP$-complete even in this special case. If $V$ is a polynomial representation of 
a torus $H$, then all coefficients of
the system are nonnegative, and the decision problem is trivially in $P$.

\noindent{\em Composition:}

We can now  compose  the basic  explicit (polynomial) homomorphisms above functorially:

\begin{enumerate} 
\item If $\rho_i: H \rightarrow G_i$ are explicit, the product map
$\rho:H \rightarrow \prod_i G_i$ is also explicit.

\item If $\rho_i: H_i\rightarrow G_i$ are explicit, the product map
$\rho: \prod H_i \rightarrow \prod G_i$ is also explicit. 
\end{enumerate} 

Instead of products, we can also allow simple semi-direct products
such as wreath products here.
We may  also allow other functorial constructions such
as induced representations and restrictions.
For example, if 
$\rho:H\rightarrow G$ is an explicit polynomial homomorphism, and $G'\subseteq G$ is an 
explicit subgroup of $G$ such that $\rho(H)\subseteq G'$, then 
the restricted homomorphism $\rho':H \rightarrow G'$ can also be considered to be 
an  explicit polynomial homomorphism.
But for the sake of simplicity, we shall confine ourselves to the 
simple functorial constructions above.

\subsection{Input specification and  bitlengths} 
Now we describe the  specifications $[H],[\rho],[\lambda],[\mu]$, their 
bitlengths.
These are very similar to the ones in the plethysm problem.

\noindent {\bf The specification $[H]$:}

We assume that $H$ is specified as follows.

\noindent (1) If $H$ is a complex, simple, simply connected Lie group, then the specification $[H]$
consists of
the root system of $H$  or the Dynkin diagram. Let $\bitlength{H}$ be the bitlength of
this specification. Thus, if $H=SL_n(\C)$, then $\bitlength{H}=O(n)$.

\noindent (2) 
If $H$ is a simple group of Lie type (Chevalley group) then it has a similar specification 
\cite{carter}. The only finite groups of Lie type that arise in GCT are 
$SL_n(F_{p^k})$ and $GL_n(F_{p^k})$. In this case the specification $[H]$ is easy: we only have to
specify $n,p,k$. We define $\bitlength{H}$ in this case to be $n+k+\log_2 p$; not 
$\log_2 n + \log_2 k + \log_2 n$.  
 As a rule, $\bitlength{H}$ is defined to be the sum of
the rank parameters (such as $n$ and $k$ here) and bit lengths of the weight parameters (such
as $p$ here) in the specification. This is equivalent to assuming that
the rank parameters  are specified in unary.

\noindent (3) If $H$ is the alternating group $A_n$, we only
specify $n$. Let $\bitlength{H}=n$.

\noindent (4) 
The torus is specified by its dimension. We define 
$\bitlength{H}$ to be the dimension. 

\noindent (5)
If $H$ is a product of such groups, its specification is composed from the specifications of
its factors, and the bitlength  $\bitlength{H}$ is defined to be
the sum of the  bitlengths of the constituent specifcations.

\noindent {\bf The specification $[\rho]$:}

Let us first assume that $\rho$ is a basic explicit polynomial homomorphism.
In this case the specification of $\rho: H\rightarrow G=GL(V)$ is
a pair  $[\rho]=([H],[V])$ consisting of the 
specification $[H]$ of $H$ as above, and the combinatorial specification $[V]$ of the 
representation $V$ as defined  below:

\noindent (1) If $H$ is a semisimple, simply connected Lie group, and 
$V=V_\mu(H)$ its irrreducible representation 
for a dominant weight $\mu$ of $H$, then $V$ is specified by simply 
giving the coordinates of $\mu$ in terms of the fundamental weights of $H$. Thus $[V]=\mu$, and its
bitlength $\bitlength{V}$ is the total bitlength of all coordinates of $\mu$,
and the combinatorial bit length 
$\comb{V}$ is the total number of coordinates of $\mu$.

\noindent (2) If $H=S_n$, and $V=S_\gamma$ its irreducible representation (Specht module),
then $[V]$ is the partition $\gamma$ labelling this Specht module. 
We define $\bitlength{V}$ to be the bitlength of this partition,
and $\comb{V}=\bitlength{V}$.

\noindent (3) 
If $H$ is a finite general linear group $GL_n(F_{p^k})$, and $V$ its  irreducible representation,
as classified by  Green \cite{macdonald}, then $[V]$ is the combinatorial classifying label of $V$ 
as given in \cite{macdonald}. It is a certain partition-valued function, which can be 
specified by listing the places where the function is nonzero and the  nonzero partition
values at these places. Let $\bitlength{V}$ be the bitlength of this specification;
it is $O(\poly(n,k,\bitlength{p}))$.  We let $\comb{V}=\bitlength{V}$. 
More generally, if $H$ is a finite  group of Lie type, and $V$ its irreducible
representation, then $[V]$ is the combinatorial classifying label of $V$ as given by
Lusztig \cite{lusztig}. 

\noindent (4) If $H$ is a torus and $V$ is a 
polynomial character, then $[V]$ is the specification of
the character. Its bitlength is the bitlength of the specification,
and combinatorial bit length  is the dimension of
$H$.

\noindent (5) If $V$ is composed from irreducible representations, then $[V]$ is composed from the 
specifications of the irreducible representations in an obvious way. Bitlengths and combinatorial bitlengths are defined additively.

The bitlength $\bitlength{\rho}$ is defined to be 
$\bitlength{H}+\bitlength{V}$, where $\bitlength{V}$ is the bitlength of $[V]$.

If  $\rho$ is a composite homomorphism,
its  specification $[\rho]$  is composed from the 
specifications of its basic constituents in an obvious way.
The bitlength $\bitlength{\rho}$ is defined to be the sum of the bitlengths 
of these basic specifications.

\noindent {\bf The specifications $[\lambda]$ and $[\pi]$:}

$V_\pi(H)$ is the tensor product of the irreducible representations of the factors
of $H$.  We let $[\pi]$ be the tuple of the combinatorial classifying labels 
of each of these irreducible representations, as specified above. Let $\bitlength{\pi}$ 
be their total bit length, and $\comb{\pi}$ the total combinatorial bit length. Similarly, $V_\lambda(G)$ is the 
 tensor product of the irreducible representations of the factors
of $G$.  When $G=GL_m(\C)$, $\lambda$ is a partition, which we specify by only giving its
nonzero parts, whose number is equal to the height of $\lambda$.
This is crucial since the height of $\lambda$ can be much less than than the rank $m$ of $G$,
as in  the plethysm problem (Problem~\ref{pintroplethysm}). We shall leave a similar compact
specification $[\lambda]$ for a general 
connected, reductive $G$ to the reader. Let $\bitlength{\lambda}$ be its bitlength and 
$\comb{\lambda}$ its combinatorial bit length.

\subsection{Stretching function and quasipolynomiality} \label{sstretching}
Let $f(x)=m_\lambda^\pi$ as above, with $x=([H],[\rho],[\lambda],[\pi])$. 
Here $\lambda$ is the dominant weight of $G$. 
First, assume that $H$ is connected, reductive. Then $\pi$ is the dominant weight of $H$.
For a given $x$, let us define the stretching function as
\begin{equation} 
\tilde f(x,n)=\tilde m_{\lambda}^\pi(n)=m_{n \lambda}^{n \pi},
\end{equation}
which is the multiplicity of $V_{n \pi}(H)$ in $V_{n \lambda}(G)$, considered as an
$H$-module via $\rho: H \rightarrow G$. 
Let $M_\lambda^\pi(t)=\sum_{n \ge 0} \tilde m_{\lambda}^\pi(n) t^n$ be the 
generating function of this stretching quasi-polynomial.

The following is the generalization of Theorem~\ref{tquasiplethysm} in this setting.

\begin{theorem} \label{tquasisubgroup}

\noindent (a) (Rationality) The generating function 
$M_{\lambda}^\pi(t)$ is rational.

\noindent (b) (Quasi-polynomiality) The stretching function
$\tilde m_{\lambda}^\pi(n)$ is a quasi-polynomial function of
$n$. 

\noindent (c) 
There exist   graded, normal 
$\C$-algebras 
 $S=S(m_{\lambda}^\pi)=\oplus_n S_n$ and  $T=T(m_{\lambda}^\pi)=\oplus_n T_n$ such that:
\begin{enumerate} 
\item The schemes $\spec(S)$ and  $\spec(T)$
are normal and have  rational singularities.
\item $T=S^H$, the subring of $H$-invariants in $S$.
\item The quasi-polynomial 
$\tilde m_{\lambda}^\pi(n)$ is the Hilbert function of $T$.
\end{enumerate}

\noindent (d) (Positivity) The rational function $M_{\lambda}^\pi(t)$
can be expressed in a positive form: 

\begin{equation} 
M_{\lambda}^\pi(t)=\f{h_0+h_1 t+ \cdots+h_d t^d}{\prod_j(1-t^{a(j)})^{d(j)}},
\end{equation} 
where $a(j)$'s and $d(j)$'s are positive integers, 
 $\sum_j d(j)=d+1$, where $d$ is the degree of the quasi-polynomial, 
$h_0=1$, and $h_i$'s  are nonnegative integers.
\end{theorem}

The specific rings $S(m_\lambda^\pi)$ and  $T(m_\lambda^\pi)$
 constructed in the proof of this result
are called the  {\em canonical rings} 
associated with the structrural  constant $m_{\lambda}^\pi$. 
The projective schemes $Y(m_{\lambda}^\pi)=\proj(S(m_{\lambda}^\pi))$,
and $Z(m_{\lambda}^\pi)=\proj(T(m_{\lambda}^\pi))$ are called the canonical models associated 
with  $m_{\lambda}^\pi$. 

\ignore{The minimal positive form of
$M_\lambda^\pi(t)$ is defined very much as in Section~\ref{sintroplethysm}, and 
an analogue of Conjecture~\ref{cminimalplethysm} can be made, which 
would imply PH3 (Hypothesis~\ref{phph3}) for the structural constant $m_\lambda^\pi$.
}

Theorem~\ref{tquasisubgroup}
and its generalization, when $H$ can be disconnected, is proved in Chapter~\ref{cquasipoly};
cf. Theorem~\ref{tquasimain1}. 

\subsubsection{Finitely generated semigroup} 

The following is an analogue of Theorem~\ref{tconeplethysm}. 

\begin{theorem} \label{tfinitegensubgroup} 
Assume that $H$ is connected. For a fixed $\rho:H \rightarrow G$, 
let $T(H,G)$ be the set of pairs $(\mu,\lambda)$ of dominant weights 
of $H$ and $G$  such that the irreducible
representation 
$V_\pi(H)$ of $H$ occurs in the irreducible 
representation $V_\lambda(G)$ of $G$
with nonzero multiplicity. Then
$T(H,G)$ is a finitely generated semigroup with respect to addition.
\end{theorem}

This is proved in  Section~\ref{sconesub}.

\subsubsection{PSPACE} 

The following is a generalization of Theorem~\ref{tpspaceplethysm}.

\begin{theorem} \label{tpspacesubgroup} 
Assume that $H$ in Problem~\ref{pintrosubgroup} is a direct product, whose each factor is
a complex simple, simply connected Lie group, or an alternating (or 
symmetric) group,  or $SL_n(F_{p^k})$ (or $GL_n(F_{p^k})$), or  a torus.
Then  $f(x)=m_\lambda^\pi$ can be computed in $\poly(\bitlength{x})$ space,
with $x$ as specified above.
\end{theorem}

This is proved in  Chapter~\ref{cpspace}. 
It may be conjectured that Theorem~\ref{tpspacesubgroup} holds even when
the composition factors of $H$ are allowed to be general finite simple groups of Lie type. 
This will be so if Lusztig's algorithm \cite{lusztigchar}  for computing the characters of 
finite simple groups of Lie type can be parallelized; cf. Section~\ref{sfinitesimple}.

\subsubsection{Positivity hypotheses}
\label{sclassical}

Theorem~\ref{tquasisubgroup}-\ref{tpspacesubgroup}, along with the experimental  results
in  special cases (cf. Chapter~\ref{cevidence}),
constitute the main evidence in support of the positivity 
Hypotheses~\ref{phph1}-\ref{phph3} for the subgroup restrition problem.

\section{The decision problem in geometric invariant theory} \label{sspgit}
Finally, let us turn to the most general Problem~\ref{pintrogit}. 

\subsection{Reduction from Problem~\ref{pintrosubgroup} to Problem~\ref{pintrogit}} 
\label{sreduction}
First, let us note that the subgroup restriction problem (Problem~\ref{pintrosubgroup}) 
is a special case of Problem~\ref{pintrogit}. 
To see this, let $H,\rho$ and $G$ be as in Problem~\ref{pintrosubgroup}, and let
$X$ be the closed $G$-orbit of the point $v_\lambda$ corresponding to the highest weight vector 
of $V_\lambda(G)$ in 
 the projective space $P(V_\lambda(G))$. Then
\begin{equation} \label{eqXsubgroup}
X=G v_\lambda \cong G/P_\lambda,
\end{equation}
where the  $P=P_\lambda=G_{v_\lambda}$ is the parabolic stabilizer of $v_\lambda$. 
We have  a natural
action of $H$  on $X$ via $\rho$.  Let $R$ be the homogeneous coordinate ring of $X$. 
By \cite{hashimoto,mehta,ramanathan,smith}, the singularities of $\mbox{spec}(R)$
are rational. By Borel-Weil \cite{fultonrepr},
the degree one component $R_1$  of the homogeneous coordinate ring $R$
of $X$ is $V_\lambda(G)$. 
Hence, $s_1^\pi$ in this special case of Problem~\ref{pintrogit} is precisely 
$m^\pi_\lambda$ in Problem~\ref{pintrosubgroup}. The results in Section~\ref{sssubgroup} for
$s_1^\pi$ generalize in a natural way for $s_d^\pi$. 

\subsection{Input specification}
The variety $X$ in the above example is completely specified by $H,\rho$ and $\lambda$.
Hence its specification $[X]$ can be given in the form a tuple $([H],[\rho],[\lambda])$, 
where $[H],[\rho]$ and $[\lambda]$ are the specifications of $H$, $\rho$ and $\lambda$
 as in Section~\ref{sssubgroup},
The input specification $x$ for Problem~\ref{pintrogit}  in the special case above 
is the tuple 
$([X], d,[\pi])=([H],[\rho], [\lambda],d,[\pi])$,
where $[\pi]$ is the specification 
of  $\pi$ as in Section~\ref{sssubgroup}.

We now describe a class of varieties $X$
which have similar compact specifications.

Let $G$ be a connected, reductive group, $H$ a reductive, possibly disconnected,
reductive group, and $\rho: H \rightarrow G$ an explicit polynomial homomorphism as in 
Section~\ref{sssubgroup}.  Let $V=V_\lambda(G)$ be an irreducible representation of $G$ for a 
dominant weight $\lambda$. Let $P(V)$ be the projective space associated with $V$.
It has a natural action of $H$ via $\rho$. Let $v \in P(V)$ be a point that is characterized 
by its stabilizer $G_v \subseteq G$. This means it is the only point in $P(V)$ that is
stabilized by $G_v$.  For example, the point $v_\lambda$ above is characterized by its
parabolic stabilier. 
We assume that we know the Levi decompositioon of $G_v$ explicity, 
and its compact specification $[G_v]$, like that of $H$, 
and also an explicit compact specification of the embedding $\rho':G_v \rightarrow G$,
aking to that of the explicit homomorphism $\rho:H \rightarrow G$. 
Let $X\subseteq P(V)$ be the projective closure of the $G$-orbit of $v$ in $P(V)$. 
Then $X$ as well as the action of $H$ on $X$ are 
 completely specified by $\lambda,H,\rho,G_v$ and $\rho'$. Hence, 
we can let $[X]$ be the tuple  $(\lambda,[H],[\rho],[G_v],[\rho'])$. 
The input specification $x$ for Problem~\ref{pintrogit} with the $X$ of this form is
the tuple $([X], d,[\pi])$. The  bitlengths $\bitlength{x}$ and $\comb{x}$ 
are defined additively. The $\rank(x)$ is defined to be the sum of 
the ranks of $H$ and $G$, $\dim(V)$ and $\comb{\pi}$. 
Since the point $v_\lambda$ above is characterized by its
stabilizer, $G/P$ is a variety of this form. 

The class varieties \cite{GCT1,GCT2} are either of this form, or a slight extension of this form,
and admit such  compact specifications.
The algebraic geometry of  an $X$ of the above form is completely determined by the representation 
theories of the two homomorphisms $\rho:H \rightarrow G$ and $\rho':G_v\rightarrow G$. Furthermore, the results in \cite{GCT2} say that Problem~\ref{pintrogit} for a class
variety is intimately linked with  the subgroup restriction problem and its 
variants  for the
homomomorphisms $\rho$ and $\rho'$. Hence it 
is qualitatively similar to the subgroup restriction problem in this case; cf. \cite{GCT10} for
further  elaboration of the connection between these two problems.

\subsection{Stretching function and quasi-polynomiality}
Now let $H,X,R$ and $s_d^\pi$ be as in Problem~\ref{pintrogit}, with $H$ therein assumed to be
connected.
We associate with $f(x)=s_d^\pi$  the following stretching fucntion:
\begin{equation} 
\tilde f(x,n)=\tilde s_d^\pi(n)= s_{n d}^{n \pi}, 
\end{equation}
where $s_{n d}^{n \pi}$ is the multiplicity of the irrreducible representation $V_{n \pi}(H)$ 
of $H$ in $R_{n d}$, the componenent of the homogeneous coordinate ring $R$ of $X$ with degree
$n d$.
Let $S(t)=\sum_{n\ge 0} \tilde s_d^\pi(n) t^n$. 

\begin{theorem} 
\label{tquasigit}

Assume that the singularities of $\spec(R)$ are rational. 

\noindent (a) (Rationality) The generating function 
$S_{d}^\pi(t)$ is rational.

\noindent (b) (Quasi-polynomiality) The stretching function
$\tilde s_{d}^\pi(n)$ is a quasi-polynomial function of
$n$. 

\noindent (c) 
There exist   graded, normal 
$\C$-algebras 
 $S=S(s_{d}^\pi)=\oplus_n S_n$ and  $T=T(s_{d}^\pi)=\oplus_n T_n$ such that:
\begin{enumerate} 
\item The schemes $\spec(S)$ and  $\spec(T)$
are normal and have  rational singularities.
\item $T=S^H$, the subring of $H$-invariants in $S$.
\item The quasi-polynomial 
$\tilde s_{d}^\pi(n)$ is the Hilbert function of $T$.
\end{enumerate}

\noindent (d) (Positivity) The rational function $S_{d}^\pi(t)$
can be expressed in a positive form: 

\begin{equation} 
S_{d}^\pi(t)=\f{h_0+h_1 t+ \cdots+h_k t^k}{\prod_j(1-t^{a(j)})^{k(j)}},
\end{equation} 
where $a(j)$'s and $k(j)$'s are positive integers, 
 $\sum_j k(j)=k+1$, where $k$ is the degree of the quasi-polynomial $\tilde s_d^\pi(n)$, 
$h_0=1$, and $h_i$'s  are nonnegative integers.
\end{theorem}
This is proved in Chapter~\ref{cquasipoly}. 
Theorem~\ref{tquasisubgroup} is a special case of this theorem, in view 
 of the reduction in Section~\ref{sreduction}. 
Theorem~\ref{tquasigit}
is applicable when $X$ is a class variety, assuming that its singularities are
rational. %; cf. \cite{GCT2} for the results which support this.

\subsection{Positivity hypotheses}
Even though Theorem~\ref{tquasigit} holds for any $X$, with $\spec(R)$ having rational
singularities,  the positivity hypotheses PH1, SH, PH2, and PH3
can be expected to
hold for only very special $X$'s. 
In  general,  characterizing the   $X$'s with compact specification 
for which these hypotheses   hold is
a delicate problem.
Hypotheses~\ref{phph1}-\ref{phph3}  say that these 
hold
 when $X$ in Problem~\ref{pintrogit} is $G/P$ (as in Section~\ref{sreduction}) 
or a class variety, with 
the input specification $x$  as described above. 
For future reference, we shall reformulate these hypotheses 
purely  in geometric terms.

For this we need a definition.

Let $T=\sum_n T_n$ be a graded complex $\C$-algebra so that 
the  singularities of $\spec(T)$ rational.
Let $Z=\proj(T)$. Assume that $Z$ has a compact specification $[Z]$;  we shall specify it below
for the  $Z$'s  of interest to us. 
We let $[T]$, the specification of $T$, to be $[Z]$. This will play the role of the input 
in the definition below. Let $\bitlength{T}$ denote its bitlength,
and $\comb{T}$ combinatorial bit length.
Let $h_T(n)=\dim(T_n)$ be its Hilbert function, which is a quasipolynomial, since the 
singularities of $\spec(T)$ are rational; cf.  Lemma~\ref{ldehy}.

\begin{defn} \label{dphcanonicalz}

We say that PH1 holds for $T$ (or $Z$) if the Hilbert quasi-polynomial $h_T(n)$ is
convex. This means there exists a polytope  $P=P_T$ depending on the input $[T]$,
whose Ehrhart quasipolynomial $f_P(n)$ coincides with 
the Hilbert function $h_T(n)$, and whose membership function $\chi_P(y)$ can be
computed in $\poly(\bitlength{T},y)$ time. We assume that a separating hyperplane can also
be computed in polynomial time   if $y \not \in P$ (Section~\ref{sseporacle}).
\end{defn} 

If PH1 holds we can also ask if analogues of SH, PH2, and PH3--whose 
formulation is similar and hence omitted--hold.

\subsection{$G/P$ and Schubert varieties} \label{sschubertandgmodp}
Let us illustrate this definition with an example.
Let $X\cong G/P_\lambda$ be as in Section~\ref{sreduction} and $R$ its 
homogeneous coordinate ring.
 We have already seen that it has a compact specification: namely $[X]=\lambda$.
Since singularities of $\spec(R)$ are rational, PH1 makes sense. 
For $G/P$  it follows from the  Borel-Weil theorem. 
The Hilbert series of $R$
is of the form 
\[ \f {h_0 + \cdots + h_d t^d} {(1-t)^{d+1}},\] 
with $h_0=1$ and $h_i$'s nonnegative. This is so because $R$ is Cohen-Macauley \cite{ramanathan}
 and is generated by its degree one component. Hence, the modular index 
of the Hilbert function is one (PH3).
PH2 turns out to be nontrivial. Experimental evidence in its support for the classical
$G/P$  is given in Section~\ref{sevigmodp}.
Considerations for the Schubert subvarieties are similar. Experimental evidence for
PH2 for the classical Schubert varieties is also given in Section~\ref{sevigmodp}.

Now let $s=s_d^\pi$ be the multiplicity as Problem~\ref{pintrogit}, with
$X$ having a compact specification $[X]$ as above.
Let $T=T(s)$ be the ring associated with $s$ as in Theorem~\ref{tquasigit} (c). 
Let $Z=Z(s)=\proj(T)$.
We let the specification $[Z]=([X],d,\pi)$. Let $\bitlength{Z}$ be its bitlength.

So Theorem~\ref{tindexquasi} in this context implies:

\begin{theorem} 
If PH1 and SH holds  for $Z(s)$ then nonvanishing of
$s$, modulo small relaxation, can be decided in $\poly(\bitlength{Z})$  time.
\end{theorem}

We also have the following reformulation:

\begin{prop}
Hypotheses~\ref{phph1}-\ref{phph3} are equivalent to
PH1,SH,PH2,PH3  for  $Z(s)$, where 
$s$ is a stucture constant that corresponds the structure constant $f(x)$ in 
Hypotheses~\ref{phph1}.
Thus, in the case of the subgroup restriction problem,  $s=s_1^\pi=m^\pi_\lambda$
as  in Section~\ref{sreduction}. 
\end{prop} 
This is just a consequence of definitions.

\section{PH3 and existence of  a simpler algorithm} \label{ssissimpler}
As we  remarked in Section~\ref{sistheresimpler}, the use of 
the ellipsoid method and basis reduction in lattices makes the 
the  algorithm for  saturated integer programming (cf.
 Theorem~\ref{tindexquasi})  fairly intricate.
For the flip (cf. \cite{GCTflip} and Chapter~\ref{cobstruction}),
it is desirable to have simpler algorithms for the relaxed forms
of the decision problems under consideration,
akin to the the  polynomial time  combinatorial algorithms
in combinatorial optimization \cite{schrijver} 
that do not rely on the elliposoid method or basis reduction.
We briefly examine in this section the role of PH3 in this context.

The simple combinatorial algorithms in combinatorial optimization 
work only when the problem under consideration 
is  unimodular--in which case the vertices of the underlying polytope $P$ 
are integral--or 
almost unimodular--e.g. when the vertices of $P$ are half integral.
Edmond's algorithm for finding minimum weight
perfect matching in nonbipartite graphs \cite{schrijver}
is a classic example of the second case.

In the unimodular case, Stanley's positivity result \cite{stanleyenu} implies
that the rational function $F_P(t)$ has a positive form 
\[F_P(t)=\f {h(d) t^d + \cdots + h(0)} {(1-t)^{d+1}}.\]

If PH3 (Hypothesis~\ref{phph3}) holds 
for a structural function $f(x)$ under consideration  then the Ehrhart 
series $F_{P_x}(t)$ of the polytope $P_x$ associated with $x$ in
PH1 (Hypothesis~\ref{phph1}) has a minimal positive form in which
each root of the denominator has $O(\poly(\comb{x}))$ order.
Roughly, this says that the situation is ``close'' to the unimodular case. 
Hence, in such a case we can 
expect a purely combinatorial polynomial-time algorithm for
deciding nonvanishing of $f(x)$, modulo small relaxation,
that does not need the ellipsoid method or basis reduction.

\section{Other structural constants} \label{sother}
The paradigm of saturated and positive integer programming 
in this paper, along with   appropriate analogues of PH1,SH,PH2,PH3,
may be applicable 
several other fundamental structural constants in representation theory
and algebraic geometry, in addition 
to   the ones in Problems~\ref{pintrokronecker}-\ref{pintrogit} treated above, such as
\begin{enumerate} 
\item the value of a Kazhdan-Lusztig polynomial at $q=1$, \cite{kazhdan};
\item the values at $q=1$ of the  well behaved special cases 
of the parabolic Kostka polynomials and their $q$-analogues \cite{kirillov};
\item 
the structural coefficients 
of the multiplication of Schubert polynomials, and so on.
\end{enumerate}

\chapter{Quasi-polynomiality and canonical models} \label{cquasipoly}
In this chapter we prove quasipolynomiality of the stretching functions 
associated with the various structural constants under consideration (Section~\ref{squasiproof}), 
describe the associated canonical models
(Section~\ref{scanonicalmodel}),
describe the role of nonstandard quantum groups
in \cite{GCT4,plethysm,canonical}  in the deeper study 
of  these models
 (Section~\ref{squantumgroup}),
 prove finite generation of 
the semigroup of weights (Theorem~\ref{tfinitegensubgroup}) (Section~\ref{sconesub}),
and  give an elementary proof of 
rationality in Theorem~\ref{tquasisubgroup} (a)  (Section~\ref{selement}).

\section{Quasi-polynomiality} \label{squasiproof}
Here we prove Theorem~\ref{tquasigit}; Theorems~\ref{tquasiplethysm} and \ref{tquasisubgroup} 
are its special cases in view of the reduction in Section~\ref{sreduction}. 
This, in turn, follows from the following more general result.

Let $R=\oplus_k R_d$ be a normal graded $\C$-algebra  with an action of a reductive group $H$. 
Assume that $\spec(R)$ has rational singularities. 
Let $H_0$ be the connected component of $H$ containing the identity. 
Let $H_D=H/H_0$ be its discrete component. 
Given a dominant weight $\pi$ of $H_0$, we consider 
the module $V_\pi=V_\pi(H_0)$, an $H$-module with trivial action of $H_D$.
Let $s_d^\pi$ denote the multiplicity of 
the $H$-module $V_\pi$ in $R_d$. 
Let $\tilde s_d^\pi(n)$ be the multiplicity of the $H$-module 
$V_{n \pi}$ in $R_{n d}$. This is a stretching function
associated with the mulitplicity $s_d^\pi$. 
Let $S_d^\pi(t)=\sum_{n\ge 0} \tilde s_d^\pi(n) t^n$.

\begin{theorem} \label{tquasimain1}
\noindent (a) (Rationality) The generating function 
$S_{d}^\pi(t)$ is rational.

\noindent (b) (Quasi-polynomiality) The stretching function
$\tilde s_{d}^\pi(n)$ is a quasi-polynomial function of
$n$. 

\noindent (c) 
There exist   graded, normal 
$\C$-algebras 
 $S=S(s_{d}^\pi)=\oplus_n S_n$ and  $T=T(s_{d}^\pi)=\oplus_n T_n$ such that:
\begin{enumerate} 
\item The schemes $\spec(S)$ and  $\spec(T)$
are normal and have  rational singularities.
\item $T=S^H$, the subring of $H$-invariants in $S$.
\item The quasi-polynomial 
$\tilde s_{d}^\pi(n)$ is the Hilbert function of $T$.
\end{enumerate}

\noindent (d) (Positivity) The rational function $S_{d}^\pi(t)$
can be expressed in a positive form: 

\begin{equation} \label{eqtquasipnew}
S_{d}^\pi(t)=\f{h_0+h_1 t+ \cdots+h_k t^k}{\prod_j(1-t^{a(j)})^{k(j)}},
\end{equation} 
where $a(j)$'s and $k(j)$'s are positive integers, 
 $\sum_j k(j)=k+1$, where $k$ is the degree of the quasi-polynomial $\tilde s_d^\pi(n)$, 
$h_0=1$, and $h_i$'s  are nonnegative integers.
\end{theorem} 
Theorem~\ref{tquasigit} follows from this by letting $R$ be the homogeneous coordinate
ring of $X$.

More generally, if $W$ is an irreducible representation of $H_D$, we can 
consider the $H$-module $V_\pi\otimes W$. Let $s_d^{\pi,W}$ be its multiplicity
in $R_d$. Let $\tilde s_d^{\pi,W}(n)$ be the multiplicity of the trivial $H$-representation
in the $H$-module $R_{n d} \otimes V_{n \pi}^* \otimes \sym^n(W^*)$. Then

\begin{theorem} \label{tquasimain2} 
Analogue of Theorem~\ref{tquasimain1} holds for $\tilde s_d^{\pi,W}(n)$.
\end{theorem} 
For the purposes of the flip, Theorem~\ref{tquasimain1} suffices.

\proof 
We shall only prove Theorem~\ref{tquasimain1}, the proof of Theorem~\ref{tquasimain2} being
similar.
The proof is an extension of  M. Brion's proof (cf. \cite{dehy})  of quasi-polynomiality of the
stretching function associated with a Littlewood-Richardson coefficient
of any semisimple Lie algebra.

Clearly (a) follows from (b); cf. \cite{stanleyenu}.

\noindent (b) and (c):

Let $C_d$ be the cyclic group generated by the primitive root $\zeta$ of unity of order $d$.
It has a natural action on $R$: $ x \in C_d$ maps $z \in R_k$ to $x^k z$. 
Let $B=R^{C_d}=\sum_{n\ge 0} R_{nd} \subseteq R$ be the subring of $C_d$-invariants.
By Boutot \cite{boutot}, $B$  is a normal $\C$-algebra and $\spec(B)$ has rational singularities.

Assume that $H_0$ is semisimple; extension to the reductive case 
being easy. Let $\pi^*$ be the dominant weight of $H_0$ such that 
$V_\pi^*=V_{\pi^*}$. By Borel-Weil \cite{fultonrepr},
\[
C_{\pi^*}=\oplus_{n \ge 0} V_{n \pi}^*= 
\oplus_{n \ge 0} V_{n \pi^*},\] 
is the homogeneous coordinate ring of the $H_0$-orbit of the point 
$v_{\pi^*} \in P(V_{\pi^*})$  corresponding to the highest weight vector. This $H_0$-orbit
is isomorphic to $H_0/P_{\pi^*}$, where $P_{\pi^*} \subseteq H_0$ is the 
parabolic stabilizer of $v_{\pi^*}$. Hence $C_{\pi^*}$ is 
 normal and $\spec(C_{\pi^*})$ has rational signularities;
cf. \cite{hashimoto,mehta,ramanathan,smith}. 
It follows that 
 $B\otimes C_{\pi^*}$ is also  normal, and $\spec(B\otimes C_{\pi^*})$ has
rational singularities. Consider the action of $\C^*$ on 
$B\otimes C_{\pi^*}$ given by: 
\[ x (b \otimes c)=(x\cdot b) \otimes (x^{-1}
\cdot c),\] 
where $x \in \C^*$ maps  $b \in B_n$ to $x^n b$, the action on $\C_{\pi^*}$ being similar.
Consider   the invariant ring
\begin{equation} \label{eqS}
S=(B\otimes C_{\pi^*})^{\C^*}= \oplus_n S_n=
\otimes_{n \ge 0} R_{n d} \otimes V_{n \pi}^*.
\end{equation}
By Boutot \cite{boutot}, it is a normal, and $\spec(D)$ has rational singularities.

Since $V_{n \pi}$ is an $H$-module,
the algebra $S$  has an action of $H$. Let 
\begin{equation} \label{eqT}
 T=T(s_d^\pi)=S^{H}=\oplus_{n \ge 0} T_n 
\end{equation}
be its subring of $H$-invariants.
By Boutot \cite{boutot}, it is 
normal, and  $\spec(T)$ has  rational singularities--this is the crux of the proof. 
By Schur's lemma,
the multiplicity of the trivial $H$-representation in $S_n=R_{n d}\otimes V_{n \pi}^*$ is 
precisely the multiplicity $\tilde s_d^\pi(n)$  of the $H$-module $V_{n \pi}$ in $R_{n d}$. Hence,
the Hilbert function of $T$, i.e., $\dim(T_n)$,   is precisely 
$\tilde s_d^\pi(n)$, and the Hilbert series $\sum_{n\ge 0} \dim(T_n) t^n$ is 
$S_d^\pi(t)$.
Quasipolynomiality of $\tilde s_d^\pi(n)$ follows by applying the following lemma:

\begin{lemma} \label{ldehy} (cf. \cite{dehy})
If $T=\oplus_{n=0}^\infty T_n$ is a graded  $\C$-algebra,
such that $\spec(T)$ is normal and has rational simgularites, then
$\dim(T_n)$, the Hilbert function of $T$,  is a quasi-polynomial function of $n$.
\end{lemma} 

\noindent (d) 
Since $\spec(T)$ has rational singularities, $T$ is Cohen-Macaualey. Let $t_1,\ldots,t_u$ be
its homogeneous sequence of parameters (h.s.o.p.), where $u=k+1$ is the Krull dimension of $T$.
By the theory of Cohen-Macauley rings \cite{stanleycomb}, it follows that its Hilbert 
series $S_d^\pi(t)$ is of the form 
\begin{equation}  \label{eqposform1}
\f{h_0+ h_1 t + \cdots +h_k t^k} {\prod_{i=1}^{k+1}  (1-t^{d_i})},
\end{equation}
where (1) $h_0=1$, (2) $d_i$ is the degree of $t_i$, and (3) $h_i$'s are nonnegative integers. 
This proves (d).
\qed

\begin{remark} \label{rfinitegen}
A careful examination of the proof above shows that   rationality of $S_d^\pi(t)$,
and more strongly, asymptotic quasi-polynomiality of $\tilde s_d^\pi(n)$ as $n\rightarrow 
\infty$,  can be proved using just 
 Hilbert's result  on finite generation of the algebra of invariants of 
a reductive-group action.
Boutot's result is necessary to prove  quasi-polynomiality for all $n$. This 
is crucial   for saturated  and positive integer programming (Chapter~\ref{csatpos}).
\end{remark}

\subsection{The minimal positive form and modular index} \label{sminimal}
The form (\ref{eqposform1}) of $S_d^\pi(t)$  is not unique because it depends on the 
degrees $d_i$'s of the paramters $t_i$'s. 
For future use, let us record the following consequences of the proof.
Let $T$ be the ring constructed in the proof above.

\begin{cor} 
Suppose $T$ has an h.s.o.p. $t=(t_1,\ldots,t_u)$ with $d_i=\deg(t_i)$.
Then $S_d^\pi(T)$ has a positive rational form (\ref{eqposform1}) with 
$d_i=\deg(t_i)$ therein.
\end{cor}

The proof above is  lets us define
a minimal  positive form of the rational function $S_d^\pi(t)$ associated
with a  structural constant $s$. 
For this, let us order h.s.o.p.'s of $T$ lexicographically as per their degree sequences.
Here the degree seqeunce of an h.s.o.p. $t=(t_1,\ldots,t_u)$ is defined to be
$(d_1,\ldots,d_u)$, where $d_i=\deg(t_i)$.  
The form (\ref{eqposform1}) is the same  for any  h.s.o.p. of lexicographically minimum degree sequence. 
We call it the {\em minimal positive form} of $S_d^\pi(t)$. 
The {\em modular index} of $s^\pi_d$ is defined to be
$\max\{d_i\}$, where $(d_1,\ldots,d_u)$ is the degree sequence of a
lexicographically minimal h.s.o.p.
Since Problems~\ref{pintrokronecker}, \ref{pintroplethysm},\ref{pintrosubgroup}, 
\ref{pintrolittle} are special cases of Problem~\ref{pintrogit}, this 
defines  minimal  positive forms of the rational generating functions of the
stretching quasi-polynomials (cf. Theorem~\ref{tquasisubgroup})
associated with the structural constants in these problems,
and also the modular indices of these structural constants.

\subsection{The rings associated with a structural constant}\label{ssringsassoc}
The preceding proof  also associates with the structural constant $s$ a few rings which will
be important later. 
Specifically, let $S=S(s)$ and 
$T=T(s)$ be the rings as in Theorem~\ref{tquasimain1} (c) associated with
the structural constant $s=s_d^\pi$. 
Let $R=R(s)$ be the homogeneous coordinate ring of $X$ as in Theorem~\ref{tquasimain1}.
We call $R(s),S(s)$ and $T(s)$ the
rings associated with the structure constant $s$.

When $s=m_\lambda^\pi$, as in the subgroup restriction problem (Problem~\ref{pintrosubgroup}), 
$X\cong G/P$  as given in eq.(\ref{eqXsubgroup}. Then
these rings are explicitly as follows:
\begin{equation}
\begin{array} {lcl} 
R(m_\lambda^\pi)&=&\oplus_{n\ge 0} V_{n \lambda} (G), \\
S(m_\lambda^\pi)&=&\oplus_{n\ge 0} V_{n \lambda}(G) \otimes V_{n \pi}(H)^*, \\
T(m_\lambda^\pi)&=& \oplus_{n\ge 0} (V_{n \lambda} (G) \otimes V_{n \pi}(H)^*)^H.
\end{array}
\end{equation}

By specializing the  subgroup restriction problem further to 
 the Littlewood-Richardson problem (Problem~\ref{pintrolittle}), we get the 
following rings associated by Brion (cf. \cite{dehy}) 
with the Littlewood-Richardson coefficient $c_{\alpha,\beta}^\lambda$:
\begin{equation} \label{eqringslittle}
\begin{array}{lcl}
R(c_{\alpha,\beta}^\lambda)&=&\oplus_{n\ge 0} V_{n \alpha} (H) \otimes V_{n \beta} (H), \\
S(c_{\alpha,\beta}^\lambda)&=&
\oplus_{n\ge 0} V_{n \alpha} (H) \otimes V_{n \beta} (H) \otimes V_{n \lambda}(H)^*, \\
T(c_{\alpha,\beta}^\lambda)&= &\oplus_{n\ge 0} (V_{n \alpha} (H) \otimes V_{n \beta} (H) \otimes 
V_{n \lambda}(H)^*)^H.
\end{array}
\end{equation}

\section{Canonical models} \label{scanonicalmodel}
There are  several rings other than $T(c_{\alpha,\beta}^\lambda)$
whose Hilbert function
coincides with the Littlewood-Richardson stretching quasi-polynomial 
$\tilde c_{\alpha,\beta}^\lambda (n)$. For example,
let $P=P_{\alpha,\beta}^\lambda$ be the $BZ$-polytope \cite{berenstein}
 whose Ehrhart quasi-polynomial coincides with $\tilde c_{\alpha,\beta}^\lambda (n)$.
We can associate with $P$ a ring $T_P$ as in  Stanley \cite{stanleytoric} 
whose Hilbert function coincides with 
$\tilde c_{\alpha,\beta}^\lambda (n)$. There are many other choices for $P$.
For example, in type $A$, we can consider a hive polytope or a honeycomb polytope \cite{knutson} instead of
the BZ-polytope. The rings $T_P$'s associated with  different $P$'s will, in general,
be different, and there is nothing canonical about them.
In contrast,  the ring 
$T(c_{\alpha,\beta}^\lambda)$ is special because:

\begin{prop} \label{plusztigkashi} {\bf (PH0)}
The rings $R(c_{\alpha,\beta}^\lambda), S(c_{\alpha,\beta}^\lambda), T(c_{\alpha,\beta}^\lambda)$
have  quantizations 
$R_q(c_{\alpha,\beta}^\lambda), S_q(c_{\alpha,\beta}^\lambda), T_q(c_{\alpha,\beta}^\lambda)$
endowed with  canonical bases in the terminology of Lusztig \cite{lusztigbook}. 
Furthermore, the canonical bases of $R_q(c_{\alpha,\beta}^\lambda), S_q(c_{\alpha,\beta}^\lambda)$
are compatible with the action of the Drinfeld-Jimbo quantum group associated with
$H=GL_n(\C)$, and the canonical basis of $S_q(c_{\alpha,\beta}^\lambda)$ is an 
extension of the canonical basis of $T_q(c_{\alpha,\beta}^\lambda)$ in a natural way.
\end{prop} 
This  follows from the work of Lusztig (cf. \cite{lusztigpnas}, Chapter 27 in \cite{lusztigbook})
 and Kashiwara (cf. Theorem2 in \cite{kashiwaraglobal}). 
Specializations of these canonical bases at $q=1$ will be  called canonical bases of
$R(c_{\alpha,\beta}^\lambda), S(c_{\alpha,\beta}^\lambda), T(c_{\alpha,\beta}^\lambda)$.
Lusztig \cite{lusztigbook} has conjectured that the structural constants
associated with the  canonical 
bases in Proposition~\ref{plusztigkashi} 
are polynomials in $q$  with nonnegative integral coefficients
as in the case of  the canonical basis of the (negative part of the)
Drinfeld-Jimbo enveloping algebra. We refer to Proposition~\ref{plusztigkashi}
as PH0 in view of this (conjectural) positivity property.

In view of this proposition, we  call the rings 
$R(c_{\alpha,\beta}^\lambda),S(c_{\alpha,\beta}^\lambda)$ and $T(c_{\alpha,\beta}^\lambda)$
the  {\em canonical
rings} associated with the Littlewood-Richardson coefficient $c_{\alpha,\beta}^\lambda$,
and $X=\proj(R(c_{\alpha,\beta}^\lambda)), Y=\proj(S(c_{\alpha,\beta}^\lambda))$
and $Z=\proj(T(c_{\alpha,\beta}^\lambda))$ the {\em canonical models} associated with
$c_{\alpha,\beta}^\lambda$.

\subsection{From PH0 to PH1,3} \label{ssignilittle}
Now we study the relevance of 
PH0 above in the context of PH1,SH,PH2, and PH3 for
Littlewood-Richardson coefficients (Section~\ref{sintroLR}).

\subsubsection{PH1}
As already remarked in Section~\ref{sapproach}, 
PH1 for Littlewood-Richardson coefficients 
is a formal  consequence of the properties of Kashiwara's crystal operators
on   the canonical bases in PH0 (Proposition~\ref{plusztigkashi});
\cite{dehy,kashiwara2,littelmann,lusztigbook}.

Specifically,  the canonical basis of the ring $R_q(c_{\alpha,\beta}^\lambda)$
also yields a canonical basis for the tensor product $V_{q,\alpha} \otimes
V_{q,\beta}$ of the irreducible 
$H_q$ modules with highest weights $\alpha$ and $\beta$.
The  Littlwood-Richardson rule for arbitrary types follows from the 
study  of Kashiwara's crystal operators on this  canonical basis for the
tensor product; \cite{lusztigbook}.
 This rule is equivalent to the one in \cite{littelmann} 
based on combinatorial interpretation of the
crystal operators in the path model therein.
The article \cite{dehy} derives a convex polyhedral formula for 
Littlewood-Richardson coefficients (of arbitrary type) using this
combinatorial interpretation.
Though the complexity-theoretic issues are not addressed in \cite{dehy},
it can be verified that the polyhedral formula therein is a 
convex $\#P$-formula. This yields PH1 for Littlewood-Richardson coefficients
of arbitrary types using PH0. 

\subsubsection{SH}
Now let us see the relevance of PH0 in the context of SH for 
Littlewood-Richardson coefficients of arbitrary type.

The polytope in \cite{dehy}, mentioned above, for type $A$ is equivalent 
to the hive polytope in \cite{knutson} in the sense that the number
integer points in both the polytopes is the same.
Knutson and Tao prove SH for type $A$ 
by showing that the hive polytope always has in
integral vertex. 
To extend this proof to an arbitrary type,  one has to 
convert the 
polytope in \cite{dehy} into a polytope that is guaranteed to contain an integral vertex 
if the index of the stretching quasipolynomial $\tilde c_{\alpha,\beta}^\lambda(n)$ is one.
The main difficulty here  is that we do not have a nice mathematical interpretation for
the index.
Algorithm in Theorem~\ref{tindexquasi} applied to the polytope in \cite{dehy}
 computes this index
in polynomial time.
 But it does not give a nice interpretation  that can be used in
a proof as above. 

This index is simply the largest integer dividing the
degrees of all  elements
in any basis of the  canonical ring 
$T(c_{\alpha,\beta}^\lambda)$--in particular, the 
canonical basis.  This follows by applying Proposition~\ref{pcorsatint} to the
polytope in \cite{dehy}.
This leads us to ask: is there an interpretation for
 the index based on Lusztig's
topological construction of the canonical basis
in Proposition~\ref{plusztigkashi}? If so, this may be used to extend 
the known polyhedral proof for SH in type $A$ to arbitrary types. 
Alternatively, it may be possible to prove SH using topological properties of the
canonical basis in the spirit of the
 topological (intersection-theoretic) proof 
\cite{belkale} of SH in type $A$.

\subsubsection{PH3} 
Now let us see the relevance of PH0 in the context of PH3 for
Littlewood-Richardson coefficients.

First, let us consider the minimal  positive form (Section~\ref{sminimal}) 
associated with
a Littlewood-Richardson coefficient $c_{\alpha,\beta}^\lambda$ of type $A$.
Let $T=T(c_{\alpha,\beta}^\lambda)$ denote the ring that arises
in this case; cf. eq.(\ref{eqringslittle}).
Now we can ask:

\begin{question} \label{qintrointegral}
Are all $d_i$'s  occuring in  the minimal   positive form (cf.  (\ref{eqposform1})) 
one in this special case? 
This is equivalent to asking if the ring 
$T=T(c_{\alpha,\beta}^\lambda)$  in this case is integral over 
$T_1$, the degree one component of $T$. 
\end{question} 
If so, this would provide an explanation for the conjecture of King at al \cite{king}
(cf. eq.(\ref{eqintro1})) in the theory of Cohen-Macauley rings:

\begin{prop} 
Assuming yes,
the conjecture of King et al \cite{king} (Hypothesis~\ref{hintrolittleph3}) holds.
\end{prop} 

\begin{remark} In contrast,  the ring $T_P$ associated with the hive polytope 
(cf. beginning of Section~\ref{scanonicalmodel}) 
need not be integral over
its degree one component, in view of the fact that the hive polytope can have nonintegral
vertices \cite{deloeravertices}.
\end{remark} 

\begin{remark}  $T=T(c_{\alpha,\beta}^\lambda)$ need not be 
generated by its degree
one component $T_1$. If this were always so,
the $h$-vector $(h_d,\cdots,h_0)$ in eq.(\ref{eqintro1}) would be an M-vector (Macauley-vector)
\cite{stanleycomb}. But one can construct  $\alpha,\beta$ and $\lambda$ for which  this does not
hold. 
\end{remark} 

\proof (of the proposition)
Since $T$ is integral over $T_1$, it has an h.s.o.p., all of whose elements have degree $1$.
By Theorem~\ref{tquasisubgroup},  the singularities of $\spec(T)$ are rational. Hence $T$
is Cohen-Macaulay. Now  the result
 immediately follows from the theory of Cohen-Macauley
rings \cite{stanleycomb}. \qed

In view of this Proposition, the conjecture of King et al will follow if all
canonical basis elements of $T(c_{\alpha,\beta}^\lambda)$ can be shown to be integral
over the basis elements of degree one. This requires a further study of the
multiplicative structure of this canonical basis.
Considerations  for PH3 (Hypothesis~\ref{hintrolittleph3gen})
 for  Littlewood-Richardson 
coefficients of arbitrary type are similar.

\subsubsection{PH2} 
Similarly,
 the positivity property (PH2)
of the stretching quasipolynomial
associated with   Littlewood-Richardson coefficients may possibly follow
from  a deep study of the multiplicative structure of the canonical basis 
as per PH0 (Proposition~\ref{plusztigkashi}),
 just as positivity of the multiplicative structural
coefficients of the canonical basis for the (negative part of the)
Drinfeld-Jimbo enveloping algebra  follows from a 
deep study of the  multiplicative structure of this basis  \cite{lusztigbook}.

\subsection{On PH0 in general} \label{sph0general}
The discussion above indicates that for Littlewood-Richardson coefficients
PH1,SH,PH3, and plausibly PH2 as well are intimately related to PH0
(Proposition~\ref{plusztigkashi}). 
This leads us to ask if the  rings associated in Section~\ref{ssringsassoc}
 with
 other structural constants under
consideration in this paper  have quantizations which satisfy appropriate
forms of PH0. If so, this PH0 may be used to derive 
PH1, SH, PH3, and  PH2 (Hypotheses~\ref{phph1}-\ref{phph3}) 
for these structural constants. Note that
SH (a)  follows from PH3 (see the remark after Hypothesis~\ref{phph3});
PH2 may also follow
from PH3. Thus PH1 and PH3 are the ones to focus on.

To formalize this,
let $s$ be a structural constant which is either the Kronecker coefficient 
as in Problem~\ref{pintrokronecker}, or the plethysm constant as in Problem~\ref{pintroplethysm},
or the multiplicity $m_\lambda^\pi$ in Problem~\ref{pintrosubgroup}, or 
the multiplicity $s_d^\pi$, as in Problem~\ref{pintrogit}, when
 $X$ therein is  a class variety.
Let  $R(s),S(s),T(s)$ be the rings  associated with $s$ (Section~\ref{ssringsassoc}). 
Let $X(s)=\proj(R(s)), Y(s)=\proj(S(s))$ and $Z(s)=\proj(R(s))$.
We  call $R=R(s), S=S(s), T=T(s)$
  the {\em canonical rings} associated with $s$,
and $X(s), Y(s), Z(s)$ the {\em canonical models} associated with $s$,
because we expect these rings and models
to be special as in the case of the Littlewood-Richardson coefficients. 

Let $H$ be as in Problem~\ref{pintrosubgroup} or Problem~\ref{pintrogit}.
Assume that $H$ is connected. Let $H_q$
denote the Drifeld-Jimbo quantization of $H$.
Now we  ask:

\begin{question} \label{queph0main} {\bf (PH0??)}
Are there 
quatizations $R_q$, $S_q$ of $R,S$, with $H_q$-action,
and a quantization 
$T_q$ of $T$
with  ``canonical'' bases (in some appropriate sense)
 $B(R_q),B(S_q),B(T_q)$,
where $B(R_q)$ and $B(S_q)$ are compatible with the $H_q$-action 
and $B(S_q)$ is an extension of $B(T_q)$? Furthermore, do these 
canonical bases have appropriate positivity properties? 

In other words, are there quantizations  of $R,S$ and $T$ for which
PH0 (Proposition~\ref{plusztigkashi})  can be extended in a natural way? 
\end{question} 

If so, this extended PH0 may be used
to prove PH1 and SH for $s$ 
just as in the case of Littlewood-Richardson coefficients (of type $A$).

\section{Nonstandard quantum group for the Kronecker and the plethysm problems} \label{squantumgroup}
We now consider this question when
$s$  is the kronecker or
the plethysm constant (cf. Problems~\ref{pintrokronecker} and 
\ref{pintroplethysm}).
PH0 for Littlewood-Richardson coefficients (Proposition~\ref{plusztigkashi})
depends critically  on the theory of
Drinfeld-Jimbo quantum groups. This  is intimately related 
(in type $A$) \cite{gro} to the 
representation theory of Hecke algebras. 
To extend PH0 in the context of the kronecker and the plethysm constants, 
one needs  extensions of these theories
in the context of Problems~\ref{pintrokronecker}-\ref{pintroplethysm}.
In this section, we briefly review the results in 
\cite{GCT4,canonical,plethysm} in this direction 
and the theoretical and
experimental evidence it provides in support of PH0--that is,
affirmative answer to Question~\ref{queph0main}--in this context.

So let us consider
the generalized plethysm problem (Problem~\ref{pintroplethysm}).
As expected,  the representation theory of Drinfeld-Jimbo quantum groups
and Hecke algebras  does not work in the  context of this general problem.
Briefly, the problem is that if $H$ is 
a connected, reductive group and $V$ its representation,
then the homomorphism $H\rightarrow G=GL(V)$ does not quantize in the
setting of Drinfeld-Jimbo quantum groups. 
That is, there is no quantum group homomorphism from $H_q$, the 
Drinfeld-Jimbo quantization of $H$, to $G_q$, the Drinfeld-Jimbo quantization
of $G$. In \cite{GCT4,plethysm}, a new {\em nonstandard} 
quantization $G^H_q$ of $G$-- called a {\em nonstandard quantum group}--is
 constructed so that
there is a quantum group homomorphism $H_q \rightarrow  G^H_q$.
When $H=G$, $G^H_q$ coincides with the Drinfeld-Jimbo quantum group.
The article  \cite{canonical} gives a conjectural scheme 
for constructing a {\em nonstandard canonical basis} for the matrix
coordinate ring of
$G^H_q$ that is akin to the  canonical basis for the
matrix coordinate ring  of the Drinfeld-Jimbo
quantum group \cite{lusztigbook,kashiwaraglobal}.

It is known that the Drinfeld-Jimbo quantum group 
$G_q=GL_q(V)$ and the Hecke algebra $H_n(q)$ are dually paired: i.e.,
they have commuting actions on $V_q^{\otimes n}$ from the
left and the right that determine each other, where $V_q$ denotes the standard
quantization of $V$. Furthermore,
the Kazhdan-Lusztig basis for $H_n(q)$ is intimately 
related  to the canonical basis for $G_q$ \cite{gro}. Similarly, 
\cite{plethysm} constructs  a {\em nonstandard} 
generalization  $B^H_n(q)$ of the Hecke algebra which is 
(conjecturally) 
dually paired  to $G^H_q$. The article \cite{canonical} 
gives a conjectural scheme for constructing a {\em nonstandard canonical basis}
 of $B^H_n(q)$ akin to the Kazhdan-Lusztig basis of the Hecke algebra
$H_n(q)$.

The nonstandard quantum group $G^H_q$ and the nonstandard algebra 
$B^H_n(q)$ turn out to be fundamentally different from the standard
Drinfeld-Jimbo quantum group $G_q$ and the Hecke algebra $H_n(q)$. 
For example, the nonstandard quantum group $G^H_q$ is a 
nonflat deformation of $G$ in general.
This means the Poincare series of the matrix coordinate ring of $G^H_q$ 
is different from the Poincare series of the matrix coordinate ring of
$G$. Specifically,
the terms of the first series can be smaller than the respective terms
of the second series. Similarly, $B^H_n(q)$
is a nonflat deformation of the group algebra $\C[S_n]$ of the 
symmetric group $S_n$; i.e., its dimension can be bigger than that of
$\C[S_n]$.

Nonflatness of  $G^H_q$ intuitively means that it is ``smaller''
than  $G$ in general. Hence,  it  may seem that
there is a loss of information when one goes from $G$ to $G^H_q$.
Fortunately, there is none, as per  the reciprocity conjecture 
in \cite{plethysm}. This roughly says that the information which
is lost in the transition from $G$ to $G^H_q$ simply gets transfered 
to $B^H_n(q)$, which is bigger than $H_n(q)$. In other words, there
is no information loss overall. Hence
analogues of the properties in the standard setting should also hold
in the nonstandard setting, though in a far more complex way.

That is what seems to happen to positivity. Specifically, experimental
evidence suggests that the conjectural nonstandard canonical bases 
 in \cite{canonical} have nonstandard positivity 
properties  which are 
complex versions of the positivity properties
in the standard setting. See \cite{plethysm,canonical,GCT10} for a 
detailed story.

\section{The cone associated with the subgroup restriction problem}\label{sconesub}
In this section, we prove Theorem~\ref{tfinitegensubgroup},
 by extending the proof of Brion and Knop (cf. \cite{elashvili}) for
the Littlewood-Richardson problem. The proof is in the spirit of the 
proof of quasipolynomiality in Section~\ref{squasiproof}. 
 
Let $G$ be a connected, reductive group,
 $H$ a connected, reductive subgroup, and $\rho:H\rightarrow G$ a homomorphism.
Theorem~\ref{tfinitegensubgroup}  has the following equivalent formulation. 
 Let $S(H,G)$ be the set of pairs $(\mu,\lambda)$  such that 
$V_\mu(H) \otimes V_\lambda(G)$ has a nonzero $H$-invariant.
Then,
\begin{theorem} 
The set $S(H,G)$ is a finitely generated semigroup with 
respect to addition. 
\end{theorem} 

When $G=H\times H$ and the embedding $H\subseteq G$ is diagonal, this
specializes  to the Brion-Knop result mentioned above.
The proof follows by
an  extension the technique therein.

\proof
Let $B$ be a Borel subgroup of $G$, $U$ the unipotent radical of $B$ and
$T$ the maximal torus in $B$. Similarly, let
$B'$
 be a Borel subgroup of $H$, $U'$ the unipotent radical of $B'$ and
$T'$ 
the maximal torus in $B'$. Without loss of generality, we can assume that
$B'\subseteq B$, $U'\subseteq U$, $T'\subseteq T$. 
 Let $A=\C[G]^U$ be the algebra of regular
functions on $G$ that are invariant with respect to the right multiplication
by $U$. It is known to be finitely generated \cite{elashvili}.
The groups  $G$ and $T$ act on $A$ via
left and right multiplication, respectively.  As a $G\times T$-module,
\begin{equation} 
A=\oplus_\lambda V_\lambda(G),
\end{equation}
where  the torus $T$ acts on $V_\lambda(G)$ via
multiplication by the highest weight $\lambda^*$ of the dual module.
Similarly, 
\begin{equation} 
A'=\C[H]^{U'}=\oplus_\lambda V_\mu(H),
\end{equation}
where  the torus $T'$ acts on $V_\mu(H)$ via
multiplication by the highest weight $\mu^*$ of the dual module.

Now  $A\otimes  A'$ is finitely generated since
$A$ and $A'$ are. Let $X=(A\otimes A')^H$ be  the ring of invariants of $H$
acting diagonally on $A \otimes A'$. The torus 
$T\times T'$ acts on $X$ from the right.
Since $H$ is reductive, $X$ is finitely
generated \cite{popov}. 
Hence,  the semigroup of the weights of the right action of $T\times T'$ 
on $X$ is finitely generated.
We have
\[ 
X=(A\otimes  A')^H=((\oplus V_\lambda(G))\otimes (\oplus V_\mu(H)))^H
=\oplus(V_\lambda(G)\otimes V_\mu(H))^H,\]
and the weights of the algebra $X$ are of the form $(\lambda^*,\mu^*)$ 
such that $V_\lambda(G)\otimes V_\mu(H)$ contains a nontrivial $H$-invariant.
Therefore these pairs form a finitely generated semigroup.
\qed

For the sake of simplicity, assume that $G$ and $H$ are semisimple 
in what follows.
Let $T_\R(H,G)$ denote the polyhedral convex cone in
the weight space of $H\times G$ generated by  $T(H,G)$, as defined in 
Theorem~\ref{tfinitegensubgroup}. 
This is a generalization of the Littlewood-Richardson cone (Section~\ref{slittlecone}).

The following generalization of Corollary~\ref{csatlit} 
is a consequence of Theorem~\ref{tindexquasi} and its proof.

\begin{theorem} \label{tconesatform}
Assume that the positivity hypothesis PH1 (Section~\ref{sphypo}) holds for the 
subgroup restriction  problem for the pair $(H,G)$, where both $H$ and
$G$ are classical.
Given dominant weights $\mu,\lambda$ of $H$ and $G$, the polytope 
$P_{\mu,\lambda}$ as in PH1 has a 
specification of the form 
\begin{equation}  \label{eqtconesat}
A x \le b
\end{equation} 
where $A$ depends only on $H$ and $G$, but not on $\mu$ or $\lambda$,
and $b$ depends homogeneously and linearly on $\mu,\lambda$.
Let $n$ be the total number of columns in $A$.

Then, there exists a decomposition of $T_\R(H,G)$ into a set of polyhedral
cones, which form a cell complex ${\cal C}(H,G)$,
and, for each chamber $C$ in this
complex, 
a set $M(C)$ of $O(n)$ modular equations, each of 
the form 
\[ \sum_i a_i \mu_i + \sum_i b_i \lambda_i  = 0 \quad (mod \ d),\] 
such that 
\begin{enumerate} 
\item Saturation hypothesis SH   is equivalent to saying that:
$(\mu,\lambda) \in T(H,G)$ iff 
$(\mu,\lambda) \in T_\R(H,G)$ and 
$(\mu,\lambda)$ satisfies the modular equations in the 
set $M(C_{\mu,\lambda})$ associated with the smallest cone 
$C_{\mu,\lambda}\in {\cal C}(H,G)$ containing $(\mu,\lambda)$.
\item Given $(\mu,\lambda)$, 
whether $(\mu,\lambda) \in T_\R(H,G)$ can
be determined in  polynomial time.
\item If so, whether $(\mu,\lambda)$ satisfies the 
modular equations associated with the smallest cone in ${\cal C}(H,G)$ 
containing it can also be determined 
in  polynomial time.
\end{enumerate} 
\end{theorem}
\proof 
Given a point $p=(\mu',\lambda')$
in the weight space of $H\times G$, where $\mu'$ and
$\lambda'$ are arbitrary rational points,  let $S(p)$ 
denote the constraints (half-spaces)  in the sytem (\ref{eqtconesat}) whose
bounding hyperplanes contain the polytope $P_{\mu',\lambda'}$.
We can decompose $T_\R(H,G)$ into a conical, polyhedral 
 cell complex, so that given a cone $C$ in this complex, and a point 
$p$ in its interior,  the set $S(p)$  does not depend on $p$. 
We shall denote this set by $S(C)$. Thus the affine span of
$P_{\mu,\lambda}$, for any $(\mu,\lambda) \in C$, is 
determined by the linear system
\[ A'x=b',\] 
where $[A',b']$ consists of the rows of $[A,b]$ in (\ref{eqtconesat}) 
corresponding to the set $S(C)$. 
By finding the Smith normal form of $A'$, 
we can associate with $C$ a 
set of modular equations that the entries of $b'$ must satisfy 
for this affine span to contain an integer point; see the proof of
Theorem~\ref{tindexquasi}. Since the entries 
of $A'$ depend only on $H$ and $G$, these equations depend only on
$C$. If $(\mu,\lambda)\in T(H,G)$, then $(\mu,\lambda)$ is integral, and
hence these equations are satisfied. 
Conversely, if $(\mu,\lambda) \in T_\R(H,G)$ and
these equations are satisfied,  then the saturation
property implies that $(\mu,\lambda) \in T(H,G)$, as seen by examining  the
proof of Theorem~\ref{tindexquasi}.
Furthermore, given $(\mu,\lambda)$, the algorithm in the proof
of Theorem~\ref{tindexquasi} implicitly determines if $(\mu,\lambda) \in T_\R(H,G)$
and if these modular equations are satisfied in polynomial time. \qed

\section{Elementary proof of rationality} \label{selement}
In this section we give an elementary proof of rationality in Theorem~\ref{tquasisubgroup} (a),
when $H$ therein is connected--actually of   a slightly stronger statement: namely,
the stretching function $\tilde m_{\lambda}^\pi(n)$ is 
asymptotically a quasipolynomial, as $n \rightarrow \infty$;  cf. Remark~\ref{rfinitegen}.
But this proof  cannot be extended to prove  quasipolynomiality for all $n$.
%One advantage of this proof is that it suggests a method for proving a polynomial bound 
%on the order of the poles of the rational generating function of the stretching 
%function associated with  a Kronecker coefficient (Section~\ref{spole}). 
The proof here  is motivated by the work of Rassart \cite{rassart}, De Loera and McAllister on
the stretching function associated with a Littlewood-Richardson coefficient.

First, we recall some standard results that we will need.

\subsection*{Vector partition functions}\label{ssvectorpart}
Given an integral $s\times n$ matrix $B$ and integral $n$-vector $c$, 
consider the vector paritition function $\phi_B(c)$,
 which is the number of integer 
solutions to the integer programming problem 
\begin{equation} \label{eqposint1}
 By=c, \quad y\ge 0.
\end{equation}
For a fixed $c,b$, let
\begin{equation} 
\begin{array} {l}
\phi_{B,c}(n)=\phi_B(n c) \\
\phi_{B,c,b}(n)=\phi_B(n c+b).
\end{array} 
\end{equation}

By Sturmfels \cite{sturmfels} and Szenes-Vergne residue formula \cite{szenes},
 $\phi_B(c)$  is a piecewise quasipolynomial 
function of $c$. That is, 
$\R^n$ can be decomposed into polyhedral cones, called 
chambers, so that the restriction of $\phi_B(c)$ to each  chamber $R$
is a multivariate  quasipolynomial 
function of the coordinates of $c$.
This implies that 
$\phi_{B,c}(n)$ is a quasipolynomial function of $n$. It also implies that 
the function
$\phi_{B,c,b}(n)$ is asymptotically  a quasipolynomial function of $n$, as $n\rightarrow \infty$,
because the points $n c +b$, as $n\rightarrow \infty$, lie in just one chamber.

The Szenes-Verne residue formula \cite{szenes}  for vector partition functions also
implies that there is a constant $d(B)$, depending only on $B$, such that
the period of $\phi_{B,c}(n)$, for any $c$, divides $d(B)$.

\subsection*{Klimyk's formula}
Let $H \subseteq G$ and $m_\lambda^\pi$  be as in Theorem~\ref{tquasisubgroup} 
(a), with $H$ connected. Let us assume that $H$ is semisimple, the general case being similar.
Let ${\cal H}$ and ${\cal G}$  be the Lie algebras of $H$ and $G$ respectively.
We  recall Klimyk's formula for  $m_\lambda^\pi$.
Without loss of generality, we can assume that the Cartan subalgebra 
${\cal C} \subseteq {\cal H}$ is a subalgebra of the Cartan subalgebra 
${\cal D} \subseteq {\cal G}$. So we have a restriction from ${\cal D}^*$
to ${\cal C}^*$, and we assume that the half-spaces determining positive
roots are compatible. We denote weights of ${\cal H}$ by symbols such as $\mu$ and 
of ${\cal G}$ by  symbols such as  $\bar \mu$.
To be consistent, we shall use the notation 
$m^\pi_{\bar \lambda}$ instead of $m_\lambda^\pi$ in this proof.
We write $\bar \mu\downarrow \mu$ if
the  weight $\bar \mu$ of ${\cal G}$
restricts to the weight $\mu$ of ${\cal H}$.
We denote a typical element of the Weyl group of ${\cal H}$ by $W$,
and a  typical element of the Weyl group of ${\cal G}$ by $\bar W$.
Given a dominant weight $\pi$ of ${\cal G}$ and a weight 
$\bar \mu$ of ${\cal G}$, let $n_{\bar \mu}({\bar \lambda})$ denote the 
dimension of the weight space for $\bar \mu$ in 
$B_{\bar \lambda}=V_{\bar \lambda}(G)$. 

We  assume  that:

\noindent (A): For any weight $\mu$ of ${\cal H}$,
the number of $\bar \mu$'s such that $\bar \mu \downarrow  \mu$ is
finite. 

For example, this is so in the plethysm problem (Problem~\ref{pintroplethysm}). 
We shall see later how this assumption can be removed.

By Klimyk's formula  (cf.  page 428, \cite{fultonrepr}), 
\begin{equation} \label {eqklimyk}
 m^\pi_{\bar \lambda}=\sum_W (-1)^W \sum_{\bar \mu\downarrow \pi-\rho -W(\rho)}
n_{\bar \mu}(V_{\bar \lambda}),
\end{equation}
where $\rho$ is half the sum of positive roots of ${\cal H}$.
We  allow $\bar \mu$ in the inner  sum to range over all weights $\bar \mu$ 
of ${\cal G}$ such that $\bar \mu\downarrow \pi-\rho -W(\rho)$ 
by defining $n_{\bar \mu}(V_{\bar \lambda})$ to be zero 
if $\bar \mu$ does not occur in $V_{\bar \lambda}$.

\subsection*{Proof of Theorem~\ref{tquasisubgroup} (a)}
The goal is to  express 
$\tilde m^\pi_{\bar \lambda}(n)$ as a linear combination of vector partition functions 
$\phi_{B,c,b}(n)$'s, for suitable $B,c,b$'s,
using Klimyk's formula for $m^\pi_{\bar \lambda}$. After this, we can deduce 
asymptotic quasipolynomiality of $\tilde m^\pi_{\bar \lambda}(n)$
 from asymptotic quasipolynomiality of 
$\phi_{B,c,b}(n)$'s.

By Kostant's multiplicity formula (cf. page 421 \cite{fultonrepr}),
\begin{equation} \label{eqkostantform}
n_{\bar \mu}(V_{\bar \lambda})=\sum_{\bar W} (-1)^{\bar W}
 P(\bar W({\bar \lambda}+\bar \rho)-(\bar \mu+\bar \rho)),
\end{equation}  
where $P(\bar \lambda)$, for a weight $\bar \lambda$ of ${\cal G}$,
denotes the Kostant partition function; i.e., the number of ways to write
$\bar \lambda$ as a sum of positive roots of ${\cal G}$. 
It is important for the proof that Kostant's  formula (\ref{eqkostantform}) holds 
even if $\bar \mu$ is not a weight that occurs in the representation 
$V_{\bar \lambda}$--in this case, $n_{\bar \mu}(V_{\bar \lambda})=0$,
and the right hand side of (\ref{eqkostantform})  vanishes.

By eq.(\ref{eqklimyk}) and (\ref{eqkostantform}),

\begin{equation} \label{eqkliko}
 m^\pi_{\bar \lambda}=\sum_W \sum_{\bar W} (-1)^W (-1)^{\bar W} 
 \sum_{\bar \mu\downarrow \pi-\rho -W(\rho)}
 P(\bar W({\bar \lambda}+\bar \rho)-(\bar \mu+\bar \rho)).
\end{equation}

Let $D$ denote the dominant Weyl chamber in the weight space of
${\cal G}$. 
Let ${\cal C}$ denote the Weyl chamber complex associated with 
the weight space of ${\cal G}$. The cells in this complex are
closed polyhedral cones. Each cone is either
the chamber $\bar W(D)$, for some  Weyl group
element $\bar W$, or a closed face of $\bar W(D)$ of any dimension.

Using M\"obius inversion,  the inner sum 
\[ \sum_{\bar \mu\downarrow \pi-\rho -W(\rho)}
 P(\bar W({\bar \lambda}+\bar \rho)-(\bar \mu+\bar \rho)) \] 
in eq.(\ref{eqkliko}) can be written as a  linear combination 
\[ \sum_C a(C) \sum_{\bar \mu \in C:
\bar \mu\downarrow \pi-\rho -W(\rho) } 
 P(\bar W({\bar \lambda}+\bar \rho)-(\bar \mu+\bar \rho)), \] 
where $C$ ranges over chambers in the Weyl chamber complex ${\cal C}$,
$a(C)$ is an appropriate constant for each $C$.

Hence,

\begin{equation}
 m^\pi_{\bar \lambda}=\sum_W \sum_{\bar W} (-1)^W (-1)^{\bar W} 
\sum_C a(C) \sum_{\bar \mu \in C:
\bar \mu\downarrow \pi-\rho -W(\rho) } 
 P(\bar W({\bar \lambda}+\bar \rho)-(\bar \mu+\bar \rho)).
\end{equation}

Now think of $\pi$ and ${\bar \lambda}$ as variables. But
${\cal H}$ and ${\cal G}$ are fixed, and hence also the quantities 
such as $\rho$ and $\bar \rho$. 
\begin{claim} \label{claimelem}
For fixed Weyl group elements $W,\bar W$ and a fixed $C$, the sum 
\begin{equation} \label{eqsum1}
\sum_{\bar \mu \in C:
\bar \mu\downarrow \pi-\rho -W(\rho) } 
 P(\bar W({\bar \lambda}+\bar \rho)-(\bar \mu+\bar \rho)) 
\end{equation} 
can be expressed as a vector partition function associated with an
appropriate 
linear system 
\begin{equation} \label{eqlinsystemkliko}
B y = c,  \quad y\ge 0,
\end{equation} 
where the matrix 
\[ B=B_{{\cal H},{\cal G},C},\] 
depends only on $C$
and the root systems of ${\cal H}$ and ${\cal G}$, but not on $\pi$ and
$\bar \lambda$,
and the coordinates of the vector 
\[c= m_{W,\bar W, C}(\bar \lambda,\pi,\rho,\bar \rho),\] 
depend on 
$W,\bar W,C,\rho,\bar \rho,\pi,\pi$, and furthermore, their dependence 
on $\pi,\bar \lambda,\rho,\bar \rho$ is linear.
\end{claim} 
Here assumption (A) is crucial. Without it, the sum (\ref{eqsum1}) can
diverge. 
Of course, without assumption (A), we can still 
make the sum  finite,
by requiring that $\bar \mu$ lie within the convex hull 
$H_{\bar \lambda}$ generated
by the points $\{\bar W(\bar \lambda)\}$, where $\bar W$ ranges over
all Weyl group elements. This means we have to add constraints to the
system (\ref{eqlinsystemkliko}) 
corresponding to the facets of $H_{\bar \lambda}$. 
But the entries  of the resulting $B$ would depend on 
$\bar \lambda$, and  the theory of vector partition functions
 will no longer apply.

\noindent {\em Proof of the claim:} 
Let $\bar \mu_i$'s denote the integer coordinates of $\bar \mu$ in the basis of 
fundamental weights. We denote the integer vector $(\bar \mu_1,\bar \mu_2,\cdots)$
by $\bar \mu$ again.
The  Kostant partition function $P(\nu)$ is a vector partition function
associated with an integer programming problem:

\[B_P v = \nu, \quad v \ge 0,\]
where the columns of $B_P$ correspond to positive roots of ${\cal G}$.
The sum in (\ref{eqsum1}) is equal to the number of integral pairs 
$(\bar \mu,v)$ such that
\begin{enumerate} 
\item $\bar \mu \in C$,
\item $\bar \mu \downarrow \pi -\rho-W(\rho)$,
\item $B_P v= \bar W (\bar \lambda+\bar \rho)-(\bar \mu+\bar \rho)$, $v \ge 0$.
\end{enumerate} 

The first two condititions here 
can be expressed in terms
of  linear constraints (equalities and inequalities) on the coordinates
$\bar \mu_i$'s.
Thus the three conditions together can be expressed in terms of
linear constraints on $(\bar \mu,v)$. 
By the finiteness assumption (A),
the polytope determined by these constraints
is a bounded polytope. The number of integer points in such a polytope
can be expressed as a vector partition function (cf. \cite{baldoni}).
This proves the claim.

Let us  denote the vector partition associated with the 
integer programming problem (\ref{eqlinsystemkliko}) in the claim by
$\phi_{W,\bar W,C}(c(\bar \lambda,\pi,\rho,\bar \rho))$.
Then
\begin{equation} \label{eqmlambda}
 m^\pi_{\bar \lambda}=\sum_W \sum_{\bar W} (-1)^W (-1)^{\bar W} 
\sum_C a(C) \phi_{W,\bar W,C}(c({\bar \lambda},\pi,\rho,\bar \rho)).
\end{equation}

Hence, 
\begin{equation}
\tilde m_{\bar \lambda}^\pi(n)=
m_{n \pi}^{n {\bar \lambda}}=\sum_W \sum_{\bar W} (-1)^W (-1)^{\bar W} 
\sum_C a(C)
 \phi_{W,\bar W,C}(c(n {\bar \lambda},n \pi,\rho,\bar \rho)).
\end{equation}

It follows from Claim~\ref{claimelem} and 
 the standard results on vector partition functions mentioned in the begining of this section
that 
\[
g_{W,\bar W,C}(n)=
 \phi_{W,\bar W,C}(c(n {\bar \lambda},n \pi,\rho,\bar \rho)),
\]
is asymptitically a  quasipolynomial function of $n$.
Hence,  $\tilde m_{\bar \lambda}^\pi(n)$ is also asymptotically 
a quasipolynomial function of $n$.
 This implies  (cf.  \cite{stanleyenu}) 
that 
\begin{equation} \label{eqrationaltemp}
M_{\bar \lambda}^\pi(t)=\sum_{n \ge 0} \tilde m^{\pi}_{\bar \lambda}(n) t^n
\end{equation}
is rational function of $t$. 

This proves Theorem~\ref{tquasisubgroup} (a)   under the finiteness assumption (A).

It remains to remove the  assumption (A).
Let ${\cal G'} \supseteq {\cal H}$ be the smallest Levi subalgebra of
${\cal G}$ containing ${\cal H}$. Then 
\begin{equation} \label{eqfiniassu}
m_{\bar \lambda}^\pi=
\sum_{\pi'} m_{\bar \lambda}^{\pi'} m_{\pi'}^{\pi},
\end{equation} 
where $\pi'$ ranges over dominant weights of ${\cal G'}$,
$m_{\bar \lambda}^{\pi'}$ denotes the multiplicity of 
$V_{\pi'}({\cal G}')$ in $V_{\bar \lambda}({\cal G})$, and
$m_{\pi'}^{\pi}$ the multiplicity of 
$V_{\pi}({\cal H})$ in $V_{\pi'}({\cal G}')$.
Furthermore, 
\begin{enumerate}
\item the finiteness asssumption (A) is now satisfied 
for the pair $({\cal G}',{\cal H})$: i.e., for any weight $\mu$ of
${\cal H}$, the number of weights $\mu'$'s of ${\cal G}'$ 
such that $\mu' \downarrow \mu$ is finite.
\item There is a polyhedral expression for $m_{\bar \lambda}^{\pi'}$; this follows from
\cite{littelmann,dehy}.
\end{enumerate} 

By the first condition and the argument above, we get an expression for
$m_{\pi'}^{\pi}$ akin to (\ref{eqmlambda}). 
Substituting this expression and the polyhedral expression for
$m_{\bar \lambda}^{\pi'}$ 
in (\ref{eqfiniassu}), leads to a formula for $\tilde m_{\bar \lambda}^\pi(n)$ as a 
linear combination of $\phi_{B,c,b}(n)$'s for appropriate $B,c,b$'s.
After this, we proceed as before.

This proves Theorem~\ref{tquasisubgroup} (a). \qed

We also note down the following consequence of the proof.

\begin{prop} \label{pboundorder}
There is a constant $D$ depending only
${\cal G}$ and ${\cal H}$, such that for any $\bar \lambda,\pi$, 
orders of the poles of $M_{\bar \lambda}^\pi(t)$ (cf. (\ref{eqrationaltemp}), as roots of unity, 
divide $D$. 
\end{prop}
A bound on $D$ provided by the proof below is very weak:
$D=O(2^{O(\rank({\cal G}))})$.

\proof 
It suffices to 
to bound the  period of the quasipolynomial
$\tilde m_{\bar \lambda}^\pi(n)$. For this,
it suffices to let $n\rightarrow \infty$.
For a fixed $W,\bar W, C$,
the chamber containing $c(n {\bar \lambda},n \pi,\rho,\bar \rho))$
is completely determined  by $\bar \lambda$ and $\pi$ as
$n \rightarrow \infty$.
Under these conditions,
the degree of 
$\phi_{W,\bar W,C}(c(n {\bar \lambda},n \pi,\rho,\bar \rho))$
is equal to the dimension of the 
polytope associated with this vector partition function.
This dimension is clearly $O(\rank({\cal G})^2)$. 

By Szenes-Vergne residue formula \cite{szenes},
there is a constant $D$
depending on only ${\cal G},{\cal H},W,\bar W,C$, such that 
the period of the quasipolynomial 
$h(n)=\phi_{W,\bar W,C}(c(n {\bar \lambda},n \pi,0,0))$ divides $D$
 for every $\bar \lambda,\pi$; here we are putting $\rho$ and $\bar \rho$ 
equal to zero, since we are interested in what happens as 
$n\rightarrow \infty$.
\qed

\chapter{Parallel and PSPACE algorithms}  \label{cpspace}
In this chapter  we give PSPACE algorithms (cf. Theorem~\ref{tpspacesubgroup})
for computing the  various 
structural constants under consideration . 
We shall only  prove Theorem~\ref{tpspacesubgroup},
when $H$ is therein is either a complex, semisimple
 group, or a symmetric group, or a general linear group over a finite
field, the extension to the general case being routine. 

We recall two standard results in parallel complexity theory \cite{karp}, which will be
used repeatedly. 

Let $NC(t(N),p(N)$ denote the class of problems 
that can be solved in $O(t(N))$ parallel time using $O(p(N))$ processors, where $N$ denotes
the bitlength of the input.
Let \[NC=\cup_i NC(\log^i(N),\poly(N)).\]
This  is the class of problems having
efficient parallel algorithms.

\begin{prop} \label{pparallel} \cite{csanky,karp}
Let $A$ be an $n\times n$-matrix with entries in a ring $R$ of  characteristic
zero. Then the determinant of $A$, and $A^{-1}$, if $A$ is nonsingular,
can be computed in $O(\log^2 n)$ parallel steps using $\poly(n)$ 
processors; here each operation in the ring is considered one step.
Hence, if $R=\Q$, the problems of computing the determinant, the inverse and
solving linear systems belong to $NC$.
\end{prop} 

\begin{prop}  \label{pinclusion}
The class $NC(t(N),2^{t(N)})\subseteq SPACE(O(t(N)))$.
In particular, $NC(\poly(N),2^{O(\poly(N))}) \subseteq PSPACE$.
\end{prop}

\section{Complex semisimple Lie group} 
In this section  we prove a special case of 
Theorem~\ref{tpspacesubgroup}
for the generalized plethym problem (Problem~\ref{pintroplethysm}). Accordingly,
let  $H$ be  a complex,
semisimple, simply connected Lie group,
$G=GL(V)$, where $V=V_\mu(H)$ is an
irreducible representation of $H$ with dominant weight $\mu$,
$\rho: H \rightarrow G$ the homomorphism corresponding to the representation,
and 
$m_\lambda^\pi$  the multiplicity   of $V_\pi(H)$
in $V_\lambda(G)$, considered as an $H$-module via $\rho$; cf. Problem~\ref{pintrosubgroup}.
Then:

\begin{theorem} \label{tpspacegenplethysm}
The multiplicity $m_\lambda^\pi$ can be computed in $\poly(\bitlength{\lambda},
\bitlength{\mu}, \bitlength{\pi}, \dim(H))$ space.
\end{theorem}

Here it is assumed that the partition $\lambda=\lambda_1\ge \lambda_2 \ge \cdots \lambda_r >0$
is represented in a compact form by specifying only its 
nonzero parts $\lambda_1,\ldots, \lambda_r$.
This is important since $\dim(G)$ can be exponential in $\dim(H)$ and $\bitlength{\mu}$.
A compact representation allows  $\bitlength{\lambda}$ to be small, say $\poly(\dim(H),
\bitlength{\mu})$, in this case.

We begin with a simpler special case.

\begin{prop} 
If $\dim(V)=\poly(\dim(H))$, then $m_{\lambda}^\pi$ can be computed in 
$PSPACE$; i.e., in $\poly(\bitlength{\lambda},\bitlength{\mu},\bitlength{\pi},\dim(H))$ space.
\end{prop} 
This implies that the Kronecker coefficient
(Problem~\ref{pintrokronecker}) can be computed in PSPACE.

\proof
Let us use the notation $\bar \lambda$ instead of $\lambda$ to be consistent with the
notation used in 
Klimyk's formula (\ref{eqklimyk}). By the latter,
$m^{\pi}_{\bar \lambda}$ can be computed in PSPACE if
$n_{\bar \mu}(V_{\bar \lambda})$ in that formula  can be computed in PSPACE
for every $\bar \mu$ and $\bar \lambda$.
In type $A$, this is just the number of Gelfand-Tsetlin  tableau with  
the shape $\bar \lambda$ and weight $\bar \mu$. 
If  $\dim(V)=\poly(\dim(H))$, the size of such a tableau is
$O(\dim(V)^2)=\poly(\dim(H))$. So we can count the number of such tableu
in PSPACE as follows: Begin with a zero count, and
cycle through all tableaux of shape $\bar \lambda$  in polynomial space  one
by one,  increasing  the count by one everytime the tableau satisfies
all  constraints for Gelfand-Tsetlin tableau and has weight $\bar \mu$. 
In general, the role of Gelfand-Tsetlin tableaux is played by 
Lakshmibai-Seshadri (LS) paths \cite{littelmann,dehy}.
\qed

The argument above does not work 
if $\dim(V)$ is not $\poly(\dim(H))$, as in the plethym problem (Problem~\ref{pintroplethysm}),
where $\dim(V)=\dim(V_\mu)$ can be exponential in $n=dim(H)$ and the bitlength
of $\mu$.
In this case, the algorithm cannot even afford to write down a tableau
since its  size need not be
polynomial.

Next we turn to Theorem~\ref{tpspacegenplethysm}. 
For the sake of simplicity, we shall prove it only for  $H=SL_n(\C)$, or rather $GL_n(\C)$--i.e.,
the usual plethysm problem.
This  illustrates all the basic ideas.  The general case is similar.
We shall prove a slightly stronger result in this case:

\begin{theorem} \label{tpspaceplethysm2}
The plethysm constant $a_{\lambda,\mu}^\pi$ can be 
can be computed in $\poly(\bitlength{\lambda},\bitlength{\mu}, \bitlength{\pi})$ space. 
\end{theorem}

Here the dependence on $n=\dim(H)$ is not there. This makes a difference if the
heights of $\mu$ and $\pi$ are less than $n=\dim(H)$--remember that we are
using a compact representation of a partition in which only  nonzero parts are specified.
This  is really not a big issue.
Because $a_{\lambda,\mu}^\pi$ depends only on the partitions 
$\lambda,\mu,\pi$ and not $n$. Hence, without loss of generality, we can assume that
$n$ is the maximum of the heights of $\mu$ and $\pi$.
It is possible to strengthen Theorem~\ref{tpspacegenplethysm} similarly. 

To prove Theorem~\ref{tpspaceplethysm2},
we shall  give an efficient parallel algorithm 
to compute $\tilde a_{\lambda,\mu}^\pi$ that works in
$\poly(\bitlength{\lambda},\bitlength{\mu},\bitlength{\pi})$ parallel time using
$O(2^{\poly(\bitlength{\lambda},\bitlength{\mu},\bitlength{\pi})})$ processors. This will show that
the problem of computing $\tilde a_{\lambda,\mu}^\pi$ is in
the complexity class 
$NC(\poly(\bitlength{\lambda},\bitlength{\mu},\bitlength{\pi}),
2^{\poly(\bitlength{\lambda},\bitlength{\mu},\bitlength{\pi})})$, which is contained in
PSPACE by Proposition~\ref{pinclusion}. 
The basic idea is to parallelize the classical 
character-based algorithm for computing $a_{\lambda,\mu}^\pi$ by
using efficient parallel algorithm for inverting a matrix and solving a linear 
system (Proposition~\ref{pparallel}).

We begin by   recalling  the standard facts concerning the characters of the
general linear group.
Given a representation  $W$ of $GL_m(\C)$, let $\rho:GL_m(\C)\rightarrow GL(W)$ be 
the representation map. 
Let $\chi_\rho(x_1,\ldots,x_m)$
denote the formal character of this representation $W$. This is the 
trace of $\rho(\mbox{diag}(x_1,\ldots,x_m))$,
where  $\mbox{diag}(x_1,\ldots,x_n)$ denotes the generic diagonal matrix 
with variable entries $x_1,\ldots,x_m$ on its diagonal.
If $W$ is an irreducible representation $V_\lambda(GL_m(\C))$, then 
$\chi_\rho(x_1,\ldots,x_m)$ is the Schur polynomial $S_\lambda(x_1,\ldots,x_m)$.
By the Weyl character formula,
\begin{equation} \label{eqweylchar}
S_\lambda=\f{|x_j^{\lambda_i+m-i}|}{|x_j^{m-i}|},
\end{equation} 
where $|a^i_j|$ denotes the determinant of an $m\times m$-matrix $a$.
The Schur polynomials form a basis of the ring of symmetric polynomials
in $x_1,\ldots,x_m$. The simplest basis of this ring consists of
the complete symmetric polynomials $M_\beta(x_1,\ldots,x_m)$ defined 
by 
\[M_\beta(x_1,\ldots,x_m)= \sum_\gamma t^\gamma,\] 
where $\gamma$ ranges over all permutations of $\beta$ and 
$t^\gamma=\prod_i x_i^{\gamma_i}$.
Schur polynomials are related to $M_\beta$ by: 
\begin{equation} \label{eqschurmon}
S_\lambda=\sum_\beta k_\lambda^\beta M_\beta,
\end{equation} 
where $k_\lambda^\beta$ is the  Kostka number. This is the number of 
semistandard tableau of shape $\lambda$ and weight $\beta$. 

If the representation $W$ is reducible, 
its decomposition into irreducibles is given by:
\begin{equation} 
W=\sum_\pi m(\pi) V_\pi(GL_n(\C)),
\end{equation}
where $m(\pi)$'s are the coefficients of the formal character 
$\chi_\rho(x_1,\ldots,x_m)$ in the Schur basis:
\[\chi_\rho=\sum_\pi m(\pi) S_\pi.\]

\subsubsection*{Proof of Theorem~\ref{tpspaceplethysm2}}
Let  $\lambda,\mu,\pi$ be  as in Theorem~\ref{tpspaceplethysm2}.  Let 
$H=GL_n(\C)$, $V=V_\mu(H)$, $G=GL(V)$. 
Let $s_\lambda(x_1,\ldots,x_m)$ be the formal character of
the representation $V_\lambda(G)$ of $G$. Here 
$m=\dim(V_\mu)$ can be exponential in $n$ and $\bitlength{\mu}$. 
The basis of $V_\mu(H)$ is indexed by semistandard tableau 
of shape $\mu$ with entries in $[1,n]$. Let us order these tableau, say
lexicographically, and let $T_i$, $1\le i \le m$, denote the 
$i$-th tableau in this order. With each tableau $T$,
we associate a monomial  
\[t(T)=\prod_{i=1}^n t_i^{w_i(T)},\]
where $w_i(T)$ denotes the number of $i$'s in $T$. 
Given a polynomial $f(x_1,\ldots,x_m)$, let us define
$f_\mu=f_\mu(t_1,\ldots,t_n)$ to be the polynomial obtained by
 substituting $x_i=t(T_i)$ in $f(x_1,\ldots,x_m)$. 
Then the formal character of $V_\lambda(G)$, considered as 
an $H$-representation of via the homomorphism
 $H \rightarrow G=GL(V_\mu(H))$, is 
the symmetric polynomial $S_{\lambda,\mu}(t_1,\ldots,t_n)
=(S_\lambda)_\mu$.
The plethysm constant $a_{\lambda,\mu}^\pi$ is defined by:
\begin{equation} 
S_{\lambda,\mu}(t_1,\ldots,t_n)
=\sum_\pi a_{\lambda,\mu}^\pi S_\pi(t_1,\ldots,t_n).
\end{equation} 

An efficient parallel algorithm to compute $a_{\lambda,\mu}^\pi$
is as follows. Here by an  efficient parallel algorithm, we mean 
an algorithm that works in $\poly(\bitlength{\lambda},\bitlength{\mu},\bitlength{\pi})$ 
time using $2^{\poly(\bitlength{\lambda},\bitlength{\mu},\bitlength{\pi})}$ processors.
We will repeatedly use Proposition~\ref{pparallel}.

\subsubsection*{Algorithm} 
\noindent (1) Compute $S_{\lambda,\mu}(t_1,\ldots,t_n)$. By the Weyl
character formula (\ref{eqweylchar}), 
\[S_{\lambda,\mu}(t_1,\ldots,t_n)=\f {A_{\lambda,\mu}(t_1,\ldots,t_n)} 
 {B_{\lambda,\mu}(t_1,\ldots,t_n)},\] 
where $A_\lambda(x_1,\ldots,x_m)$ and $B_\lambda(x_1,\ldots, x_m)$
 denote the numerator
and denominator in (\ref{eqweylchar}), and $A_{\lambda,\mu}=(A_\lambda)_\mu$,
and $B_{\lambda,\mu}=(B_\lambda)_\mu$.
Let $R=\C[t_1,\ldots,t_n]$. Then
\[A_{\lambda,\mu}(t_1,\ldots,t_n)=|t(T_j)^{\lambda_i+m-i}|.\] 
This is the determinant of an $m\times m$ 
matrix with entries in $R$, where $m=\dim(V)$ can be exponential in
$n$ and $\bitlength{\mu}$. It
can be evaluated  in $O(\log^2m)$ parallel ring operations
using $\poly(m)$ processors. Each ring element that arises in
the course of this algorithm is a polynomial in $t_1,\ldots, t_n$ of
total degree $O(|\lambda| m)$, where $|\lambda|$ denotes the size of $\lambda$.
The total number of its coefficients
is $r=O((|\lambda| m)^n)$. 
Hence  each ring operation  can be carried out efficiently
in $O(\log^2(r))$
 parallel time using $\poly(r)$ processors. 
Since $\log m= \poly(n,\bitlength{\mu})$ and
$\log r=\poly(n,\bitlength{\lambda},\bitlength{\mu})$, it follows that
$A_{\lambda,\mu}$ can be evaluated in $\poly(n,\bitlength{\mu},\bitlength{\lambda})$
parallel time using $2^{\poly(n,\bitlength{\mu,\lambda})}$ processors.
The determinant $B_{\lambda,\mu}$ can also be computed efficiently in parallel
in a similar fashion. To compute $S_{\lambda,\mu}$, we have to
divide $A_{\lambda,\mu}$ by $B_{\lambda,\mu}$. This can be 
done by solving an $r\times r$ linear system, which, again, can be done
efficiently in parallel. This computation yields  representation
of $S_{\lambda,\mu}$ 
in the monomial basis $\{M_\beta\}$ of the ring of symmetric polynomials
in $t_1,\ldots, t_n$. 

\noindent (2) To get the coefficients $a_{\lambda,\mu}^\pi$, we 
have to get the representation of $S_{\lambda,\mu}(t)$ in the 
Schur basis. This change of basis 
requires inversion of the matrix in the    linear system (\ref{eqschurmon}). 
The entries of the matrix $K$ occuring in this linear system 
are Kostka numbers. 
Each Kostka number can be computed efficiently in parallel. Hence, 
all entries of this matrix can be computed efficiently in parallel. 
After this, the matrix can be inverted efficiently in parallel, and 
the coefficients $a_{\lambda,\mu}^\pi$'s of $S_{\lambda,\mu}$ in the Schur basis
can  be computed   efficiently in parallel. 
Finally, we use Proposition~\ref{pinclusion} to conclude that $a_{\lambda,\mu}^\pi$ 
can be computed in $PSPACE$. \qed

\section{Symmetric group}
Next we  prove Theorem~\ref{tpspacesubgroup}   when 
$H=S_m$. Let $X=V_\mu(S_m)$ be an irreducible representation  (the Specht module) of
$S_m$ corresponding to a partition $\mu$ of size $m$.
Let  $\rho: H \rightarrow G=GL(X)$ be the corresponding homomorphism.

\begin{theorem}  \label{tpspacesymmetric}
Given
partitions $\lambda,\mu,\pi$, where $\mu$ and $\pi$ have size $m$,
the multiplicity $m_{\lambda,\mu}^\pi$
of the Specht module $V_\pi(S_m)$ in $V_\lambda(G)$ can be computed in 
$\poly(m,\bitlength{\lambda})$ space.
\end{theorem} 
				   
\begin{remark} 
The bitlengths $\bitlength{\mu}$ and $\bitlength{\pi}$ are not mentioned in 
the complexity bound because
 they are   bounded by $m$. 
\end{remark}

For this, we need  three lemmas.

\begin{lemma} \label{lsymchar}
The character of a symmetric group can be computed in $PSPACE$. 
Specifically, given a partition $\pi$  of size $m$,
and a sequence $i=(i_1,i_2,\ldots)$ of nonnegative integers
such that $\sum j i_j=m$, 
the value of the character $\chi_\pi$ of $S_m$ on the conjugacy 
class $C_i$ of permutations indexed by $i$
can be computed in $\poly(m)$ parallel time using $2^{\poly(m)}$ processors.
Hence it can be computed in $\poly(m)$ space (cf. Proposition~\ref{pinclusion}).
\end{lemma} 
Here the conjugacy class $C_i$ consists of those permutations that
have $i_1$ $1$-cycles, $i_2$ $2$-cycles, and so on.

\proof 
Let $k$ be the height of the partition $\pi$.
Let $x=(x_1,\ldots,x_k)$ be the tuple of variables $x_i$'s.
Given a formal series $f(x)$ and a tuple $(l_1,\ldots,l_k)$ of
nonnegative integers,
let $[f(x)]_{(l_1,\ldots,l_k)}$ denote the coefficient of $x_1^{l_1}\cdots 
x_k^{l_k}$ in $f$.

By the Frobenius character formula \cite{fultonrepr},
\begin{equation} 
\chi_\lambda(C_i)=[f(x)]_{(l_1,\ldots,l_k)},
\end{equation} 
where 
\[
l_1=\pi_1+k-1,l_2=\pi_2+k-2,\ldots, l_k=\pi_k,
\]
and 
\[f(x)=\Delta(x) \prod_{j=1}^m P_j(x)^{i_j},\] 
with 
\begin{equation} 
\begin{array}{lcl} 
\Delta(x)&=&\prod_{i<j} (x_i-x_j),\\
P_j(x)&=&x_1^j+\cdots+x_k^j.\\
\end{array}
\end{equation}
Since $\deg(f)=\poly(m)$ and $k\le m$,
the total number of coefficients of $f(x)$ is $2^{\poly(m)}$.
Hence, we can evaluate $f(x)$ in PSPACE by setting up appropriate 
recurrence relations.

Alternatively, we can  easily evaluate  $f(x)$  
in $\poly(m)$ parallel time using $2^{\poly(m)}$ processors,
and then extract its required coefficient. 
After this, the result follows from Proposition~\ref{pinclusion}.
\qed

\begin{lemma}\label{lsymmultiplicity}
Suppose $\phi$ is a character of $S_m$ whose value on 
any conjugacy class $C_i$ can be computed in $O(s)$ space,
for some parameter $s$. Then,
the multiplicity of the representation $V_\pi(S_m)$ in the
representation $V_\phi(S_m)$ corresponding $\phi$ can be 
computed in  $O(\poly(m)+s)$ space.
\end{lemma} 
\proof
The multiplicity is given by the inner product 
\begin{equation}
\langle \phi,\chi_\pi\rangle
=\f 1{m!} \sum_{\sigma \in S_m}  \bar{\phi(\sigma)} \chi_\pi(\sigma).
\end{equation}
By assumption,
$\phi(\sigma)$ can be computed in $O(s)$ space, and 
by Lemma~\ref{lsymchar}, $\chi_\pi(\sigma)$ can be computed in
$\poly(m)$ space. Hence, the result follows  from the preceding formula.
\qed

Given an irreducible representation $X=V_\mu(S_m)$ and an
irreducible representation $W=V_\lambda(G)$ of $G=GL(X))$, 
let $\rho_\mu$ denote the representation map
$S_m\rightarrow G$, $\rho_\lambda$ the representation map
$G \rightarrow GL(W)$, and
\[\rho:S_m \rightarrow G \rightarrow GL(W)\]
their composition. This is a representation of $S_m$.
Let $\chi_\rho$ be the  character of $\rho$. 

\begin{lemma} \label{lsymcomposition}
For any $\sigma\in S_m$,  $\chi_\rho(\sigma)$
can be computed in $\poly(m,\bitlength{\lambda})$ in 
$\poly(m,\bitlength{\lambda})$ space.
\end{lemma}

The bitlength $\bitlength{\mu}$ is not mentioned  in the complexity bound because 
it is bounded by $m$.

\proof
Let $r=\dim(X)$.
The formal character of the representation 
$V_\lambda(G)$ of $G=GL(X)$ is the Schur polynomial
$S_\lambda(x_1,\dots,x_r)$, $r=\dim(X)$.
Hence, 
\[
\chi_\rho(\sigma)=S_\lambda(\alpha)
\]
where  $\alpha=(\alpha_1,\ldots,\alpha_r)$ is the tuple 
of  eigenvalues of $\rho_\mu(\sigma)$. 
We shall   compute the right hand side fast in parallel--i.e., in
$\poly(m,\bitlength{\lambda})$ parallel time using
$2^{\poly(m,\bitlength{\lambda})}$ processors--and then
use Proposition~\ref{pinclusion} to conclude the proof.

This is done as follows.

\noindent 
(1) Let $\chi_\mu$ denote the character of the representation $\rho_\mu$. 
Let $p_i(\alpha)=\alpha_1^i+\cdots+\alpha_r^i$ denote the
$i$-th power sum of the eigenvalues. 
For any $i$,
\[p_i(\alpha)=\chi_\mu(\sigma^i).\] 
We can compute  $\sigma^i$, for $i\le |\lambda|$, where $|\lambda|$ denotes the 
size of $\lambda$, 
in $\poly(\log i,m)=\poly(m,\bitlength{\lambda})$ time
using repeated squaring. 
After this $\chi_\mu(\sigma^i)$ can be computed fast in parallel
in $\poly(m)$ time using Lemma~\ref{lsymchar}. 
Thus each $p_i(\alpha)$ can be computed in $\poly(m,\bitlength{\lambda})$
time in parallel using $2^{\poly(m,\bitlength{\lambda})}$ processors.
We calculate $p_i(\alpha)$ in parallel for all $i\le |\lambda|$,
and all  $p_\gamma(\alpha)=\prod_j p_{\gamma_j}(\alpha)$  in parallel for all partitions
$\gamma$ of size at most  $m$.

(2) After this, we calculate 
the complete symmetric function $h_i(\alpha)$, for each $i\le |\lambda|$,
fast in parallel, 
by using the relation \cite{macdonald}:
\[ h_i=\sum_{|\gamma|=i} z_\gamma^{-1} p_\gamma,\] 
where $z_\gamma=\prod_{i\ge 1} i^{m_i}m_i!$, and 
$m_i=m_i(\gamma)$ denotes the number of parts of $\gamma$ equal to $i$.
Thus we can calculate  $h_\gamma(\alpha)=\prod_j h_{\gamma_j}(\alpha)$, for all partitions
$\gamma$ of size $m$, fast in parallel.

(3) To compute $S_\lambda(\alpha)$, we
recall that the transition matrix between the Schur basis $\{S_\lambda\}$ 
and the complete symmetric basis $\{h_\gamma\}$ of the
ring of symmetric functions is $K^*$, the transpose inverse of the
Kostka matrix $K=[K_{\lambda,\gamma}]$, where 
$K_{\lambda,\gamma}$ denote the Kostka number; cf. \cite{macdonald}.
As we noted in the proof of Theorem~\ref{tpspaceplethysm2}, each Kostka number
can be computed in fast in parallel. Hence, $K$ can be
computed fast in parallel.
After this,  its inverse $K^{-1}$ can be
computed fast in parallel by Proposition~\ref{pparallel}--this is the crux of the proof--and 
finally  $K^*$ as well.
Thus  $S_\lambda(\alpha)$ can be computed fast in parallel, since
each $h_\gamma(\alpha)$ can be computed fast in parallel.
\qed

Theorem~\ref{tpspacesymmetric} follows from Lemma~\ref{lsymchar},\ref{lsymmultiplicity} 
and \ref{lsymcomposition}.  \qed

\section{General linear group over a finite field} 
In this section we prove Theorem~\ref{tpspacesubgroup},
 when $H$ therein is the general linear group 
$GL_n(F_{p^k})$ over a finite field $F_{p^k}$. Irreducible representations of
$H=GL_n(F_{p^k})$ have been classified by Green \cite{macdonald}. They are labelled
by certain   partition-valued functions. See \cite{macdonald} for a precise 
description of these labelling  functions. 
It is clear from the description therein that each labelling function 
has  a compact representation of bitlength $O(n+k+\bitlength{p})$,
where $\bitlength{p}=\log_2 p$; we specify a function by giving  its partition values 
at the places where it is nonzero. 
Let $\mu$ denote any such label.
Let $X=V_\mu(H)$ be the corresponding irreducible representation  of $H$,
and   $\rho: H \rightarrow G=GL(X)$ the corresponding homomorphism.

\begin{theorem}  \label{tpspacefinitefield}
Given a
partition $\lambda$ and labelling functions  $\mu$ and $\pi$ as above,
the multiplicity $m_{\lambda,\mu}^\pi$
of the irreducible representation  $V_\pi(H)$ in $V_\lambda(G)$ can be computed in 
$\poly(n,k,\bitlength{p},\bitlength{\lambda})$ space.
\end{theorem}

The proof is similar to 
that of Theorem~\ref{tpspacesymmetric}
 for the symmetric group with the following result playing the role of
Lemma~\ref{lsymchar}:

\begin{lemma} \label{lfinitefieldchar}
Given a label $\gamma$ of an irreducible 
character $\chi_\gamma$ of $H=GL_n(F_{p^k})$ and a 
label $\delta$ of a 
conjugacy class in $H$,
the value $\chi_\gamma(\delta)$ can be
computed in $\poly(n,k,\bitlength{p})$ parallel time using 
$2^{\poly(n,k,\bitlength{p})}$ processors, and hence by Proposition~\ref{pinclusion}, in
$\poly(n,k,\bitlength{p})$ space. 
\end{lemma} 

The label  $\delta$ of a conjugacy class in $H$ is also a partition valued function
\cite{macdonald}, which admits a compact representation of bitlength  $\poly(n,k,\bitlength{p})$.

\proof
We shall parallelize Green's algorithm \cite{macdonald}  for 
computing the character values, 
 and then conclude by Proposition~\ref{pparallel}.
Green shows that $\chi_\gamma(\delta)$'s are 
entries of a transition matrix between a two polynomial bases:
the first constructed using Hall-Littlewood polynomials, and the
second using Schur polynomials. 
We have  construct this transition matrix fast in
parallel.
We shall only indicate here how the transition matrix between the 
basis of Hall-Littlewood polynomials and the Schur basis for the ring
symmetric functions over $Z[t]$ 
can be constructed fast in parallel. This idea can then be 
easily extended to complete the proof.

First, we recall the definition of the Hall-Littlewood polynomial
$P_\pi(x;t)=P_\pi(x_1,\ldots,x_k;t)$ \cite{macdonald}. This is a symmetric polynomial in
$x_i$'s with coefficients in $\Z[t]$. 
It interpolates between the Schur function $s_\pi(x)$ and
the monomial symmetric function $m_\pi(x)$ because 
$P_\pi(x;0)=s_\pi(x)$ and 
$P_\pi(x;1)=m_\pi(x)$. The formal definition is as follows:

For a given partition $\pi$, let
$v_\pi(t)=\prod_{i\ge 0} v_{m_i}(t)$, where $m_i$ is the number of 
 parts of $\pi$ equal to $i$, and
\[v_m(t)=\prod_{i=1}^m \f{1-t^i}{1-t}.\] 

Then 
\begin{equation} \label{eqppi}
P_\pi(x;t)=\f {A_\pi(x,t)} {B_\pi(x,t)},
\end{equation} 
where 
\begin{equation} \label{eqAhall}
\begin{array}{lcl} 
A_\pi(x,t)=
 \sum_{\sigma \in S_k} \mbox{sgn}(\sigma) \sigma(x_1^{\pi_1} \cdots x_k^{\pi_k} \prod_{i<j}
  {x_i-t x_j}, \\
B_\pi(x,t)={v_\pi(t)} \prod_{i<j} (x_i-x_j).
\end{array}
\end{equation}
Here $\mbox{sgn}(\sigma)$ denotes the sign of $\sigma$.

Let $w_{\pi,\alpha}(t)$'s be the coeffcients of $P_\pi(x,t)$ in
the Schur basis:
\begin{equation} 
P_\pi(x;t)=\sum_\alpha w_{\pi, \alpha}(t) s_\alpha(x).
\end{equation}

We want to calculate the matrix $W=[w_{\pi,\alpha}]$ fast in
parallel.
Using formula (\ref{eqAhall}), we calculate $A_\pi(x;t)$ 
fast in parallel; i.e., we calculate the coefficients of 
$A_\pi(x;t)$ in the basis of monomials in $x$ and $t$.
We calculate $B_\pi(x;t)$ similarly. After this
the division in (\ref{eqppi}) can be carried out by solving a 
an appropriate linear system. This can be done  fast in 
parallel by Proposition~\ref{pparallel}.
Since, $P_\pi(x;t)$ is symmetric in  $x_i$'s, this yields its coefficients
in the monomial symmetric basis $\{m_\alpha(x)\}$ with the  coefficients
being in $\Z[t]$. 
The transition matrix \cite{macdonald} from the monomial symmetric basis to the
Schur basis is given by the inverse of the Kostka matrix.
This inverse can be computed fast in parallel by Proposition~\ref{pparallel}. After this, 
the coefficients $w_{\pi,\alpha}$'s of $P_\pi(x;t)$ in the
Schur basis can be computed fast in parallel.

Furthermore, the inverse of $W=[W_{\pi,\alpha}]$ can also be
computed fast in parallel by Proposition~\ref{pparallel}.
\qed

\subsection{Tensor product problem}
Analogue of the Kronecker problem (Problem~\ref{pintrokronecker}) 
for $H=GL_n(F_{p^k})$ is:

\begin{problem} 
Given  partition valued functions $\lambda,\mu,\pi$, 
decide if the multiplicity 
$b_{\lambda,\mu}^\pi$ of $V_\pi(H)$ in the tensor
product $V_\lambda(H)\otimes V_\mu(H)$ is nonzero.
\end{problem} 

In this context:

\begin{theorem} \label{tpspacetensor}
The multiplicity  $m_{\lambda,\mu}^\pi$ can be computed in
PSPACE; i.e., in $\poly(n,k,\bitlength{p})$ space.
\end{theorem} 
\proof 
This follows from Lemma~\ref{lfinitefieldchar} and analogues of Lemmas~\ref{lsymmultiplicity} and
\ref{lsymcomposition} in this
setting. \qed

A possible  canditate for a stretching function assoociated with $b_{\lambda,\mu}^\pi$ is:
\[ 
\tilde b_{\lambda,\mu}^\pi(n)=b_{n \lambda,n\mu}^{n \pi},\] 
where $n \lambda$ denotes the stretched partition-valed  function obtained by stretching each
partition value of $\lambda$  by a  factor of $n$.
In other words $\tilde b_{\lambda,\mu}^\pi(n)$  is the multiplicity of 
$V_{n \pi}(H(n))$ in $V_{n \lambda}(H(n)) \otimes V_{n \mu}(H(n))$, where
$H(n)=GL_{n m}(F_{p^k})$ is the stretched group.
Is it a quasi-polynomial? 
If so, we can also ask for a good bound on  its saturation and positivity 
indices.

\section{Finite simple groups of Lie type} \label{sfinitesimple}
The work of Deligne-Lusztig \cite{deligne} and Lusztig\cite{lusztigchar}  yield 
an algorithm for computing the character values for finite simple groups
of Lie type.

\begin{question} 
Can this algorithm be parallelized?
\end{question}
If so, Lemma~\ref{lfinitefieldchar}, and hence Theorem~\ref{tpspacefinitefield},
  can be extended to
finite simple groups of Lie type.

\chapter{Experimental evidence for positivity} \label{cevidence}
In this chapter we give experimental evidence for positivity (PH2,3).

\section{Littlewood-Richardson problem} \label{sevilittle}
Experimental evidence for PH2 in the context of the Littlewood-Richardson problem 
(Problem~\ref{pintrolittle})
 has been given in \cite{loera}, and for PH3 in type $A$ in \cite{king}.
We give experimental evidence for PH3 in types $B,C,D$ here.
Let $C_{\alpha,\beta}^\lambda(t)$ be as in eq.(\ref{eqlittlerational}).
Its reduced positive form for various values of $\alpha,\beta,\lambda$ is shown
in Figure~\ref{fb3} for  type  $B$, in Figure~\ref{fc3} for type $C$, and
Figure~\ref{fd3} for type $D$. The rank of the Lie algebra is three in all cases. 
In these types, the period of the stretching quasipolynomial $\tilde c_{\alpha,\beta}^\lambda(n)$
is at most two. Accordingly, the period of every  pole of 
$C_{\alpha,\beta}^\lambda(t)$ is at most two. The tables were computed from the 
tables in \cite{loera} for the stretching quasi-polynomial
 $\tilde c_{\alpha,\beta}^\lambda(n)$ in these cases.

\begin{sidewaysfigure}[h!] 
\[
\left [\begin {array}{cccc}
\alpha&\beta&\lambda&C_{\alpha,\beta}^\lambda(t) \\ \hline \\
(0,15,5)&(12,15,3)&(6,15,6)& {\frac {350\,{t}^{8}+19121\,{t}^{7}+123576\,{t}^{6}+297561\,{t}^{5}+342064\,{t}^{4}+192779\,{t}^{3}+46992\,{t}^{2}+2641\,t+1}{\left (1-t\right )^{3}\left (1-t^2\right )^{3}}}\\ \\
(4,8,11)&(3,15,10)&(10,1,3)&{\medskip}{\frac {1+5\,t+6\,{t}^{2}+{t}^{3}}{\left (1-t^2\right )^{3}}}\\ \\
(8,1,3)&(11,13,3)&(8,6,14)&{\medskip}{\frac {2\,{t}^{8}+45\,{t}^{7}+259\,{t}^{6}+591\,{t}^{5}+773\,{t}^{4}+522\,{t}^{3}+198\,{t}^{2}+29\,t+1}{\left (1-t\right )^{3}\left (1-t^2\right )^{4}}}\\ \\
(8,9,14)&(8,4,5)&(1,5,15)&{\medskip}{\frac {136\,{t}^{9}+3422\,{t}^{8}+20204\,{t}^{7}+53608\,{t}^{6}+76076\,{t}^{5}+60986\,{t}^{4}+26674\,{t}^{3}+5568\,{t}^{2}+345\,t+1}{\left (1-t\right )^{3}\left (1-t^2\right )^{4}}}\\ \\
(10,5,6)&(5,4,10)&(0,7,12)&{\medskip}{\frac {219\,{t}^{8}+12135\,{t}^{7}+79231\,{t}^{6}+193003\,{t}^{5}+223919\,{t}^{4}+127907\,{t}^{3}+31704\,{t}^{2}+1870\,t+1}{\left (1-t\right )^{6}\left (1+t\right )^{3}}}\end {array}\right ]
\]
\caption{$C_{\alpha,\beta}^\lambda(t)$ for $B_3$}
\label{fb3}
\end{sidewaysfigure}

\begin{sidewaysfigure} [h!] 
\[
\left [\begin {array}{cccc}
\alpha&\beta&\lambda&C_{\alpha,\beta}^\lambda(t) \\ \hline \\
(1,13,6)&(14,15,5)&(5,11,7)& {\frac {18145\,{t}^{8}+267151\,{t}^{7}+1070716\,{t}^{6}+1917716\,{t}^{5}+1735692\,{t}^{4}+778184\,{t}^{3}+144596\,{t}^{2}+5538\,t+1}{\left (1-t\right )^{4}\left (1-t^2\right )^{3}}}\\ \\
(4,15,14)&(12,12,10)&(4,9,8)&{\medskip}{\frac {2280\,{t}^{9}+267658\,{t}^{8}+2746131\,{t}^{7}+9276935\,{t}^{6}+14682332\,{t}^{5}+11903923\,{t}^{4}+4746803\,{t}^{3}+751126\,{t}^{2}+21249\,t+1}{\left (1-t\right )^{4}\left (1-t^2\right )^{3}}}\\ \\(9,0,8)&(8,12,9)&(7,7,3)&{\medskip}{\frac {3\,{t}^{2}+4\,t+1}{\left (1-t\right )^{6}}}\\ \\
(10,2,7)&(8,10,1)&(7,5,5)&{\medskip}{\frac {8984\,{t}^{8}+132826\,{t}^{7}+534183\,{t}^{6}+960491\,{t}^{5}+873227\,{t}^{4}+394045\,{t}^{3}+74067\,{t}^{2}+2941\,t+1}{\left (1-t\right )^{4}\left (1-t^2\right )^{3}}}\\ \\
(10,10,15)&(11,3,15)&(10,7,15)&{\medskip}{\frac {7162\,{t}^{9}+736327\,{t}^{8}+7178960\,{t}^{7}+23540366\,{t}^{6}+36359642\,{t}^{5}+28788904\,{t}^{4}+11166361\,{t}^{3}+1693696\,{t}^{2}+43515\,t+1}{\left (1-t\right )^{7}\left (1+t\right )^{3}}}\end {array}\right ]
\]
\caption{$C_{\alpha,\beta}^\lambda(t)$ for $C_3$}
\label{fc3}
\end{sidewaysfigure}

\begin{sidewaysfigure} [h!] 
\[
\left [\begin {array}{cccc}
\alpha&\beta&\lambda&C_{\alpha,\beta}^\lambda(t) \\ \hline \\
(0,15,5)&(12,15,3)&(6,15,6)& {\frac {633\,{t}^{7}+24259\,{t}^{6}+142236\,{t}^{5}+252113\,{t}^{4}+168220\,{t}^{3}+36131\,{t}^{2}+1414\,t+1}{\left (1-t\right )^{7}\left (1-t^2\right )}}\\ \\
(4,8,11)&(3,15,10)&(10,1,3)&{\medskip}{\frac {7962\,{t}^{8}+503679\,{t}^{7}+4525372\,{t}^{6}+11944350\,{t}^{5}+12218255\,{t}^{4}+4879052\,{t}^{3}+586370\,{t}^{2}+10862\,t+1}{\left (1-t\right )^{8}\left (1-t^2\right )}}\\ \\
(8,1,3)&(11,13,3)&(8,6,14)&{\medskip}{\frac {81\,{t}^{7}+19407\,{t}^{6}+211964\,{t}^{5}+513585\,{t}^{4}+426652\,{t}^{3}+110317\,{t}^{2}+4609\,t+1}{\left (1-t\right )^{6}\left (1-t^2\right )}} \\ \\
(8,9,14)&(8,4,5)&(1,5,15)&{\medskip}{\frac {9\,{t}^{2}+8\,t+1+2\,{t}^{3}}{\left (1-t\right )^{3}}}\\ \\
(10,5,6)&(5,4,10)&(0,7,12)&{\medskip}{\frac {3647\,{t}^{7}+111208\,{t}^{6}+570739\,{t}^{5}+920201\,{t}^{4}+560336\,{t}^{3}+106748\,{t}^{2}+3435\,t+1}{\left (1-t\right )^{8}\left (1+t\right )}}\end {array}\right ]
\]
\caption{$C_{\alpha,\beta}^\lambda(t)$ for $D_3$}
\label{fd3}
\end{sidewaysfigure}

%Experimental evidence for (strict) PH2 (Hypothesis~\ref{}) 
%in the context of the Littlewood-Richardson problem 
%(Problem~\ref{pintrolittle})  has already been given in \cite{loera}. 

\section{Kronecker problem, $n=2$} \label{sevikron}
Let $k_{\lambda,\mu}^\pi$ be the Kronecker coefficient in Problem~\ref{pintrokronecker}. 
Let 
$\tilde k_{\lambda,\mu}^\pi(n) = \tilde k_{n \lambda, n \mu}^{n \pi}$ be the 
associated stretching quasi-polynomial, and 
\[ K_{\lambda,\mu}^\pi(t) = \sum_{n \ge 0} \tilde k_{\lambda,\mu}^\pi(n) t^n,\] 
the associated rational function.
An explicit formula (with alternating signs)  for the Kronecker coefficient, when $n=2$, 
has  given by Remmel and Whitehead \cite{remmel}
and Rosas \cite{rosas}, and a positive formula in \cite{algcomb}.
This case turns out to be nontrivial. For example, the number of chambers (domains)  of 
quasi-polynomiality in this case turns out to be more than a million. Their
explicit description 
can be found out using the formula for the Kronecker coefficient in \cite{remmel}. 

We implemented Rosas' formula to check PH2 for 
the quasipolynomial $\tilde k_{\lambda,\mu}^\pi(n)$
for a few thousand values of $\mu,\nu$ and
$\lambda$  with the help of a computer.  A large number of samples was chosen to ensure
that a significant fraction of the chambers were sampled. 
The quasi-polynomial $\tilde k_{\lambda,\mu}^\pi(n)$ 
 and a  positive form of the rational function $C_{\lambda,\mu}^\pi(t)$ are
shown  Figures~\ref{fkron1} and \ref{fkron2}  for few sample  values of
$\lambda=(\lambda_1,\lambda_2)$, $\mu=(\mu_1,\mu_2)$, and
$\pi=(\pi_1,\pi_2,\pi_3,\pi_4)$. 
It may be noted that  $\tilde k_{\lambda,\mu}^\pi(n)$ need not be
a polynomial; this answers Kirillov's question \cite{kirillov} in the negative. But its period is 
at most two for $n=2$. This follows from the formula in \cite{remmel}.
For the $\lambda,\mu$ and $\pi$ that we sampled, positivity index 
of $\tilde k_{\lambda,\mu}^\pi(n)$ is always zero.  But it turns 
out \cite{rosas2} that there are some $\lambda,\mu$ and $\pi$ for which the
saturation and positivity indices of $\tilde k_{\lambda,\mu}^\pi(n)$ 
are nonzero (one), but very small and thus consistent with SH and
PH2 (Hypothesis\ref{phph2nonstdplethysmrevised}) in this paper; in the earlier 
version of this paper, SH and  PH2 
stipulated  that the saturation and positivity indices are always zero.
These $(\lambda,\mu,\pi)$ escaped our random sampling, because their
density is extremely small \cite{rosas2}.

\begin{sidewaysfigure} [h!] 
\[
\left [\begin {array}{ccccccccccc}
\lambda_1&\lambda_2&\mu_1&\mu_2&\pi_1&\pi_2&\pi_3&\pi_4&\tilde k_{\lambda,\mu}^\pi(n);\ n\ \mbox{odd}&\tilde k_{\lambda,\mu}^\pi(n);\ n\ \mbox{even}&K_{\lambda,\mu}^\pi(t) \\\hline \\ 87&62&97&52&64&39&24&22&1/2+4\,n+11/2\,{n}^{2}&1+4\,n+11/2\,{n}^{2}&{\frac {1+8\,t+11\,{t}^{2}+2\,{t}^{3}}{\left (1-t\right )^{2}\left (1-{t}^{2}\right )}}\\\noalign{\medskip}104&95&149&50&95&78&15&11&1/2+13/2\,n+18\,{n}^{2}&1+13/2\,n+18\,{n}^{2}&{\frac {1+23\,t+36\,{t}^{2}+12\,{t}^{3}}{\left (1-t\right )^{2}\left (1-{t}^{2}\right )}}\\\noalign{\medskip}101&85&102&84&78&72&24&12&17/2\,n+{\frac {71}{2}}\,{n}^{2}&1+17/2\,n+{\frac {71}{2}}\,{n}^{2}&{\frac {1+42\,t+72\,{t}^{2}+27\,{t}^{3}}{\left (1-t\right )^{2}\left (1-{t}^{2}\right )}}\\\noalign{\medskip}79&63&93&49&88&37&14&3&3/4+{\frac {27}{2}}\,n+{\frac {303}{4}}\,{n}^{2}&1+{\frac {27}{2}}\,n+{\frac {303}{4}}\,{n}^{2}&{\frac {1+88\,t+151\,{t}^{2}+63\,{t}^{3}}{\left (1-t\right )^{2}\left (1-{t}^{2}\right )}}\\\noalign{\medskip}97&93&114&76&77&66&47&0&1/2+15/2\,n+21\,{n}^{2}&1+15/2\,n+21\,{n}^{2}&{\frac {1+27\,t+42\,{t}^{2}+14\,{t}^{3}}{\left (1-t\right )^{2}\left (1-{t}^{2}\right )}}\\\noalign{\medskip}88&56&113&31&99&35&7&3&1/2+11/2\,n+10\,{n}^{2}&1+11/2\,n+10\,{n}^{2}&{\frac {1+14\,t+20\,{t}^{2}+5\,{t}^{3}}{\left (1-t\right )^{2}\left (1-{t}^{2}\right )}}\\\noalign{\medskip}134&82&140&76&91&72&49&4&3/4+21\,n+{\frac {669}{4}}\,{n}^{2}&1+21\,n+{\frac {669}{4}}\,{n}^{2}&{\frac {1+187\,t+334\,{t}^{2}+147\,{t}^{3}}{\left (1-t\right )^{2}\left (1-{t}^{2}\right )}}\\\noalign{\medskip}133&69&149&53&98&55&43&6&1+6\,n+8\,{n}^{2}&1+6\,n+8\,{n}^{2}&{\frac {15\,{t}^{2}+13\,t+1+3\,{t}^{3}}{\left (1-t\right )^{3}}}\\\noalign{\medskip}80&63&111&32&88&38&10&7&1&1&{\frac {1+t}{1-t}}\\\noalign{\medskip}118&69&151&36&95&63&20&9&1+4\,n+4\,{n}^{2}&1+4\,n+4\,{n}^{2}&{\frac {7\,{t}^{2}+7\,t+1+{t}^{3}}{\left (1-t\right )^{3}}}\\\noalign{\medskip}96&51&103&44&90&53&3&1&1/2+{\frac {39}{2}}\,n+36\,{n}^{2}&1+{\frac {39}{2}}\,n+36\,{n}^{2}&{\frac {1+54\,t+72\,{t}^{2}+17\,{t}^{3}}{\left (1-t\right )^{2}\left (1-{t}^{2}\right )}}\\\noalign{\medskip}117&72&133&56&82&57&41&9&1+9\,n+18\,{n}^{2}&1+9\,n+18\,{n}^{2}&{\frac {35\,{t}^{2}+26\,t+1+10\,{t}^{3}}{\left (1-t\right )^{3}}}\\\noalign{\medskip}72&63&77&58&49&38&28&20&1/2+7\,n+{\frac {55}{2}}\,{n}^{2}&1+7\,n+{\frac {55}{2}}\,{n}^{2}&{\frac {1+33\,t+55\,{t}^{2}+21\,{t}^{3}}{\left (1-t\right )^{2}\left (1-{t}^{2}\right )}}\\\noalign{\medskip}48&37&49&36&34&24&16&11&1/2+6\,n+{\frac {37}{2}}\,{n}^{2}&1+6\,n+{\frac {37}{2}}\,{n}^{2}&{\frac {1+23\,t+37\,{t}^{2}+13\,{t}^{3}}{\left (1-t\right )^{2}\left (1-{t}^{2}\right )}}\\\noalign{\medskip}108&56&113&51&73&50&29&12&1+4\,n+4\,{n}^{2}&1+4\,n+4\,{n}^{2}&{\frac {7\,{t}^{2}+7\,t+1+{t}^{3}}{\left (1-t\right )^{3}}}\end {array}\right ]
\]
\caption{The quasipolynomial  $\tilde k_{\lambda,\mu}^\pi$
and the rational function $K_{\lambda,\mu}^\pi(t)$ for the 
Kronecker problem, $n=2$.}
\label{fkron1}
\end{sidewaysfigure}

\begin{sidewaysfigure}  
\[
\left [\begin {array}{ccccccccccc}
\lambda_1&\lambda_2&\mu_1&\mu_2&\pi_1&\pi_2&\pi_3&\pi_4&\tilde k_{\lambda,\mu}^\pi(n);\ n\ \mbox{odd}&\tilde k_{\lambda,\mu}^\pi(n);\ n\ \mbox{even}&K_{\lambda,\mu}^\pi(t) \\\hline \\ 77&40&78&39&58&29&24&6&1+19/2\,n+{\frac {57}{2}}\,{n}^{2}&1+19/2\,n+{\frac {57}{2}}\,{n}^{2}&{\frac {56\,{t}^{2}+37\,t+1+20\,{t}^{3}}{\left (1-t\right )^{3}}}\\\noalign{\medskip}153&81&157&77&96&63&61&14&1+3\,n+2\,{n}^{2}&1+3\,n+2\,{n}^{2}&{\frac {3\,{t}^{2}+4\,t+1}{\left (1-t\right )^{3}}}\\\noalign{\medskip}90&89&102&77&90&42&30&17&1/2+13/2\,n+6\,{n}^{2}&1+13/2\,n+6\,{n}^{2}&{\frac {1+11\,t+12\,{t}^{2}}{\left (1-t\right )^{2}\left (1-{t}^{2}\right )}}\\\noalign{\medskip}145&102&160&87&96&84&39&28&1+10\,n+25\,{n}^{2}&1+10\,n+25\,{n}^{2}&{\frac {49\,{t}^{2}+34\,t+1+16\,{t}^{3}}{\left (1-t\right )^{3}}}\\\noalign{\medskip}109&95&136&68&78&60&46&20&1+3\,n+2\,{n}^{2}&1+3\,n+2\,{n}^{2}&{\frac {3\,{t}^{2}+4\,t+1}{\left (1-t\right )^{3}}}\\\noalign{\medskip}100&42&104&38&85&27&27&3&1+8\,n&1+8\,n&{\frac {8\,t+1+7\,{t}^{2}}{\left (1-t\right )^{2}}}\\\noalign{\medskip}74&51&86&39&52&34&26&13&1&1&{\frac {1+t}{1-t}}\\\noalign{\medskip}98&90&124&64&92&67&22&7&1/2+23/2\,n+60\,{n}^{2}&1+23/2\,n+60\,{n}^{2}&{\frac {1+70\,t+120\,{t}^{2}+49\,{t}^{3}}{\left (1-t\right )^{2}\left (1-{t}^{2}\right )}}\\\noalign{\medskip}57&38&75&20&52&25&17&1&1+3\,n+2\,{n}^{2}&1+3\,n+2\,{n}^{2}&{\frac {3\,{t}^{2}+4\,t+1}{\left (1-t\right )^{3}}}\\\noalign{\medskip}159&140&170&129&89&82&73&55&1+3/2\,n+1/2\,{n}^{2}&1+3/2\,n+1/2\,{n}^{2}&{\frac {1+t}{\left (1-t\right )^{3}}}\\\noalign{\medskip}144&122&157&109&88&86&74&18&3/4+n+1/4\,{n}^{2}&1+n+1/4\,{n}^{2}&{\frac {1}{\left (1-t\right )^{2}\left (1-{t}^{2}\right )}}\\\noalign{\medskip}90&68&92&66&88&37&23&10&1/4+12\,n+{\frac {351}{4}}\,{n}^{2}&1+12\,n+{\frac {351}{4}}\,{n}^{2}&{\frac {1+98\,t+176\,{t}^{2}+76\,{t}^{3}}{\left (1-t\right )^{2}\left (1-{t}^{2}\right )}}\\\noalign{\medskip}89&42&100&31&76&28&19&8&1+6\,n+8\,{n}^{2}&1+6\,n+8\,{n}^{2}&{\frac {15\,{t}^{2}+13\,t+1+3\,{t}^{3}}{\left (1-t\right )^{3}}}\\\noalign{\medskip}88&56&107&37&71&39&20&14&1+9/2\,n+9/2\,{n}^{2}&1+9/2\,n+9/2\,{n}^{2}&{\frac {8\,{t}^{2}+8\,t+1+{t}^{3}}{\left (1-t\right )^{3}}}\\\noalign{\medskip}124&111&133&102&98&89&27&21&1/2+7\,n+{\frac {53}{2}}\,{n}^{2}&1+7\,n+{\frac {53}{2}}\,{n}^{2}&{\frac {1+32\,t+53\,{t}^{2}+20\,{t}^{3}}{\left (1-t\right )^{2}\left (1-{t}^{2}\right )}}\end {array}\right ]
\]
\caption{Continuation of Figure~\ref{fkron1}}
\label{fkron2}
\end{sidewaysfigure}

\section{$G/P$ and Schubert varieties} \label{sevigmodp}
Let $V=V_\lambda(G)$ be an irreducible representation of $G=SL_k(\C)$ corresponding to a 
partition $\lambda$. Let $v_\lambda$ be the point in $P(V)$ corresponding to the highest 
weight vector, and $X=G v_\lambda \cong G/P_\lambda$ its closed orbit. 
Let $h_{k,\lambda}(n)$ be the 
 Hilbert function of the homogeneous coordinate ring $R$
 of $X$.  It  is a quasipolynomial
since $\spec(R)$ has rational singularities. In fact, it is a polynomial, since 
$t=1$ is the only pole of the Hilbert series
\[ H_{k,\lambda}(t)=\sum_{n\ge 0} h_{k,\lambda}(n) t^n.\]
Figure~\ref{fgmodp} gives 
experimental evidence for strict  positivity (PH2) of $h_{k,\lambda}(n)$
(as discussed in Section~\ref{sschubertandgmodp})  for 
a few sample values of $k$ and $\lambda$.
Figure~\ref{fschubert} gives experimental evidence for strict
positivity of the  Hilbert 
polynomial  of the Schubert subvarieties of the Grassmanian;
there $G_{n,k}$ denotes the Grassmannian of $k$-planes in $V=\C^n$, and 
$\Omega_a$, $a=(a(1),\ldots,a(d))$  its Schubert subvariety consisting of the $k$-subspaces $W$ 
such that $\dim(W \cap V_{n-k+i-a(i)}) \ge i$ for all $i$, where 
$V=V_n \supset \cdots V_1 \supset 0$ is a complete flag of subspaces in $V$. 
The Hilbert polynomials were computed using the explicit polyhedral interpretation for them
deduced from the theory of algebras with straightening laws (Hodge algebras) 
\cite{hodge}.

\begin{sidewaysfigure} [h!] 
\[
\begin{array}{lll}
k&\lambda & h_{k,\lambda}(n) \\ \\
\hline \\
3&(21,  19) & 
399\,{n}^{3}+{\frac {35527969472513}{137438953472}}\,{n}^{2}+{\frac {4329327034365}{137438953472}}\,n+1\\ \\
5&(21,  19) &
{\frac {3700378042361}{4194304}}\,{n}^{7}+{\frac {575575719967}{524288}}\,{n}^{6}+{\frac {2157156441}{4096}}\,{n}^{5}+{\frac {266554253}{2048}}\,{n}^{4}+{\frac {4643843}{256}}\,{n}^{3}+{\frac {1468423}{1024}}\,{n}^{2}+{\frac {7619}{128}}\,n+1 \\ \\
3& (21,  9, 6) &
270\,{n}^{3}+{\frac {40819369181185}{274877906944}}\,{n}^{2}+{\frac {3092376453119}{137438953472}}\,n+1 \\ \\
3&(12,  9,  5)& 42\,{n}^{3}+{\frac {40132174413825}{1099511627776}}\,{n}^{2}+{\frac {11544872091645}{1099511627776}}\,n+1 \\ \\
3&(21,  9, 6)&
{\frac {27396522639355}{536870912}}\,{n}^{6}+{\frac {463063744509}{8388608}}\,{n}^{5}+{\frac {6265700353}{262144}}\,{n}^{4}+{\frac {5577375771}{1048576}}\,{n}^{3}+{\frac {84246529}{131072}}\,{n}^{2}+{\frac {20971505}{524288}}\,n+{\frac {1048573}{1048576}} \\ \\
3&(21,  19,  16) &
15\,{n}^{3}+{\frac {81363860455425}{4398046511104}}\,{n}^{2}+{\frac {8246337208319}{1099511627776}}\,n+1 \\ \\
4& (9, 7, 5) & 
{\frac {7215545057279}{17179869184}}\,{n}^{6}+{\frac {4183298146289}{4294967296}}\,{n}^{5}+{\frac {247765925897}{268435456}}\,{n}^{4}+{\frac {1914699777}{4194304}}\,{n}^{3}+{\frac {4160749567}{33554432}}\,{n}^{2}+{\frac {587202553}{33554432}}\,n+{\frac {67108863}{67108864}} \\ \\
4&(21, 12, 9)&
{\frac {16437913583613}{268435456}}\,{n}^{6}+{\frac {132498063359}{2097152}}\,{n}^{5}+{\frac {109509083155}{4194304}}\,{n}^{4}+{\frac {1462763527}{262144}}\,{n}^{3}+{\frac {171442179}{262144}}\,{n}^{2}+{\frac {10485755}{262144}}\,n+{\frac {524287}{524288}}
\\  \\
4&(21,  9, 5) &
{\frac {32469952757755}{536870912}}\,{n}^{6}+{\frac {129805320191}{2097152}}\,{n}^{5}+{\frac {108129157137}{4194304}}\,{n}^{4}+{\frac {2926313487}{524288}}\,{n}^{3}+{\frac {86638593}{131072}}\,{n}^{2}+{\frac {10616825}{262144}}\,n+{\frac {262143}{262144}}
\\ \\
4&(21,  9, 6) &
{\frac {27396522639355}{536870912}}\,{n}^{6}+{\frac {463063744509}{8388608}}\,{n}^{5}+{\frac {6265700353}{262144}}\,{n}^{4}+{\frac {5577375771}{1048576}}\,{n}^{3}+{\frac {84246529}{131072}}\,{n}^{2}+{\frac {20971505}{524288}}\,n+{\frac {1048573}{1048576}} \\ \\
4& (31,  19, 5) &
{\frac {35969680015355}{33554432}}\,{n}^{6}+{\frac {1424674346311}{2097152}}\,{n}^{5}+{\frac {22705493343}{131072}}\,{n}^{4}+{\frac {46973953}{2048}}\,{n}^{3}+{\frac {3423915}{2048}}\,{n}^{2}+{\frac {65365}{1024}}\,n+{\frac {16383}{16384}}
\end{array}
\]
\caption{Hilbert polynomial for $G/P_\lambda$, $G=SL_k(\C)$. There is a slight
rounding error caused by interpolation--e.g., the constant term of each polynomial
should be one.}
\label{fgmodp}
\end{sidewaysfigure}

\begin{sidewaysfigure} [h!] 
\[
\begin{array}{llll} 
n& k & a \\ \hline \\
7&3& (1, 3, 5) &
1/3\,{n}^{3}+{\frac {59373627899905}{39582418599936}}\,{n}^{2}+{\frac {28587302322173}{13194139533312}}\,n+1
\\ \\
7&3&(1, 2, 4) &
n+1 \\ \\
7&3&(1, 4, 6)&{\frac {22265110462465}{534362651099136}}\,{n}^{5}+{\frac {4638564679679}{11132555231232}}\,{n}^{4}+{\frac {13}{8}}\,{n}^{3}+{\frac {105942526633}{34359738368}}\,{n}^{2}+{\frac {146028888073}{51539607552}}\,n+{\frac {34359738361}{34359738368}}
\\ \\
6&2&(1, 4, 5)&
{\frac {15637498706143}{2251799813685248}}\,{n}^{6}+{\frac {3665038759245}{35184372088832}}\,{n}^{5}+{\frac {1389660529559}{2199023255552}}\,{n}^{4}+{\frac {272014595421}{137438953472}}\,{n}^{3} \\ \\ &&& +{\frac {230973796809}{68719476736}}\,{n}^{2}+{\frac {100215903571}{34359738368}}\,n+1
\\ \\
6&2&(1, 4, 6)&
{\frac {69578470195}{25048249270272}}\,{n}^{7}+{\frac {1217623228439}{25048249270272}}\,{n}^{6}+{\frac {372534725887}{1043677052928}}\,{n}^{5}+{\frac {30953963537}{21743271936}}\,{n}^{4}+{\frac {12044363351}{3623878656}}\,{n}^{3} \\ \\ &&& +{\frac {683671553}{150994944}}\,{n}^{2}+{\frac {1335466297}{402653184}}\,n+{\frac {268435457}{268435456}}
\\ \\
7&3& (1, 4, 6) &
{\frac {23456248059223}{562949953421312}}\,{n}^{5}+{\frac {7330077518505}{17592186044416}}\,{n}^{4}+{\frac {1786706395137}{1099511627776}}\,{n}^{3}+{\frac {423770106525}{137438953472}}\,{n}^{2}+{\frac {24338148015}{8589934592}}\,n+{\frac {17179869169}{17179869184}}
\\ \\
6&3&(1, 3, 6)&
1/8\,{n}^{4}+{\frac {16126170540715}{17592186044416}}\,{n}^{3}+{\frac {19}{8}}\,{n}^{2}+{\frac {710101259605}{274877906944}}\,n+1
\\ \\
8&3&(1, 3, 6)&
{\frac {171798691840001}{1374389534720000}}\,{n}^{4}+{\frac {31496426837333}{34359738368000}}\,{n}^{3}+{\frac {4080218931199}{1717986918400}}\,{n}^{2}+{\frac {443813287253}{171798691840}}\,n+1
\end{array}
\]
\caption{Hilbert polynomial of the Schubert subvariety $\Omega_a$,
$a=(a(1),\ldots,a(k))$, 
 of the Grassmannian $G_{n,k}$.}
\label{fschubert}
\end{sidewaysfigure} 

\section{The ring of symmetric functions} \label{sevisym}
Let  $V=\C^k$, $G=GL(V)$, $H=S_k$, 
with the natural embedding $H \rightarrow G$. 
Let us consider the spacial case of the subgroup restriction problem 
(Problem~\ref{pintrosubgroup}),
with $V_\lambda(G)=V$, and $V_\pi(H)$ the trivial representation of $H$. 
Then $s=m_\lambda^\pi$,  the multiplicity of the trivial representation 
in $V$, is one. Though the decision problem (Problem~\ref{pintrosubgroup})
is trivial
in this case, the canonical model associated with $s$ is nontrivial.

The canonical rings $R=R(s)$ and $S=S(s)$ associated with $s$ in this case coincide with
$\C[V]=\C[x_1,\ldots,x_k]$.
The ring $T=T(s)=S^H=\C[x_1,\ldots,x_k]^{S_k}$ 
is  the subring of symmetric functions.
Its Hilbert 
function $h(n)$ is a quasipolynomial. 
PH1 and PH3 for $Z=\proj(T)$, as per  Definition~\ref{dphcanonicalz}, 
follow easily, the latter from the
well known rational generating function for the partition function \cite{stanleyenu}. 
But PH2 turns out to be nontrivial. 
Figures~\ref{fsym1}-\ref{fsym6} give experimental evidence for strict
positivity of $h(n)$ (PH2).
In these figures, the $i$-th row of
the table  for a given $k$ shows 
$h_i(n)$, where
$h_i(n)$, $1\le i \le l$,  are such that $h(n)=h_i(n)$, when
$n=i$ modulo the period $l$ of $h(n)$.

\begin{figure} [h!] 
\[
\begin{array}{l}
k=2 \\ \\
\left [\begin {array}{c} 1/2\,n+1/2\\\noalign{\medskip}1/2\,n+1\end {array}\right ]
\\ \\
k=3 \\ \\
\left [\begin {array}{c} 1/12\,{n}^{2}+1/2\,n+{\frac {5}{12}}\\\noalign{\medskip}1/12\,{n}^{2}+1/2\,n+2/3\\\noalign{\medskip}1/12\,{n}^{2}+1/2\,n+3/4\\\noalign{\medskip}1/12\,{n}^{2}+1/2\,n+{\frac {46912496118443}{70368744177664}}\\\noalign{\medskip}1/12\,{n}^{2}+1/2\,n+{\frac {58640620148053}{140737488355328}}\\\noalign{\medskip}1/12\,{n}^{2}+1/2\,n+1\end {array}\right ]
\\ \\
k=4 \\ \\
\left [\begin {array}{c} {\frac {1}{144}}\,{n}^{3}+{\frac {5}{48}}\,{n}^{2}+{\frac {61572651155457}{140737488355328}}\,n+{\frac {15881834623431}{35184372088832}}\\\noalign{\medskip}{\frac {1}{144}}\,{n}^{3}+{\frac {117281240296107}{1125899906842624}}\,{n}^{2}+{\frac {140737488355325}{281474976710656}}\,n+{\frac {19}{36}}\\\noalign{\medskip}{\frac {1}{144}}\,{n}^{3}+{\frac {234562480592215}{2251799813685248}}\,{n}^{2}+{\frac {123145302310909}{281474976710656}}\,n+{\frac {19791209299969}{35184372088832}}\\\noalign{\medskip}{\frac {1}{144}}\,{n}^{3}+{\frac {234562480592215}{2251799813685248}}\,{n}^{2}+{\frac {70368744177667}{140737488355328}}\,n+{\frac {62549994824587}{70368744177664}}\\\noalign{\medskip}{\frac {1}{144}}\,{n}^{3}+{\frac {5}{48}}\,{n}^{2}+{\frac {61572651155453}{140737488355328}}\,n+{\frac {748278746681}{2199023255552}}\\\noalign{\medskip}{\frac {1}{144}}\,{n}^{3}+{\frac {117281240296107}{1125899906842624}}\,{n}^{2}+{\frac {70368744177665}{140737488355328}}\,n+{\frac {26388279066621}{35184372088832}}\\\noalign{\medskip}{\frac {1}{144}}\,{n}^{3}+{\frac {117281240296107}{1125899906842624}}\,{n}^{2}+{\frac {7}{16}}\,n+{\frac {7940917311717}{17592186044416}}\\\noalign{\medskip}{\frac {1}{144}}\,{n}^{3}+{\frac {117281240296107}{1125899906842624}}\,{n}^{2}+{\frac {35184372088831}{70368744177664}}\,n+{\frac {6841405683939}{8796093022208}}\\\noalign{\medskip}{\frac {1}{144}}\,{n}^{3}+{\frac {29320310074027}{281474976710656}}\,{n}^{2}+{\frac {30786325577729}{70368744177664}}\,n+{\frac {9}{16}}\\\noalign{\medskip}{\frac {1}{144}}\,{n}^{3}+{\frac {58640620148053}{562949953421312}}\,{n}^{2}+{\frac {35184372088831}{70368744177664}}\,n+{\frac {5619726097523}{8796093022208}}\\\noalign{\medskip}{\frac {1}{144}}\,{n}^{3}+{\frac {117281240296105}{1125899906842624}}\,{n}^{2}+{\frac {7}{16}}\,n+{\frac {2993114986727}{8796093022208}}\\\noalign{\medskip}{\frac {1}{144}}\,{n}^{3}+{\frac {58640620148055}{562949953421312}}\,{n}^{2}+1/2\,n+1\end {array}\right ]
 \end{array}
\]
\caption{The Hilbert quasipolynomial of $T_k=\C[x_1,\ldots,x_k]^{S_k}$; $k=2,3,4$.} 
\label{fsym1}
\end{figure}

\begin{figure} 
\[
\left [\begin {array}{c} {\frac {1}{2880}}\,{n}^{4}+{\frac {46912496118441}{4503599627370496}}\,{n}^{3}+{\frac {3787206717893}{35184372088832}}\,{n}^{2}+{\frac {469583091025}{1099511627776}}\,n+{\frac {499743305817}{1099511627776}}\\\noalign{\medskip}{\frac {1}{2880}}\,{n}^{4}+{\frac {46912496118445}{4503599627370496}}\,{n}^{3}+{\frac {3787206717891}{35184372088832}}\,{n}^{2}+{\frac {503942829399}{1099511627776}}\,n+{\frac {310001195055}{549755813888}}\\\noalign{\medskip}{\frac {1}{2880}}\,{n}^{4}+{\frac {46912496118441}{4503599627370496}}\,{n}^{3}+{\frac {3787206717897}{35184372088832}}\,{n}^{2}+{\frac {469583091031}{1099511627776}}\,n+{\frac {30279519437}{68719476736}}\\\noalign{\medskip}{\frac {400319966877379}{1152921504606846976}}\,{n}^{4}+{\frac {23456248059221}{2251799813685248}}\,{n}^{3}+{\frac {3787206717893}{35184372088832}}\,{n}^{2}+{\frac {503942829403}{1099511627776}}\,n+{\frac {47340083975}{68719476736}}\\\noalign{\medskip}{\frac {400319966877379}{1152921504606846976}}\,{n}^{4}+{\frac {5864062014805}{562949953421312}}\,{n}^{3}+{\frac {3787206717895}{35184372088832}}\,{n}^{2}+{\frac {117395772755}{274877906944}}\,n+{\frac {89955703917}{137438953472}}\\\noalign{\medskip}{\frac {400319966877379}{1152921504606846976}}\,{n}^{4}+{\frac {5864062014805}{562949953421312}}\,{n}^{3}+{\frac {3787206717893}{35184372088832}}\,{n}^{2}+{\frac {31496426837}{68719476736}}\,n+{\frac {92771293595}{137438953472}}\\\noalign{\medskip}{\frac {400319966877379}{1152921504606846976}}\,{n}^{4}+{\frac {46912496118447}{4503599627370496}}\,{n}^{3}+{\frac {1893603358947}{17592186044416}}\,{n}^{2}+{\frac {234791545515}{549755813888}}\,n+{\frac {5661005505}{17179869184}}\\\noalign{\medskip}{\frac {1}{2880}}\,{n}^{4}+{\frac {23456248059223}{2251799813685248}}\,{n}^{3}+{\frac {1893603358949}{17592186044416}}\,{n}^{2}+{\frac {62992853675}{137438953472}}\,n+{\frac {94680167945}{137438953472}}\\\noalign{\medskip}{\frac {400319966877379}{1152921504606846976}}\,{n}^{4}+{\frac {46912496118441}{4503599627370496}}\,{n}^{3}+{\frac {3787206717891}{35184372088832}}\,{n}^{2}+{\frac {117395772757}{274877906944}}\,n+{\frac {38869454029}{68719476736}}\\\noalign{\medskip}{\frac {1}{2880}}\,{n}^{4}+{\frac {46912496118441}{4503599627370496}}\,{n}^{3}+{\frac {3787206717897}{35184372088832}}\,{n}^{2}+{\frac {62992853675}{137438953472}}\,n+{\frac {52494044729}{68719476736}}\\\noalign{\medskip}{\frac {1}{2880}}\,{n}^{4}+{\frac {5864062014805}{562949953421312}}\,{n}^{3}+{\frac {946801679473}{8796093022208}}\,{n}^{2}+{\frac {234791545515}{549755813888}}\,n+{\frac {2830502753}{8589934592}}\\\noalign{\medskip}{\frac {400319966877379}{1152921504606846976}}\,{n}^{4}+{\frac {23456248059219}{2251799813685248}}\,{n}^{3}+{\frac {1893603358949}{17592186044416}}\,{n}^{2}+{\frac {251971414695}{549755813888}}\,n+{\frac {27487790695}{34359738368}}\\\noalign{\medskip}{\frac {400319966877379}{1152921504606846976}}\,{n}^{4}+{\frac {23456248059223}{2251799813685248}}\,{n}^{3}+{\frac {1893603358945}{17592186044416}}\,{n}^{2}+{\frac {234791545509}{549755813888}}\,n+{\frac {31233956605}{68719476736}}\\\noalign{\medskip}{\frac {1}{2880}}\,{n}^{4}+{\frac {11728124029611}{1125899906842624}}\,{n}^{3}+{\frac {946801679473}{8796093022208}}\,{n}^{2}+{\frac {251971414699}{549755813888}}\,n+{\frac {19375074691}{34359738368}}\\\noalign{\medskip}{\frac {1}{2880}}\,{n}^{4}+{\frac {11728124029611}{1125899906842624}}\,{n}^{3}+{\frac {473400839737}{4398046511104}}\,{n}^{2}+{\frac {29348943189}{68719476736}}\,n+{\frac {11005853695}{17179869184}}\\\noalign{\medskip}{\frac {400319966877379}{1152921504606846976}}\,{n}^{4}+{\frac {23456248059219}{2251799813685248}}\,{n}^{3}+{\frac {473400839737}{4398046511104}}\,{n}^{2}+{\frac {251971414699}{549755813888}}\,n+{\frac {5917510497}{8589934592}}\\\noalign{\medskip}{\frac {1}{2880}}\,{n}^{4}+{\frac {11728124029611}{1125899906842624}}\,{n}^{3}+{\frac {473400839737}{4398046511104}}\,{n}^{2}+{\frac {234791545511}{549755813888}}\,n+{\frac {15616978311}{34359738368}}\\\noalign{\medskip}{\frac {1}{2880}}\,{n}^{4}+{\frac {23456248059223}{2251799813685248}}\,{n}^{3}+{\frac {1893603358947}{17592186044416}}\,{n}^{2}+{\frac {62992853673}{137438953472}}\,n+{\frac {23192823403}{34359738368}}\\\noalign{\medskip}{\frac {1}{2880}}\,{n}^{4}+{\frac {11728124029611}{1125899906842624}}\,{n}^{3}+{\frac {946801679475}{8796093022208}}\,{n}^{2}+{\frac {117395772757}{274877906944}}\,n+{\frac {2830502755}{8589934592}}\\\noalign{\medskip}{\frac {400319966877379}{1152921504606846976}}\,{n}^{4}+{\frac {11728124029609}{1125899906842624}}\,{n}^{3}+{\frac {1893603358949}{17592186044416}}\,{n}^{2}+{\frac {31496426837}{68719476736}}\,n+{\frac {30541989663}{34359738368}}\\\noalign{\medskip}{\frac {400319966877379}{1152921504606846976}}\,{n}^{4}+{\frac {11728124029611}{1125899906842624}}\,{n}^{3}+{\frac {946801679473}{8796093022208}}\,{n}^{2}+{\frac {117395772759}{274877906944}}\,n+{\frac {9717363505}{17179869184}}\\\noalign{\medskip}{\frac {1}{2880}}\,{n}^{4}+{\frac {2932031007403}{281474976710656}}\,{n}^{3}+{\frac {1893603358945}{17592186044416}}\,{n}^{2}+{\frac {125985707353}{274877906944}}\,n+{\frac {4843768673}{8589934592}}\\\noalign{\medskip}{\frac {400319966877379}{1152921504606846976}}\,{n}^{4}+{\frac {23456248059223}{2251799813685248}}\,{n}^{3}+{\frac {59175104967}{549755813888}}\,{n}^{2}+{\frac {117395772755}{274877906944}}\,n+{\frac {2830502755}{8589934592}}\\\noalign{\medskip}{\frac {400319966877379}{1152921504606846976}}\,{n}^{4}+{\frac {23456248059221}{2251799813685248}}\,{n}^{3}+{\frac {946801679475}{8796093022208}}\,{n}^{2}+{\frac {62992853675}{137438953472}}\,n+{\frac {3435973837}{4294967296}}\\\noalign{\medskip}{\frac {400319966877379}{1152921504606846976}}\,{n}^{4}+{\frac {23456248059223}{2251799813685248}}\,{n}^{3}+{\frac {1893603358947}{17592186044416}}\,{n}^{2}+{\frac {58697886379}{137438953472}}\,n+{\frac {11244462985}{17179869184}}\\\noalign{\medskip}{\frac {400319966877379}{1152921504606846976}}\,{n}^{4}+{\frac {11728124029611}{1125899906842624}}\,{n}^{3}+{\frac {946801679475}{8796093022208}}\,{n}^{2}+{\frac {15748213419}{34359738368}}\,n+{\frac {9687537337}{17179869184}}\\\noalign{\medskip}{\frac {1}{2880}}\,{n}^{4}+{\frac {11728124029609}{1125899906842624}}\,{n}^{3}+{\frac {473400839737}{4398046511104}}\,{n}^{2}+{\frac {117395772753}{274877906944}}\,n+{\frac {1892469965}{4294967296}}\\\noalign{\medskip}{\frac {1}{2880}}\,{n}^{4}+{\frac {23456248059221}{2251799813685248}}\,{n}^{3}+{\frac {946801679473}{8796093022208}}\,{n}^{2}+{\frac {7874106709}{17179869184}}\,n+{\frac {11835020991}{17179869184}}\\\noalign{\medskip}{\frac {400319966877379}{1152921504606846976}}\,{n}^{4}+{\frac {11728124029611}{1125899906842624}}\,{n}^{3}+{\frac {473400839737}{4398046511104}}\,{n}^{2}+{\frac {117395772753}{274877906944}}\,n+{\frac {3904244579}{8589934592}}\\\noalign{\medskip}{\frac {400319966877379}{1152921504606846976}}\,{n}^{4}+{\frac {11728124029611}{1125899906842624}}\,{n}^{3}+{\frac {946801679473}{8796093022208}}\,{n}^{2}+{\frac {125985707349}{274877906944}}\,n+{\frac {1879048193}{2147483648}}\end {array}\right ]
\]
\caption{The Hilbert quasipolynomial of $T_k=\C[x_1,\ldots,x_k]^{S_k}$, $k=5$; the first $30$ 
rows.}\label{fsym2}
\end{figure} 

\begin{figure} 
\[
\left [\begin {array}{c} {\frac {1}{2880}}\,{n}^{4}+{\frac {11728124029609}{1125899906842624}}\,{n}^{3}+{\frac {946801679473}{8796093022208}}\,{n}^{2}+{\frac {58697886375}{137438953472}}\,n+{\frac {88453211}{268435456}}\\\noalign{\medskip}{\frac {1}{2880}}\,{n}^{4}+{\frac {11728124029611}{1125899906842624}}\,{n}^{3}+{\frac {946801679473}{8796093022208}}\,{n}^{2}+{\frac {15748213419}{34359738368}}\,n+{\frac {2958755247}{4294967296}}\\\noalign{\medskip}{\frac {400319966877379}{1152921504606846976}}\,{n}^{4}+{\frac {23456248059223}{2251799813685248}}\,{n}^{3}+{\frac {946801679475}{8796093022208}}\,{n}^{2}+{\frac {58697886375}{137438953472}}\,n+{\frac {4858681755}{8589934592}}\\\noalign{\medskip}{\frac {400319966877379}{1152921504606846976}}\,{n}^{4}+{\frac {11728124029609}{1125899906842624}}\,{n}^{3}+{\frac {473400839737}{4398046511104}}\,{n}^{2}+{\frac {62992853673}{137438953472}}\,n+{\frac {151367771}{268435456}}\\\noalign{\medskip}{\frac {1}{2880}}\,{n}^{4}+{\frac {23456248059221}{2251799813685248}}\,{n}^{3}+{\frac {946801679473}{8796093022208}}\,{n}^{2}+{\frac {58697886375}{137438953472}}\,n+{\frac {2274244835}{4294967296}}\\\noalign{\medskip}{\frac {1}{2880}}\,{n}^{4}+{\frac {11728124029611}{1125899906842624}}\,{n}^{3}+{\frac {236700419869}{2199023255552}}\,{n}^{2}+{\frac {62992853671}{137438953472}}\,n+{\frac {858993459}{1073741824}}\\\noalign{\medskip}{\frac {400319966877379}{1152921504606846976}}\,{n}^{4}+{\frac {11728124029611}{1125899906842624}}\,{n}^{3}+{\frac {946801679473}{8796093022208}}\,{n}^{2}+{\frac {14674471595}{34359738368}}\,n+{\frac {3904244571}{8589934592}}\\\noalign{\medskip}{\frac {400319966877379}{1152921504606846976}}\,{n}^{4}+{\frac {23456248059219}{2251799813685248}}\,{n}^{3}+{\frac {59175104967}{549755813888}}\,{n}^{2}+{\frac {7874106709}{17179869184}}\,n+{\frac {2421884337}{4294967296}}\\\noalign{\medskip}{\frac {400319966877379}{1152921504606846976}}\,{n}^{4}+{\frac {11728124029611}{1125899906842624}}\,{n}^{3}+{\frac {59175104967}{549755813888}}\,{n}^{2}+{\frac {29348943189}{68719476736}}\,n+{\frac {3784939927}{8589934592}}\\\noalign{\medskip}{\frac {400319966877379}{1152921504606846976}}\,{n}^{4}+{\frac {23456248059223}{2251799813685248}}\,{n}^{3}+{\frac {473400839737}{4398046511104}}\,{n}^{2}+{\frac {62992853673}{137438953472}}\,n+{\frac {1908874355}{2147483648}}\\\noalign{\medskip}{\frac {400319966877379}{1152921504606846976}}\,{n}^{4}+{\frac {5864062014805}{562949953421312}}\,{n}^{3}+{\frac {946801679473}{8796093022208}}\,{n}^{2}+{\frac {29348943189}{68719476736}}\,n+{\frac {1952122291}{4294967296}}\\\noalign{\medskip}{\frac {400319966877379}{1152921504606846976}}\,{n}^{4}+{\frac {11728124029609}{1125899906842624}}\,{n}^{3}+{\frac {946801679471}{8796093022208}}\,{n}^{2}+{\frac {62992853675}{137438953472}}\,n+{\frac {2899102929}{4294967296}}\\\noalign{\medskip}{\frac {400319966877379}{1152921504606846976}}\,{n}^{4}+{\frac {11728124029611}{1125899906842624}}\,{n}^{3}+{\frac {473400839735}{4398046511104}}\,{n}^{2}+{\frac {58697886381}{137438953472}}\,n+{\frac {707625689}{2147483648}}\\\noalign{\medskip}{\frac {400319966877379}{1152921504606846976}}\,{n}^{4}+{\frac {11728124029613}{1125899906842624}}\,{n}^{3}+{\frac {59175104967}{549755813888}}\,{n}^{2}+{\frac {15748213419}{34359738368}}\,n+{\frac {2958755253}{4294967296}}\\\noalign{\medskip}{\frac {1}{2880}}\,{n}^{4}+{\frac {11728124029609}{1125899906842624}}\,{n}^{3}+{\frac {473400839737}{4398046511104}}\,{n}^{2}+{\frac {58697886377}{137438953472}}\,n+{\frac {3288334339}{4294967296}}\\\noalign{\medskip}{\frac {100079991719345}{288230376151711744}}\,{n}^{4}+{\frac {5864062014805}{562949953421312}}\,{n}^{3}+{\frac {59175104967}{549755813888}}\,{n}^{2}+{\frac {62992853675}{137438953472}}\,n+{\frac {1210942169}{2147483648}}\\\noalign{\medskip}{\frac {1}{2880}}\,{n}^{4}+{\frac {11728124029611}{1125899906842624}}\,{n}^{3}+{\frac {946801679477}{8796093022208}}\,{n}^{2}+{\frac {29348943193}{68719476736}}\,n+{\frac {1415251375}{4294967296}}\\\noalign{\medskip}{\frac {1}{2880}}\,{n}^{4}+{\frac {5864062014805}{562949953421312}}\,{n}^{3}+{\frac {59175104967}{549755813888}}\,{n}^{2}+{\frac {31496426839}{68719476736}}\,n+{\frac {3435973835}{4294967296}}\\\noalign{\medskip}{\frac {100079991719345}{288230376151711744}}\,{n}^{4}+{\frac {1466015503701}{140737488355328}}\,{n}^{3}+{\frac {473400839737}{4398046511104}}\,{n}^{2}+{\frac {29348943195}{68719476736}}\,n+{\frac {976061147}{2147483648}}\\\noalign{\medskip}{\frac {100079991719345}{288230376151711744}}\,{n}^{4}+{\frac {11728124029611}{1125899906842624}}\,{n}^{3}+{\frac {473400839737}{4398046511104}}\,{n}^{2}+{\frac {31496426839}{68719476736}}\,n+{\frac {3280877793}{4294967296}}\\\noalign{\medskip}{\frac {1}{2880}}\,{n}^{4}+{\frac {5864062014805}{562949953421312}}\,{n}^{3}+{\frac {946801679473}{8796093022208}}\,{n}^{2}+{\frac {29348943191}{68719476736}}\,n+{\frac {946234983}{2147483648}}\\\noalign{\medskip}{\frac {100079991719345}{288230376151711744}}\,{n}^{4}+{\frac {2932031007403}{281474976710656}}\,{n}^{3}+{\frac {946801679475}{8796093022208}}\,{n}^{2}+{\frac {15748213419}{34359738368}}\,n+{\frac {1479377623}{2147483648}}\\\noalign{\medskip}{\frac {1}{2880}}\,{n}^{4}+{\frac {5864062014805}{562949953421312}}\,{n}^{3}+{\frac {59175104967}{549755813888}}\,{n}^{2}+{\frac {29348943195}{68719476736}}\,n+{\frac {122007643}{268435456}}\\\noalign{\medskip}{\frac {100079991719345}{288230376151711744}}\,{n}^{4}+{\frac {2932031007403}{281474976710656}}\,{n}^{3}+{\frac {473400839737}{4398046511104}}\,{n}^{2}+{\frac {3937053355}{8589934592}}\,n+{\frac {1449551461}{2147483648}}\\\noalign{\medskip}{\frac {100079991719345}{288230376151711744}}\,{n}^{4}+{\frac {2932031007403}{281474976710656}}\,{n}^{3}+{\frac {236700419869}{2199023255552}}\,{n}^{2}+{\frac {29348943193}{68719476736}}\,n+{\frac {71070151}{134217728}}\\\noalign{\medskip}{\frac {1}{2880}}\,{n}^{4}+{\frac {1466015503701}{140737488355328}}\,{n}^{3}+{\frac {59175104967}{549755813888}}\,{n}^{2}+{\frac {15748213419}{34359738368}}\,n+{\frac {739688813}{1073741824}}\\\noalign{\medskip}{\frac {100079991719345}{288230376151711744}}\,{n}^{4}+{\frac {11728124029611}{1125899906842624}}\,{n}^{3}+{\frac {473400839739}{4398046511104}}\,{n}^{2}+{\frac {14674471597}{34359738368}}\,n+{\frac {607335219}{1073741824}}\\\noalign{\medskip}{\frac {1}{2880}}\,{n}^{4}+{\frac {2932031007403}{281474976710656}}\,{n}^{3}+{\frac {236700419869}{2199023255552}}\,{n}^{2}+{\frac {3937053355}{8589934592}}\,n+{\frac {605471085}{1073741824}}\\\noalign{\medskip}{\frac {100079991719345}{288230376151711744}}\,{n}^{4}+{\frac {11728124029611}{1125899906842624}}\,{n}^{3}+{\frac {473400839737}{4398046511104}}\,{n}^{2}+{\frac {29348943193}{68719476736}}\,n+{\frac {88453211}{268435456}}\\\noalign{\medskip}{\frac {100079991719345}{288230376151711744}}\,{n}^{4}+{\frac {1466015503701}{140737488355328}}\,{n}^{3}+{\frac {473400839737}{4398046511104}}\,{n}^{2}+{\frac {3937053355}{8589934592}}\,n+{\frac {2147483647}{2147483648}}\end {array}\right ]
\]
\caption{The Hilbert quasipolynomial of $T_k=\C[x_1,\ldots,x_k]^{S_k}$, $k=5$; the last $30$ 
rows.} \label{fsym3}
\end{figure} 

\begin{sidewaysfigure} 
\[
\left [\begin {array}{c} {\frac {53375995583651}{4611686018427387904}}\,{n}^{5}+{\frac {21892498188609}{36028797018963968}}\,{n}^{4}+{\frac {418085902907}{35184372088832}}\,{n}^{3}+{\frac {115486898397}{1099511627776}}\,{n}^{2}+{\frac {26847522421}{68719476736}}\,n+{\frac {8448724291}{17179869184}}\\\noalign{\medskip}{\frac {53375995583651}{4611686018427387904}}\,{n}^{5}+{\frac {10946249094305}{18014398509481984}}\,{n}^{4}+{\frac {836171805815}{70368744177664}}\,{n}^{3}+{\frac {7396888121}{68719476736}}\,{n}^{2}+{\frac {15302809397}{34359738368}}\,n+{\frac {9853503363}{17179869184}}\\\noalign{\medskip}{\frac {106751991167299}{9223372036854775808}}\,{n}^{5}+{\frac {21892498188599}{36028797018963968}}\,{n}^{4}+{\frac {1672343611625}{140737488355328}}\,{n}^{3}+{\frac {57743449207}{549755813888}}\,{n}^{2}+{\frac {14060052663}{34359738368}}\,n+{\frac {975427339}{2147483648}}\\\noalign{\medskip}{\frac {13343998895913}{1152921504606846976}}\,{n}^{5}+{\frac {10946249094301}{18014398509481984}}\,{n}^{4}+{\frac {1672343611641}{140737488355328}}\,{n}^{3}+{\frac {59175104961}{549755813888}}\,{n}^{2}+{\frac {15302809423}{34359738368}}\,n+{\frac {610309549}{1073741824}}\\\noalign{\medskip}{\frac {106751991167305}{9223372036854775808}}\,{n}^{5}+{\frac {21892498188611}{36028797018963968}}\,{n}^{4}+{\frac {1672343611623}{140737488355328}}\,{n}^{3}+{\frac {115486898411}{1099511627776}}\,{n}^{2}+{\frac {13423761203}{34359738368}}\,n+{\frac {4461910959}{8589934592}}\\\noalign{\medskip}{\frac {13343998895913}{1152921504606846976}}\,{n}^{5}+{\frac {21892498188605}{36028797018963968}}\,{n}^{4}+{\frac {1672343611631}{140737488355328}}\,{n}^{3}+{\frac {29587552483}{274877906944}}\,{n}^{2}+{\frac {15939100865}{34359738368}}\,n+{\frac {1927366573}{2147483648}}\\\noalign{\medskip}{\frac {213503982334599}{18446744073709551616}}\,{n}^{5}+{\frac {10946249094305}{18014398509481984}}\,{n}^{4}+{\frac {418085902909}{35184372088832}}\,{n}^{3}+{\frac {57743449199}{549755813888}}\,{n}^{2}+{\frac {13423761217}{34359738368}}\,n+{\frac {835973461}{2147483648}}\\\noalign{\medskip}{\frac {106751991167299}{9223372036854775808}}\,{n}^{5}+{\frac {10946249094305}{18014398509481984}}\,{n}^{4}+{\frac {836171805821}{70368744177664}}\,{n}^{3}+{\frac {59175104963}{549755813888}}\,{n}^{2}+{\frac {7651404713}{17179869184}}\,n+{\frac {320001575}{536870912}}\\\noalign{\medskip}{\frac {106751991167299}{9223372036854775808}}\,{n}^{5}+{\frac {10946249094299}{18014398509481984}}\,{n}^{4}+{\frac {1672343611637}{140737488355328}}\,{n}^{3}+{\frac {115486898395}{1099511627776}}\,{n}^{2}+{\frac {14060052667}{34359738368}}\,n+{\frac {255936429}{536870912}}\\\noalign{\medskip}{\frac {213503982334597}{18446744073709551616}}\,{n}^{5}+{\frac {2736562273577}{4503599627370496}}\,{n}^{4}+{\frac {1672343611629}{140737488355328}}\,{n}^{3}+{\frac {7396888121}{68719476736}}\,{n}^{2}+{\frac {7651404703}{17179869184}}\,n+{\frac {2859997491}{4294967296}}\\\noalign{\medskip}{\frac {106751991167299}{9223372036854775808}}\,{n}^{5}+{\frac {342070284197}{562949953421312}}\,{n}^{4}+{\frac {104521475727}{8796093022208}}\,{n}^{3}+{\frac {57743449199}{549755813888}}\,{n}^{2}+{\frac {1677970153}{4294967296}}\,n+{\frac {1790721319}{4294967296}}\\\noalign{\medskip}{\frac {1}{86400}}\,{n}^{5}+{\frac {10946249094299}{18014398509481984}}\,{n}^{4}+{\frac {836171805815}{70368744177664}}\,{n}^{3}+{\frac {59175104959}{549755813888}}\,{n}^{2}+{\frac {3984775219}{8589934592}}\,n+{\frac {987842475}{1073741824}}\\\noalign{\medskip}{\frac {1}{86400}}\,{n}^{5}+{\frac {5473124547153}{9007199254740992}}\,{n}^{4}+{\frac {209042951453}{17592186044416}}\,{n}^{3}+{\frac {57743449193}{549755813888}}\,{n}^{2}+{\frac {3355940307}{8589934592}}\,n+{\frac {884291849}{2147483648}}\\\noalign{\medskip}{\frac {106751991167303}{9223372036854775808}}\,{n}^{5}+{\frac {10946249094301}{18014398509481984}}\,{n}^{4}+{\frac {836171805817}{70368744177664}}\,{n}^{3}+{\frac {118350209919}{1099511627776}}\,{n}^{2}+{\frac {1912851173}{4294967296}}\,n+{\frac {264972307}{536870912}}\\\noalign{\medskip}{\frac {213503982334605}{18446744073709551616}}\,{n}^{5}+{\frac {2736562273577}{4503599627370496}}\,{n}^{4}+{\frac {836171805827}{70368744177664}}\,{n}^{3}+{\frac {57743449201}{549755813888}}\,{n}^{2}+{\frac {7030026327}{17179869184}}\,n+{\frac {1233125369}{2147483648}}\\\noalign{\medskip}{\frac {53375995583651}{4611686018427387904}}\,{n}^{5}+{\frac {10946249094299}{18014398509481984}}\,{n}^{4}+{\frac {836171805819}{70368744177664}}\,{n}^{3}+{\frac {118350209933}{1099511627776}}\,{n}^{2}+{\frac {1912851177}{4294967296}}\,n+{\frac {1478317113}{2147483648}}\\\noalign{\medskip}{\frac {106751991167297}{9223372036854775808}}\,{n}^{5}+{\frac {1368281136787}{2251799813685248}}\,{n}^{4}+{\frac {836171805817}{70368744177664}}\,{n}^{3}+{\frac {57743449195}{549755813888}}\,{n}^{2}+{\frac {1677970151}{4294967296}}\,n+{\frac {117959881}{268435456}}\\\noalign{\medskip}{\frac {1667999861989}{144115188075855872}}\,{n}^{5}+{\frac {10946249094303}{18014398509481984}}\,{n}^{4}+{\frac {104521475727}{8796093022208}}\,{n}^{3}+{\frac {59175104961}{549755813888}}\,{n}^{2}+{\frac {3984775215}{8589934592}}\,n+{\frac {109722991}{134217728}}\\\noalign{\medskip}{\frac {106751991167297}{9223372036854775808}}\,{n}^{5}+{\frac {342070284197}{562949953421312}}\,{n}^{4}+{\frac {836171805815}{70368744177664}}\,{n}^{3}+{\frac {28871724597}{274877906944}}\,{n}^{2}+{\frac {1677970153}{4294967296}}\,n+{\frac {332087389}{1073741824}}\\\noalign{\medskip}{\frac {213503982334607}{18446744073709551616}}\,{n}^{5}+{\frac {10946249094307}{18014398509481984}}\,{n}^{4}+{\frac {836171805821}{70368744177664}}\,{n}^{3}+{\frac {59175104963}{549755813888}}\,{n}^{2}+{\frac {7651404709}{17179869184}}\,n+{\frac {384426083}{536870912}}\end {array}\right ]
\]
\caption{The Hilbert quasipolynomial of $T_k=\C[x_1,\ldots,x_k]^{S_k}$, $k=6$; the first  $20$ rows.}
\label{fsym4}
\end{sidewaysfigure} 

\begin{sidewaysfigure} 
\[
\left [\begin {array}{c} {\frac {213503982334615}{18446744073709551616}}\,{n}^{5}+{\frac {10946249094307}{18014398509481984}}\,{n}^{4}+{\frac {836171805813}{70368744177664}}\,{n}^{3}+{\frac {28871724597}{274877906944}}\,{n}^{2}+{\frac {7030026337}{17179869184}}\,n+{\frac {640721865}{1073741824}}\\\noalign{\medskip}{\frac {106751991167309}{9223372036854775808}}\,{n}^{5}+{\frac {10946249094297}{18014398509481984}}\,{n}^{4}+{\frac {209042951455}{17592186044416}}\,{n}^{3}+{\frac {29587552479}{274877906944}}\,{n}^{2}+{\frac {1912851179}{4294967296}}\,n+{\frac {157275009}{268435456}}\\\noalign{\medskip}{\frac {26687997791827}{2305843009213693952}}\,{n}^{5}+{\frac {342070284197}{562949953421312}}\,{n}^{4}+{\frac {836171805825}{70368744177664}}\,{n}^{3}+{\frac {57743449199}{549755813888}}\,{n}^{2}+{\frac {3355940301}{8589934592}}\,n+{\frac {90445245}{268435456}}\\\noalign{\medskip}{\frac {13343998895913}{1152921504606846976}}\,{n}^{5}+{\frac {5473124547153}{9007199254740992}}\,{n}^{4}+{\frac {104521475729}{8796093022208}}\,{n}^{3}+{\frac {29587552481}{274877906944}}\,{n}^{2}+{\frac {1992387605}{4294967296}}\,n+{\frac {450971557}{536870912}}\\\noalign{\medskip}{\frac {106751991167303}{9223372036854775808}}\,{n}^{5}+{\frac {10946249094307}{18014398509481984}}\,{n}^{4}+{\frac {836171805827}{70368744177664}}\,{n}^{3}+{\frac {28871724597}{274877906944}}\,{n}^{2}+{\frac {209746269}{536870912}}\,n+{\frac {17843591}{33554432}}\\\noalign{\medskip}{\frac {106751991167303}{9223372036854775808}}\,{n}^{5}+{\frac {10946249094297}{18014398509481984}}\,{n}^{4}+{\frac {836171805819}{70368744177664}}\,{n}^{3}+{\frac {924611015}{8589934592}}\,{n}^{2}+{\frac {239106397}{536870912}}\,n+{\frac {5146825}{8388608}}\\\noalign{\medskip}{\frac {1667999861989}{144115188075855872}}\,{n}^{5}+{\frac {5473124547153}{9007199254740992}}\,{n}^{4}+{\frac {418085902911}{35184372088832}}\,{n}^{3}+{\frac {7217931151}{68719476736}}\,{n}^{2}+{\frac {54922081}{134217728}}\,n+{\frac {265331665}{536870912}}\\\noalign{\medskip}{\frac {1667999861989}{144115188075855872}}\,{n}^{5}+{\frac {10946249094291}{18014398509481984}}\,{n}^{4}+{\frac {836171805825}{70368744177664}}\,{n}^{3}+{\frac {59175104963}{549755813888}}\,{n}^{2}+{\frac {478212795}{1073741824}}\,n+{\frac {326629601}{536870912}}\\\noalign{\medskip}{\frac {1667999861989}{144115188075855872}}\,{n}^{5}+{\frac {10946249094297}{18014398509481984}}\,{n}^{4}+{\frac {418085902909}{35184372088832}}\,{n}^{3}+{\frac {14435862303}{137438953472}}\,{n}^{2}+{\frac {3355940305}{8589934592}}\,n+{\frac {24121261}{67108864}}\\\noalign{\medskip}{\frac {53375995583655}{4611686018427387904}}\,{n}^{5}+{\frac {10946249094291}{18014398509481984}}\,{n}^{4}+{\frac {209042951453}{17592186044416}}\,{n}^{3}+{\frac {59175104973}{549755813888}}\,{n}^{2}+{\frac {3984775217}{8589934592}}\,n+{\frac {503316475}{536870912}}\\\noalign{\medskip}{\frac {26687997791827}{2305843009213693952}}\,{n}^{5}+{\frac {10946249094285}{18014398509481984}}\,{n}^{4}+{\frac {836171805825}{70368744177664}}\,{n}^{3}+{\frac {57743449207}{549755813888}}\,{n}^{2}+{\frac {3355940305}{8589934592}}\,n+{\frac {115234099}{268435456}}\\\noalign{\medskip}{\frac {106751991167303}{9223372036854775808}}\,{n}^{5}+{\frac {2736562273573}{4503599627370496}}\,{n}^{4}+{\frac {209042951459}{17592186044416}}\,{n}^{3}+{\frac {7396888121}{68719476736}}\,{n}^{2}+{\frac {3825702363}{8589934592}}\,n+{\frac {42684549}{67108864}}\\\noalign{\medskip}{\frac {106751991167301}{9223372036854775808}}\,{n}^{5}+{\frac {10946249094305}{18014398509481984}}\,{n}^{4}+{\frac {209042951453}{17592186044416}}\,{n}^{3}+{\frac {28871724605}{274877906944}}\,{n}^{2}+{\frac {1757506587}{4294967296}}\,n+{\frac {277411267}{536870912}}\\\noalign{\medskip}{\frac {106751991167303}{9223372036854775808}}\,{n}^{5}+{\frac {10946249094299}{18014398509481984}}\,{n}^{4}+{\frac {418085902905}{35184372088832}}\,{n}^{3}+{\frac {924611015}{8589934592}}\,{n}^{2}+{\frac {3825702353}{8589934592}}\,n+{\frac {33950041}{67108864}}\\\noalign{\medskip}{\frac {106751991167307}{9223372036854775808}}\,{n}^{5}+{\frac {1368281136787}{2251799813685248}}\,{n}^{4}+{\frac {52260737865}{4398046511104}}\,{n}^{3}+{\frac {28871724601}{274877906944}}\,{n}^{2}+{\frac {209746269}{536870912}}\,n+{\frac {61328751}{134217728}}\\\noalign{\medskip}{\frac {53375995583651}{4611686018427387904}}\,{n}^{5}+{\frac {2736562273575}{4503599627370496}}\,{n}^{4}+{\frac {104521475727}{8796093022208}}\,{n}^{3}+{\frac {29587552487}{274877906944}}\,{n}^{2}+{\frac {249048451}{536870912}}\,n+{\frac {128849017}{134217728}}\\\noalign{\medskip}{\frac {106751991167301}{9223372036854775808}}\,{n}^{5}+{\frac {5473124547145}{9007199254740992}}\,{n}^{4}+{\frac {209042951459}{17592186044416}}\,{n}^{3}+{\frac {7217931151}{68719476736}}\,{n}^{2}+{\frac {1677970157}{4294967296}}\,n+{\frac {30318473}{67108864}}\\\noalign{\medskip}{\frac {106751991167303}{9223372036854775808}}\,{n}^{5}+{\frac {342070284197}{562949953421312}}\,{n}^{4}+{\frac {418085902919}{35184372088832}}\,{n}^{3}+{\frac {14793776243}{137438953472}}\,{n}^{2}+{\frac {1912851181}{4294967296}}\,n+{\frac {143223581}{268435456}}\\\noalign{\medskip}{\frac {53375995583651}{4611686018427387904}}\,{n}^{5}+{\frac {5473124547153}{9007199254740992}}\,{n}^{4}+{\frac {209042951459}{17592186044416}}\,{n}^{3}+{\frac {1804482787}{17179869184}}\,{n}^{2}+{\frac {878753295}{2147483648}}\,n+{\frac {13898875}{33554432}}\\\noalign{\medskip}{\frac {106751991167303}{9223372036854775808}}\,{n}^{5}+{\frac {5473124547147}{9007199254740992}}\,{n}^{4}+{\frac {418085902907}{35184372088832}}\,{n}^{3}+{\frac {29587552479}{274877906944}}\,{n}^{2}+{\frac {956425585}{2147483648}}\,n+{\frac {195527051}{268435456}}\end {array}\right ]
\]
\caption{The Hilbert quasipolynomial of $T_k=\C[x_1,\ldots,x_k]^{S_k}$, $k=6$; the middle
  $20$ rows.}
\label{fsym5}
\end{sidewaysfigure} 

\begin{sidewaysfigure}  
\[
\left [\begin {array}{c} {\frac {13343998895913}{1152921504606846976}}\,{n}^{5}+{\frac {5473124547145}{9007199254740992}}\,{n}^{4}+{\frac {418085902919}{35184372088832}}\,{n}^{3}+{\frac {7217931149}{68719476736}}\,{n}^{2}+{\frac {209746269}{536870912}}\,n+{\frac {64348647}{134217728}}\\\noalign{\medskip}{\frac {53375995583651}{4611686018427387904}}\,{n}^{5}+{\frac {5473124547143}{9007199254740992}}\,{n}^{4}+{\frac {104521475729}{8796093022208}}\,{n}^{3}+{\frac {14793776237}{137438953472}}\,{n}^{2}+{\frac {498096901}{1073741824}}\,n+{\frac {28772927}{33554432}}\\\noalign{\medskip}{\frac {106751991167301}{9223372036854775808}}\,{n}^{5}+{\frac {5473124547141}{9007199254740992}}\,{n}^{4}+{\frac {209042951457}{17592186044416}}\,{n}^{3}+{\frac {3608965575}{34359738368}}\,{n}^{2}+{\frac {1677970153}{4294967296}}\,n+{\frac {11719905}{33554432}}\\\noalign{\medskip}{\frac {13343998895913}{1152921504606846976}}\,{n}^{5}+{\frac {5473124547151}{9007199254740992}}\,{n}^{4}+{\frac {418085902905}{35184372088832}}\,{n}^{3}+{\frac {3698444059}{34359738368}}\,{n}^{2}+{\frac {478212797}{1073741824}}\,n+{\frac {74631683}{134217728}}\\\noalign{\medskip}{\frac {106751991167311}{9223372036854775808}}\,{n}^{5}+{\frac {5473124547147}{9007199254740992}}\,{n}^{4}+{\frac {418085902907}{35184372088832}}\,{n}^{3}+{\frac {28871724591}{274877906944}}\,{n}^{2}+{\frac {878753299}{2147483648}}\,n+{\frac {10682367}{16777216}}\\\noalign{\medskip}{\frac {106751991167313}{9223372036854775808}}\,{n}^{5}+{\frac {5473124547151}{9007199254740992}}\,{n}^{4}+{\frac {104521475729}{8796093022208}}\,{n}^{3}+{\frac {14793776237}{137438953472}}\,{n}^{2}+{\frac {478212797}{1073741824}}\,n+{\frac {84006215}{134217728}}\\\noalign{\medskip}{\frac {106751991167307}{9223372036854775808}}\,{n}^{5}+{\frac {2736562273573}{4503599627370496}}\,{n}^{4}+{\frac {209042951459}{17592186044416}}\,{n}^{3}+{\frac {28871724589}{274877906944}}\,{n}^{2}+{\frac {104873135}{268435456}}\,n+{\frac {3161957}{8388608}}\\\noalign{\medskip}{\frac {53375995583655}{4611686018427387904}}\,{n}^{5}+{\frac {5473124547149}{9007199254740992}}\,{n}^{4}+{\frac {209042951451}{17592186044416}}\,{n}^{3}+{\frac {29587552485}{274877906944}}\,{n}^{2}+{\frac {996193803}{2147483648}}\,n+{\frac {118111589}{134217728}}\\\noalign{\medskip}{\frac {106751991167309}{9223372036854775808}}\,{n}^{5}+{\frac {1368281136787}{2251799813685248}}\,{n}^{4}+{\frac {418085902907}{35184372088832}}\,{n}^{3}+{\frac {28871724601}{274877906944}}\,{n}^{2}+{\frac {419492539}{1073741824}}\,n+{\frac {49899533}{134217728}}\\\noalign{\medskip}{\frac {106751991167305}{9223372036854775808}}\,{n}^{5}+{\frac {2736562273573}{4503599627370496}}\,{n}^{4}+{\frac {52260737863}{4398046511104}}\,{n}^{3}+{\frac {29587552479}{274877906944}}\,{n}^{2}+{\frac {1912851173}{4294967296}}\,n+{\frac {87717907}{134217728}}\\\noalign{\medskip}{\frac {106751991167309}{9223372036854775808}}\,{n}^{5}+{\frac {1368281136787}{2251799813685248}}\,{n}^{4}+{\frac {104521475729}{8796093022208}}\,{n}^{3}+{\frac {28871724595}{274877906944}}\,{n}^{2}+{\frac {878753303}{2147483648}}\,n+{\frac {8962701}{16777216}}\\\noalign{\medskip}{\frac {106751991167305}{9223372036854775808}}\,{n}^{5}+{\frac {2736562273571}{4503599627370496}}\,{n}^{4}+{\frac {209042951453}{17592186044416}}\,{n}^{3}+{\frac {29587552499}{274877906944}}\,{n}^{2}+{\frac {956425585}{2147483648}}\,n+{\frac {10878263}{16777216}}\\\noalign{\medskip}{\frac {53375995583651}{4611686018427387904}}\,{n}^{5}+{\frac {5473124547157}{9007199254740992}}\,{n}^{4}+{\frac {104521475727}{8796093022208}}\,{n}^{3}+{\frac {3608965575}{34359738368}}\,{n}^{2}+{\frac {838985083}{2147483648}}\,n+{\frac {53611225}{134217728}}\\\noalign{\medskip}{\frac {13343998895913}{1152921504606846976}}\,{n}^{5}+{\frac {5473124547145}{9007199254740992}}\,{n}^{4}+{\frac {418085902907}{35184372088832}}\,{n}^{3}+{\frac {14793776249}{137438953472}}\,{n}^{2}+{\frac {498096903}{1073741824}}\,n+{\frac {104354293}{134217728}}\\\noalign{\medskip}{\frac {106751991167299}{9223372036854775808}}\,{n}^{5}+{\frac {2736562273575}{4503599627370496}}\,{n}^{4}+{\frac {209042951447}{17592186044416}}\,{n}^{3}+{\frac {3608965575}{34359738368}}\,{n}^{2}+{\frac {838985087}{2147483648}}\,n+{\frac {7873221}{16777216}}\\\noalign{\medskip}{\frac {106751991167303}{9223372036854775808}}\,{n}^{5}+{\frac {5473124547151}{9007199254740992}}\,{n}^{4}+{\frac {418085902899}{35184372088832}}\,{n}^{3}+{\frac {14793776243}{137438953472}}\,{n}^{2}+{\frac {956425599}{2147483648}}\,n+{\frac {90737805}{134217728}}\\\noalign{\medskip}{\frac {53375995583655}{4611686018427387904}}\,{n}^{5}+{\frac {5473124547155}{9007199254740992}}\,{n}^{4}+{\frac {104521475727}{8796093022208}}\,{n}^{3}+{\frac {14435862303}{137438953472}}\,{n}^{2}+{\frac {878753295}{2147483648}}\,n+{\frac {18680385}{33554432}}\\\noalign{\medskip}{\frac {106751991167305}{9223372036854775808}}\,{n}^{5}+{\frac {2736562273573}{4503599627370496}}\,{n}^{4}+{\frac {104521475725}{8796093022208}}\,{n}^{3}+{\frac {3698444063}{34359738368}}\,{n}^{2}+{\frac {478212799}{1073741824}}\,n+{\frac {36634403}{67108864}}\\\noalign{\medskip}{\frac {106751991167311}{9223372036854775808}}\,{n}^{5}+{\frac {684140568395}{1125899906842624}}\,{n}^{4}+{\frac {209042951463}{17592186044416}}\,{n}^{3}+{\frac {7217931153}{68719476736}}\,{n}^{2}+{\frac {52436567}{134217728}}\,n+{\frac {19926951}{67108864}}\\\noalign{\medskip}{\frac {26687997791827}{2305843009213693952}}\,{n}^{5}+{\frac {1368281136789}{2251799813685248}}\,{n}^{4}+{\frac {6532592233}{549755813888}}\,{n}^{3}+{\frac {14793776243}{137438953472}}\,{n}^{2}+{\frac {498096903}{1073741824}}\,n+{\frac {67108871}{67108864}}\end {array}\right ]
\]
\caption{The Hilbert quasipolynomial of $T_k=\C[x_1,\ldots,x_k]^{S_k}$, $k=6$; the last $20$ rows.}
\label{fsym6}
\end{sidewaysfigure}

\chapter{On verification and discovery of obstructions} \label{cobstruction}
In this chapter we   give applications of the results and
positivity hypotheses in this paper to  the problem of verifying 
or discovering an  {\em obstruction}, i.e., a ``proof of
hardness'' \cite{GCT2}
in the context of the $P$ vs. $NP$ and the permanent vs. determinant
 problems in characteristic zero.

\section{Obstruction}
An obstruction in an abstract setting  of Problem~\ref{pintrogit}
is defined as follows.

Let $X$ and $Y$ be $H$-varieties with compact specifications
 (Section~\ref{sspgit}),
$H$ a connected reductive group.
Let $\bitlength{X}$ and $\bitlength{Y}$ denote the bit lengths of 
their specifications  (Section~\ref{sspgit}).
Suppose we wish to show that 
$X$ cannot be embedded as an $H$-subvariety of $Y$. Pictorially:
\begin{equation} \label{eqembedding}
X \not \hookrightarrow Y.
\end{equation}

For example, in the context of the $P$ vs. $NP$ problem in characteristic 
zero  \cite{GCT1,GCT2}, 
$X$ is a class variety $X_{NP}(n,l)$ associated
with the complexity class $NP$ for the given input size parameter
$n$ and the circuit size   parameter $l$. The variety  $Y$ is
the  class variety $X_P(l)$ associated with the class $P$ for given $l$.
And $H$ is $SL_l(\C)$. 
If $NP \subseteq P$ (over $\C$) to the contrary, then it would turn out
that 
\[X_{NP}(n,l) \hookrightarrow X_P(l)\] 
as an $H$-subvariety,
for every $l=\poly(n)$. The goal is to show that this embedding cannot
exist when $l=\poly(n)$ and $n\rightarrow \infty$.

Let $R(X)$ and $R(Y)$ be the homogeneous coordinate rings of $X$ and $Y$,
respectively. Let $R(X)_d$ and $R(Y)_d$ denote their degree $d$-compoenents.
Suppose to the contrary that an $H$-embedding as in (\ref{eqembedding})
exists. Then there exists a degre preserving $H$-equivariant surjection 
from $R(Y)_d$ to $R(X)_d$ for every $d$, and hence, 
a degree-preserving $H$-equivariant injection from 
$R(X)_d^*$ to $R(Y)_d^*$.  Hence,
every irreducible $H$-module $V_\lambda(G)$ that occurs in $R(X)_d^*$ 
also occurs in $R(Y)_d^*$. This leads to:

\begin{defn} 
An irreducible representation $V_\lambda(H)$ is called an 
obstruction for the pair (X,Y)
if it occurs (as an $H$-submodule)
in $R(X)_d^*$ but not in $R(Y)_d^*$, for some $d$.
We say that $V_\lambda(H)$ is an obstruction of degree $d$.
\end{defn}

\begin{remark} 
The obstruction as defined here is dual to the obstrution as defined in
\cite{GCT2}.
\end{remark} 

Existence of such an obstruction implies that $X$ cannot be embedded in
$Y$ as an $H$-subvariety.

Let us assume that $X$ and $Y$ are $H$-subvarieties of $P(V)$, where 
$V$ is an $H$-module, and that we are given a point
$y \in Y \subseteq P(V)$ that is  distiguished in the following 
sense. 
Let $H_y \subseteq H$ be the stabilizer of $y$. Then
$\C y$, the line in $V$ corresponding to $y$, is invariant under $H_y$. 
Let $[y]$ be the set of points in $P(V)$ stabilized by $H_y$. 
We say that $y$ is {\em characterized by its stabilizer}  $H_y$ if $y=[y]$;
i.e., $y$ is the only point in $P(V)$ stabilized by $H_y$.
Let \[ H [y]=\cup_{z \in [y]} H z\] 
be the union of the $H$-orbits of all points in $[y]$.
We say that $y$ is a distinguished point of $Y$ if 
$Y$ equals  the projective closure of $H[y]$ in $P(V)$.
If $y$ is characterized
by its stabilizer, this means $Y$ is the projective 
closure of the orbit $H y$ of $y$. 

If $V_\lambda(H)$ occurs in $R(Y)_d^*$, then  it can be shown 
(cf. Proposition 4.2 in \cite{GCT2}) that $V_\lambda(H)$ 
 contains an $H_y$-submodule 
isomorphic to $(\C y)^d$, the $d$-th tensor power of $\C y$. 
This leads to the following stronger notion of obstruction:

\begin{defn} \cite{GCT2} \label{dstrongobs}
We say that $V_\lambda(H)$ is a strong obstruction for the pair $(X,Y)$
if, for some $d$, 
it occurs in $R(X)_d^*$, but it does not contain an $H_y$-module 
isomorphic to $(\C y)^d$.
\end{defn} 
Existence of
a strong  obstruction also implies that $X$ cannot be embedded in
$Y$ as an $H$-subvariety. The results in  \cite{GCT2} 
suggest that strong obstructions exist in the context of the lower bound
problems under cosideration. 
The goal then is to show their existence. 

\section{Decision problems} 
The decisions problems that arise in this context  are the following.
Let $s_{d}^\lambda$  be the multiplicity of $V_\lambda(H)$ in $R(X)_d^*$, and
$m^{d}_\lambda$  the multiplicity of the $H_y$-module 
$(\C y)^d$ in $V_\lambda(H)$, considered an $H_y$-module via the 
 the embedding $\rho: H_y \hookrightarrow  H$. 
Thus $\lambda$ is  a strong obstruction of degree $d$ iff 
$s_d^\lambda$ is nonzero and $m^d_\lambda$ is zero.

\begin{problem} (Decision Problems) 

\noindent (a) Given $d, \lambda$ and the specification of $X$, 
decide if $s_d^\lambda$ is nonzero. 

\noindent (b) Given $d, \lambda$ and the specifications of $H, H_y$ and
$\rho$, 
decide if $m^d_\lambda$ is nonzero. 

\noindent (c) Given $d,\lambda$ and the specifications of $X, H, H_y$
 and $\rho$, decide if  $\lambda$ is a strong obstruction of degree $d$.
\end{problem} 

The first is an instance of the decision Problem~\ref{pintrogit} in
geometric invariant theory, and the second of the subgroup restriction
Problem~\ref{pintrosubgroup}.
By the results in Chapter~\ref{csatpos}-\ref{cquasipoly}, relaxed forms
of the decision problems in (a) and (b)
belong to $P$ assuming appropriate PH1 and SH; this implies that
a relaxed form of the decision problem in (c) also belongs to $P$ 
assumming PH1 and SH. We will only need a weak relaxed form of (c), 
for which the weak form 
of SH that is implied by PH1 (cf. Theorem~\ref{tweaksh}) will suffice.

\section{Verification of obstructions}
The relevant PH1 are as follows. 

Assume that that singularities of $\spec(R(X))$ are rational. 
By Theorem~\ref{tquasigit}, 
the stretching function $\tilde s_{d}^\lambda(k)
=s_{k d}^{k \lambda}$ is a quasipolynomial. Hence
PH1 for $s_{d}^\lambda$ (Hypothesis~\ref{phph1}, or rather its 
slight variant obtained by replacing $R(X)_d$ with $R(X)_d^*$) is well defined.
It is:

\begin{hypo} {\bf (PH1):} \label{hph1classvariety}

There exists 
a polytope $P_{d}^\lambda$ such that:
\begin{enumerate} 
\item The number  of 
integer points in $P_{d}^\lambda$ is equal to $s_{d}^\lambda$.
\item The Ehrhart quasi-polynomial  of $P_{d}^\lambda$ coincides with
the stretching quasi-polynomial $\tilde s_{d}^\lambda(n)$ 
(cf. Theorem~\ref{tquasigit}). 
\item The polytope $P_{d}^\lambda$ is given by a separating oracle, as in
Section~\ref{sseporacle}. Its 
encoding bitlength $\bitlength{P_d^\lambda}$ 
is  $\poly(\bitlength{d},\bitlength{\lambda}, \bitlength{X})$,
and the combinatorial size $\comb{P_d^\lambda}$ is $\poly(\height(\lambda),
\comb{X})$, where $\comb{X}$ is the combinatorial size of $X$ 
(Section~\ref{sspgit}), and $\height(\lambda)$ is the height of $\lambda$.
\end{enumerate} 
\end{hypo} 

Similarly by Theorem~\ref{tquasisubgroup}
(or rather its slight variant which can be 
proved similarly),
the stretching function $m^{k d}_{k \lambda}$
is a quasipolynomial. 
Hence
PH1 for $m^{d}_\lambda$ (cf. Hypothesis~\ref{phph1} and 
Section~\ref{sssubgroup}) is also well defined. It is:

\begin{hypo} {\bf (PH1:)} \label{hph1subg}

There exists
a polytope $Q^{d}_\lambda$ such that:
\begin{enumerate} 
\item The number  of 
integer points in $Q^{d}_\lambda$ is equal to $m^{d}_\lambda$.
\item The Ehrhart quasi-polynomial  of $Q^{d}_\lambda$ coincides with
the stretching quasi-polynomial $\tilde m^{d}_\lambda(n)$ 
(Theorem~\ref{tquasigit}). 
\item The polytope $Q^{d}_\lambda$ is given by a separating oracle. Its
encoding bitlength $\bitlength{Q^d_\lambda}$ is 
 $\poly(\bitlength{d},\bitlength{\lambda}, \bitlength{\rho}, \bitlength{H_y}, 
\bitlength{H})$, and the combinatorial size $\comb{Q^d_\lambda}$ is
$O(\poly(\height(\lambda), \bitlength{H},\bitlength{H_y}, \bitlength{\rho}))$.
Here $\bitlength{H},\bitlength{H_y}$ and  $\bitlength{\rho}$
denote the bitlengths of $H,H_y$ and $\rho$ (Section~\ref{sssubgroup}). 
\end{enumerate} 
\end{hypo} 

\begin{theorem} {\bf (Weak SH:)} \label{tweakshobs}

\noindent (a) Assuming PH1 (Hypothesis~\ref{hph1classvariety}),
the saturation index of $\tilde s_d^\lambda(n)$ is at most 
$a^{\poly(\comb{P_d^\lambda})}$, for some explicit constant $a>0$. 

\noindent (b) Assuming PH1 (Hypothesis~\ref{hph1subg}),
the saturation index of $\tilde m^d_\lambda(n)$ is at most 
$b^{\poly(\comb{Q^d_\lambda})}$, for some explicit constant $b>0$. 
\end{theorem} 

This follows from Theorem~\ref{tweaksh}.

\begin{theorem} \label{tverobs1}
Assume PH1 (Hypotheses~\ref{hph1classvariety}-\ref{hph1subg}).
Then, given $d,\lambda$, the specifications of $X,H,H_y$ and $\rho$,
and a relaxation parameter $c$ greater than the explicit bounds on 
the saturation indices in Theorem~\ref{tweakshobs}, 
whether $c \lambda$ is an obstruction of degree $d$ can be 
decided in 
\[ \poly(\bitlength{d},\bitlength{\lambda},\bitlength{X},\bitlength{H},
\bitlength{H_y},\bitlength{\rho},\bitlength{c})\]
time.
\end{theorem}
This follows by applying  Theorem~\ref{tindexquasi}
to the polytopes $P_d^\lambda$ and $Q^d_\lambda$ with
the saturation index estimates in Theorem~\ref{tweakshobs}.

\section{Robust obstruction} 
We now define a notion of
obstruction that is well behaved with respect to relaxation.

\begin{defn} \label{drobust}
Assume PH1 for both $s_{d}^\lambda$ and $m^d_\lambda$ 
(Hypotheses~\ref{hph1classvariety}-\ref{hph1subg}).
We say that $V_\lambda(H)$ is a {\em robust obstruction} for the
pair $(X,Y)$ if  one of the following hold:
\begin{enumerate} 
\item $Q^d_\lambda$ is empty, and $P_{d}^\lambda$ is 
nonempty. 
\item  Both $Q^d\lambda$ and $P_{d}^\lambda$ are nonempty, 
the affine span of $Q^d_\lambda$ does not contain an integer point and 
the affine span of $P_{d}^\lambda$ contains an integer point.
\end{enumerate}
\end{defn}

If $V_\lambda(H)$ is a robust obstruction, so is $V_{l \lambda}(H)$,
for all or most positive integral $l$, hence the name robust.

\begin{prop} \label{probustgen}
Assume PH1 for both $s_{d}^\lambda$ and $m^d_\lambda$ as above.
If $V_\lambda(H)$ is a robust obstruction for the pair $(X,Y)$, then 
for some positive integer $k$--called a relaxation 
parameter--$V_{k \lambda}(H)$ is a strong obstruction
for $(X,Y)$. In fact, this is so for most large enough $k$.
\end{prop} 
\proof

\noindent (1) Suppose 
$Q^d_\lambda$ is empty, and $P_{d}^\lambda$ is 
nonempty. Let $k$ be a large enough 
positive integer $k$ such that $k P_{d}^\lambda=
P_{d k}^{k \lambda}$ contains an integer point. Then 
$s_{k d}^{k \lambda}$ is nonzero. But
$m^{k d}_{k \lambda}$ is zero since  $Q^{k d}_{k \lambda}= k Q^d_\lambda$ 
is empty. Thus $k \lambda$ is a strong obstruction. 

\noindent (2) Suppose
both $Q^d\lambda$ and $P_{d}^\lambda$ are nonempty, 
the affine span of $Q^d_\lambda$ does not contain an integer point and 
the affine span of $P_{d}^\lambda$ contains an integer point.
We can choose a positive integer $k$ such that 
the affine span of $k Q^d_\lambda = Q^{d k}_{d \lambda}$ does not 
contain an integer point, but $k P_{d}^\lambda=P_{d k}^{k \lambda}$ 
contains an integer point; most large enough $k$ have this property.
This means 
$s_{k d}^{k \lambda}$ is nonzero, but
$m^{k d}_{k \lambda}$ is zero. Thus $k \lambda$ is a strong obstruction. \qed

\section{Verification of robust obstructions}

\begin{theorem} \label{tverifyobs}
Assume that the singularities of $\spec(R(X))$ are rational.
Assume PH1 for both $s_{d}^\lambda$ and $m^d_\lambda$ as above. 
Then, given $\lambda, d$ and the specifications of $\rho:H_y \hookrightarrow 
H$ and $X$, 
whether $V_\lambda(H)$ is a robust obstruction can be verified 
in $\poly(\bitlength{\rho}, \bitlength{H_y}, \bitlength{H}, \bitlength{X}, 
\bitlength{d}, \bitlength{\lambda})$ time.
Furthermore, a positive integral relaxation parameter 
 $k$ such that $V_{k \lambda}(G)$ is a strong 
obstruction can also be found in the same time.
\end{theorem} 
The crucial result used implicitly here
is the quasipolynomiality theorem (Theorem~\ref{tquasimain1}) because 
of which PH1 for both $s_{d}^\lambda$ and $m^d_\lambda$ are well defined.

\proof 
By linear programming \cite{lovasz},
 whether $Q^d_\lambda$ is nonempty or not can 
be determined in 
$\poly(\bitlength{Q^d_\lambda})=\poly(\bitlength{\rho}, \bitlength{H_y},
\bitlength{H}, \bitlength{d}, \bitlength{\lambda})$ 
time. If it is nonempty, the linear programming algorithm also gives 
its affine span. Whether this contains an integer point can be 
determined in polynomial time, using the polynomial time algorithm 
for computing the Smith normal form, as in the proof of 
Theorem~\ref{tindexquasi}. 

Similarly, whether 
 $P_{d}^\lambda$  is nonempty or not can 
be determined in 
$\poly(\bitlength{P_{d}^\lambda})=\poly(\bitlength{X},
\bitlength{d}, \bitlength{\lambda})$ 
time. If it is nonempty, whether its 
 affine span  contains an integer point can   be 
determined in polynomial time similarly. Furthermore, the algorithm 
can also be made to
return a vertex $v$  of the polytope $P_d^\lambda$ if it is nonempty. 

Using  these observations, whether $V_\lambda(G)$ is
a robust obstruction can be determined in polynomial time.

As far as the computation of the relaxation parameter 
$k$ is concerned, let us consider the
second case in Definition~\ref{drobust}--when 
both $Q^d_\lambda$ and $P_{d}^\lambda$ are nonempty, 
the affine span of $Q^d_\lambda$ does not contain an integer point and 
the affine span of $P_{d}^\lambda$ contains an integer point--the first
case being simpler. In this case, 
by examining the Smith normal forms of the defining 
equations of the affine spans of $P_d^\lambda$ and $Q^d_\lambda$ 
and the rational coordinates of a   vertex  $v \in P_d^\lambda$, we 
can find a large enough $k$ so that the affine span of 
$Q^{k d}_{k \lambda}$ does not contain an integer point, the affine span
of $P_{k d}^{k \lambda}$ contains an integer point, and 
$P_{k d}^{k \lambda}$ contains an integer point that is some multiple of $v$.
\qed

The value of the relaxation parameter $k$ computed above is rather
conservative. 
One may wish to compute 
as small value of $k$ as possible for which $V_{k \lambda}(G)$ 
is a strong obstruction (though in our application  this is not 
necessary). If SH for holds for the structural constant $s_d^\lambda$ 
(cf. Hypothesis~\ref{phsh} and Section~\ref{sspgit}), then we can let 
$k$ be the smallest integer larger than the saturation index (estimate) 
for $P_d^\lambda$ such that affine span of $Q^{k d}_{k \lambda}$ (if nonempty)
does not contain an integer point (as can be ensured by looking at the
Smith normal of  the  defining equations of the affine span).

\section{Arithemetic version of the
$P^{\#P}$ vs. $NC$  problem in characteristric zero} \label{sdetvsperm}
We now specialize the discussion in the preceding sections
in the context of
 the arithmetic form of the
$P^{\#P}$ vs. $NC$  problem in characteristric zero \cite{valiant}.
In concrete terms, the problem is to show that
the permanent of an $n \times n$ complex matrix $X$ cannot be 
expressed as a determinant of an $m\times m$ complex matrix, whose
 entries are  (possibly nonhomogeneous) linear combinations of the
entries of $X$.

\subsection{Class varieties}
The class varieties in this context are as follows \cite{GCT1}.
Let $Y$ be an $m\times m$ variable matrix,
which can also be thought of as a variable $l$-vector, $l=m^2$. Let
$X$ be its, say,  principal bottom-right 
$n\times n$ submatrix, $n<m$, which can be thought of as a 
variable $k$-vector, $k=n^2$. Let $V=\sym^m(Y)$ be the space of
homogeneous forms of degree $m$ in the variable entries of $Y$. 
The space $V$, and hence $P(V)$,  has a natural action of $G=GL(Y)=GL_l(\C)$
given by
\[ (\sigma f) (Y)=f(\sigma^{-1} Y),\] 
for any $f \in V$, $\sigma \in G$, and thinking of $Y$ as an $l$-vector.
Let $W=\sym^n(X)$ be the space of
homogeneous forms of degree $n$ in the variable entries of $X$. 
The space $W$, and also $P(W)$, has a similar action of $K=GL(X)=GL_{k}(\C)$.
We use  any entry $y$  of $Y$ not in 
$X$ as the homogenizing variable for embedding $W$ in $V$ 
 via the map $\phi: W \rightarrow V$ defined by:
\begin{equation} \label{ephiobs} 
\phi(h)(Y)=y^{m-n} h(X),
\end{equation}
for any $h(X) \in W$. We also think of $\phi$ as a map from $P(W)$ to $P(V)$.

Let $g=\det(Y) \in P(V)$ be the determinant form, and $f=\phi(h)$, where 
$h=\perm(X) \in   P(W)$. Let $\Delta_V[g], \Delta_V[f] \subseteq P(V)$ be
the projective closures of the orbits $G g$ and $G f$, respectively,
in $P(V)$. 
Let $\Delta_W[h] \subseteq P(W)$ be
the projective closure of the $K$-orbit $K h$  of $h$ in $P(W)$. 
Then $\Delta_V[g]$ is called the class variety associated with
$NC$ and $\Delta_V[f]$ the class variety associated with 
$P^{\#P}$; $\Delta_W[h]$ is called the base class variety associated with
$P^{\#P}$.  (The base class variety is not  used in what follows. Rather
its variant,
called a reduced class variety defined below, will be used.)
These class varieties depend on the lower bound parameters $n$ and $m$.
If we wish to make these explicit, we would  write $\Delta_V[f,n,m]$ 
and $\Delta_V[g,m]$ instead of $\Delta_V[f]$ and $\Delta_V[g]$. 

The class varieties $\Delta_V[g]=\Delta_V[g,m]$ and 
$\Delta_V[f]=\Delta_V[f,n,m]$ are $G$-subvarieties 
of $P(V)$, and  their homogeneous 
coordinate rings $R_V[g]=R_V[g,m]$ and $R_V[f]=R_V[f,n,m]$ have natural
degree-preserving $G$-action.

It is conjectured in \cite{GCT1}  that, if $m=\poly(n)$ and
 $n \rightarrow \infty$, then 
$f \not \in \Delta_V[g]$; this is equivalent to saying that 
the class variety $\Delta_V[f,n,m]$ cannot be embedded in the class 
variety
$\Delta_V[g,m]$ (as a subvariety). This implies the arithmetic form of the
$P^{\#P} \not = NC$ conjecture in characteristic zero. 

\subsection{Obstructions}
The obstruction in this context is defined as follows.
A $G$-module $V_\lambda(G)$ is called  an {\em obstruction}
for the pair $(f,g)$ if
it occurs in $R_V[f,n,m]_d^*$ but not $R_V[g,m]_d^*$ for some $d$.  
It is called a {\em strong obstruction} if, for some $d$,
it occurs in $R_V[f,n,m]_d^*$
but it does not contain $(\C g)^d$ as a $G_g$-submodule, where 
$(\C g) \subseteq V$ denotes the one dimensional line corresponding to $g$,
and $G_g \subseteq G$ is the stabilizer of $g=\det(Y) \in P(V)$. 
If $V_\lambda(G)$ is a (strong)  obstruction of degree $d$, then the size 
$|\lambda|=d m$; hence $d$ is completely determined by $\lambda$ and $m$.

Existence of an obstruction or a strong obstruction implies 
that the class variety $\Delta_V[f,n,m]$ cannot be embedded in the class 
variety
$\Delta_V[g,m]$, as sought.
The main algebro-geometric results of \cite{GCT1,GCT2} suggest that 
strong obstructions should indeed exist for all $n \rightarrow \infty$,
assuming $m=\poly(n)$; cf. Section 4, Conjecture 2.10 and Theorem 2.11 in
\cite{GCT2}. The goal then is to prove  existence of strong obstructions
for all $n$.

The definition of a strong obstruction can be simplified further as follows.
Let $X'$ denote the set of variables, which consists of the variable entries
in $X$ and the homogenizing variable $y$ above. Let $W'=\sym^m(X') \subseteq 
V=\sym^m(Y)$ be the space of homogeneous forms of degree $m$  in the
 variables of $X'$.  We have a natural action of $H=GL(X')=GL_{n^2+1}(\C)$ on
$W'$ and hence on $P(W')$.
We have a natural map $\phi': W \rightarrow W'$ given by
$\phi'(h)(X')=y^{m-n} h(X)$. The map  $\phi$  in (\ref{ephiobs}) 
is $\phi'$ followed 
by the inclusion from $W'$ to $V$. We also think of $\phi'$ as a map
from $P(W)$ to $P(W')$. 

Let $f'=\phi'(h)$, for $h=\perm(X) \in P(W)$. 
Let $\Delta_{W'}[f'] \subseteq P(W')$ be the orbit closure of $H f'$. 
It is an $H$-subvariety of $P(W')$, and hence its homogeneous coordinate ring
$R_{W'}[f']$ has the natural degree preserving $H$-action. 
We call  $\Delta_{W'}[f']$  the reduced class variety for $P^{\#P}$. 
It is known (cf.  Theorem 8.2  in \cite{GCT2}) 
that $V_\lambda(G)$ occurs in $R_V[f]_d^*$ iff $V_\lambda(H)$ occurs in 
$R_{W'}[f']_d^*$. Here the dominant weight $\lambda$  of $G$
is considered a dominant weight of $H$ 
by restriction from $G$ to $H$.

Hence $V_\lambda(G)$ is a strong obstruction for the pair $(f,g)$, iff
for some $d$, $V_\lambda(H)$ occurs in $R_{W'}[f']_d^*$ as an $H$-submodule
 and 
$V_\lambda(G)$ does not contain $(\C g)^d$ as a $G_g$-submodule.
In particular, we can assume without loss of generality that the 
height of the Young diagram for $\lambda$ is at most $n^2+1$; otherwise
$V_\lambda(H)$ would be zero.

\subsection{Robust obstructions}
It is known that the stabilizer 
$G_g$ of $g=\det(Y) \in P(V)$  consists of 
linear transformations in $G$ of the form
$Y \rightarrow A Y^{*} B^{-1}$,  thinking of $Y$ as an $m\times m$ matrix,
where $Y^*$ is either $Y$ or $Y^T$,
 $A,B \in GL_{m}(\C)$. Thus the connected component of $G_g$ is
essentially $GL_m(\C) \times GL_m(\C) \subseteq G=GL_{l}(\C)=GL_{m^2}(\C)$. 
This means the subgroup restriction problem for the embedding
$\rho: G_g \hookrightarrow G$ is essentially the Kronecker problem
(Problem~\ref{pintrokronecker}). 

Assume  PH1 (Hypothesis~\ref{hph1subg}) for the subgroup restriction 
$\rho: G_g \hookrightarrow G$; which is essentially PH1 for the 
Kronecker problem. It now assumes the following concrete form.
Let $m^d_\lambda$ denote the multiplicity of the $G_g$-module $(\C g)^d$ in
$V_\lambda(G)$. 
Assume that the height of $\lambda$ is at most
$n^2+1$ for the reasons give above.

\begin{hypo} {\bf (PH1:)} \label{hph1det}

There exists
a polytope $Q^{d}_\lambda$ such that:
\begin{enumerate} 
\item The number  of 
integer points in $Q^{d}_\lambda$ is equal to $m^{d}_\lambda$.
\item The Ehrhart quasi-polynomial  of $Q^{d}_\lambda$ coincides with
the stretching quasi-polynomial $\tilde m^{d}_\lambda(n)$ 
(cf. Theorem~\ref{tquasigit}). 
\item The polytope $Q^{d}_\lambda$ is given by a separating oracle,
and its encoding bitlength $\bitlength{Q^d_\lambda}$ is 
 $\poly(n, \bitlength{m},\bitlength{d},\bitlength{\lambda})$ time.
\end{enumerate} 
\end{hypo} 

We have to explain why $\bitlength{Q^d_\lambda}$ 
is stipulated to  depend polynomially on $n$ and $\bitlength{m}$,
rather than $m$. 
After all, the bitlengths $\bitlength{G}$, $\bitlength{G_g}$
and $\bitlength{\rho}$ are $O(\poly(m^2))$ as per the definitions 
in Section~\ref{sssubgroup}. So, as per PH1 for subgroup restriction 
in Section~\ref{sclassical}, $\bitlength{Q^d_\lambda}$  should 
depend polynomially on $m$. We are stipulating a stronger condition
for the following reason. First, as we already mentioned, 
the above hypothesis is essentially 
PH1 for the Kronecker problem, which is obtained by specializing
PH1 for the plethysm problem  (Hypothesis~\ref{hph1plethysmintro}).
In Hypothesis~\ref{hph1plethysmintro}, the encoding bitlength of the polytope 
depends polynomially on the bitlengths of the various partition parameters
$\lambda,\pi,\mu$
of the plethysm constant $a_{\lambda,\mu}^\pi$,
but is independent of  the rank of the 
group $G$ therein. 
(As explained in the remarks after 
Hypothesis~\ref{hph1plethysmintro},
 this is  justified because  the bound in 
Theorem~\ref{tpspaceplethysm} is also independent of
 the rank of $G$).  For the same reason, 
the encoding bitlength of the polytope here
should be independent of the rank of $G$ (which is $m^2$), but should 
depend polynomiallly on the total bit length of the 
 partitions parametrizing the representations $V_\lambda(G)$ and $(\C y)^d$.
This  is  $O(n+\bitlength{m}+\bitlength{d} + \bitlength{\lambda})$.
(Note that the  one dimensional representation $(\C y)^d$ of $G_g$ 
is essentially 
the $d$-th power of the 
determinant representation of $G_g$, since the connected component
of $G_g$ is isomorphic to $GL_m(\C) \times GL_m(\C)$. The Young diagram
for the 
partition corresponding to the $d$-th power of the 
determinant representation of 
$GL_m(\C)$ is a rectangle of  height $m$ and width $d$. It can be 
specified by simply giving $m$ and $d$--this specification has 
bit length  $\bitlength{m}+\bitlength{d}$.)

Next let us specialize PH1  as per Hypothesis~\ref{hph1classvariety}.
The class variety $\Delta_V[f]=\Delta[f,n,m]$ will now play the role of
$X$ in Hypothesis~\ref{hph1classvariety}.
But, for the reasons explained in the 
proof of Proposition~\ref{probust} below, we shall instead specialize 
Hypothesis~\ref{hph1classvariety}
to  the (simpler) reduced class variety $Z=\Delta_{W'}[f']$.
It now assumes that following concrete form.
Let $s_d^\lambda$ denote the multiplicity of $V_\lambda(H)$ in 
$R_{W'}[f']_d^*$. 
Putting $Z$ in place of $X$ in Hypothesis~\ref{hph1classvariety}, we get:

\begin{hypo} {\bf (PH1):}  \label{phph1Z}

There exists 
a polytope $P_{d}^\lambda$ such that:
\begin{enumerate} 
\item The number  of 
integer points in $P_{d}^\lambda$ is equal to $s_{d}^\lambda$.
\item The Ehrhart quasi-polynomial  of $P_{d}^\lambda$ coincides with
the stretching quasi-polynomial $\tilde s_{d}^\lambda(n)$ 
(cf. Theorem~\ref{tquasigit}). 
\item The polytope $P_{d}^\lambda$ is given by a separating oracle, 
and its encoding bitlength $\bitlength{P_d^\lambda}$ is 
\begin{equation} \label{eqbitlengthZ}
\poly(\bitlength{d},\bitlength{\lambda},\bitlength{Z})= 
\poly(\bitlength{d},\bitlength{\lambda},n,\bitlength{m}).
\end{equation}
\end{enumerate} 
\end{hypo} 

Here (\ref{eqbitlengthZ}) follows because $\bitlength{Z}=n + \bitlength{m}$. 
To see why, let us observe that $Z=\Delta_{W'}[f']$ is completely specified
once the point $f'=y^{m-n}h \in P(W')$ is specified. To specify $f'$, 
it sufficies to specify $m,n$ and the point $h \in P(W)$. It is known
\cite{GCT2} that the point 
$h=\perm(X) \in P(W)$
 is completely characterized by its stabilizer $K_h \subseteq 
K=GL(X)=GL_{k}(\C)$. Furthermore, $K_h$ is explicitly known \cite{minc}.
It is generated by 
the linear transformation in $K$ of the form
$X \rightarrow \lambda X \mu^{-1}$, 
thinking of $X$ as an $n \times n$ matrix,
where $\lambda$ and $\mu$ are either diagonal or permutation
matrices. So to specifiy $h$, it suffices to specify $K_h, K$ and
the embedding $\rho':K_h \hookrightarrow K$. The 
bit length of this specification
is $O(n)$ (cf. Section~\ref{sssubgroup}). To specify $f'$, and hence $Z$,
it suffices to specify
$m,n,K, K_h$ and $\rho'$. The total bit length of this specification
is $O(n + \bitlength{m})$.

Assume PH1 for both $m^d_\lambda$ and $s_{d}^\lambda$,
i.e., Hypotheses~\ref{hph1det}  and \ref{phph1Z}. 

\begin{defn} \label{drobustperm}
We say that $V_\lambda(G)$ is a {\em robust obstruction} for the
pair $(f,,g)$ if  one of the following hold:
\begin{enumerate} 
\item $Q^d_\lambda$ is empty, and $P_{d}^\lambda$ is 
nonempty. 
\item  Both $Q^d_\lambda$ and $P_{d}^\lambda$ are nonempty, 
the affine span of $Q^d_\lambda$ does not contain an integer point and 
the affine span of $P_{d}^\lambda$ contains an integer point.
\end{enumerate}

If the first condition holds, we say that $V_\lambda(G)$ is a 
{\em geometric obstruction}. If the second condition holds, it is called
a {\em modular obstruction}.
\end{defn}

\begin{prop} \label{probust}
Assume PH1 for both $m^d_\lambda$ and $s_{d}^\lambda$
(Hypotheses~\ref{hph1det}  and \ref{phph1Z}). 
If $V_\lambda(G)$ is a robust obstruction for the pair $(f,g)$, then 
for some positive integral relaxation parameter 
 $k$, $V_{k \lambda}(G)$ is a strong obstruction
for $(f,g)$. In fact, this is so for most large enough $k$. 
\end{prop} 
\proof This essentially follows from Proposition~\ref{probustgen}. 
It only remains to clarify why we can use
PH1 for the reduced class variety $\Delta_{W'}[f']$--as we are doing here--
in place of  PH1 for the class variety $\Delta_V[f]$. This is because, 
as already mentioned, $V_\lambda(G)$ occurs in $R_V[f]_d^*$ iff
$V_\lambda(H)$ occurs in $R_{W'}[f']^*_d$. \qed

\subsection{Verification of  robust obstructions}

\begin{theorem} \label{tverifyobsperm}
Assume that the singularities of $\spec(R_{W'}[f'])$ are rational.
Assume PH1 for both $m^d_\lambda$ and $s_{d}^\lambda$  as above
(Hypotheses~\ref{hph1det}  and \ref{phph1Z}). Then,
 given $n,m$, $\lambda$ and $d$,
whether $V_\lambda(H)$ is a robust obstruction can be verified 
in $\poly(n,\bitlength{m}, \bitlength{d}, \bitlength{\lambda})$ time.
Furthermore, a positive integral relaxation parameter 
$k$ such that $k \lambda$ is a strong obstruction can 
also be computed in this much time. 
\end{theorem} 

Once $n$ and $m$ are specified, the various class varieties and
$K,K_h,\rho',G,G_g,\rho$ above are automatically
specified implicitly.

\proof This follows from Theorem~\ref{tverifyobs}; cf. also the remark 
following its proof.
%It only remains to clarify the last statement. The algorithm of verification
%for robust obstruction in the proof of Theorem~\ref{tverifyobs}
%can also be made to
%return a vertex $v$  of the polytope $P_d^\lambda$ if it is nonempty. 
%We can let $k$ be the smallest integer so that $k v$ is integeral;
%i.e. l.c.m. of the denominators of the rational coordinates of $v$.
\qed

Theorem~\ref{tverobs1} can be similarly specialized in this context;
we leave that to the reader.

\subsection{On explicit construction of obstructions} 
\begin{theorem}  \label{texplicitperm}
Assume  that $m=\poly(n)$ or even $2^{\polylog(n)}$, and:
\begin{enumerate} 
\item (RH) [Rationality Hypothesis]:
The singularities of $\spec(R_{W'}[f'])$ are rational.
\item  PH1 for both $m^d_\lambda$  and $s_{d}^\lambda$ 
(Hypotheses~\ref{hph1det}  and \ref{phph1Z}).
\item OH [Obstruction Hypothesis]: 
For every (large enough)  $n$, there exists $\lambda$ 
of $\poly(n)$ bit length  such that $|\lambda|$ is divisible by $m$ and 
one of the following holds (with $d=|\lambda|/m$): 
\begin{enumerate}
\item $Q^d_\lambda$ is empty, and $P_{d}^\lambda$ is 
nonempty. 
\item  Both $Q^d\lambda$ and $P_{d}^\lambda$ are nonempty, 
the affine span of $Q^d_\lambda$ does not contain an integer point and 
the affine span of $P_{d,x}^\lambda$ contains an integer point.
\end{enumerate} 
\end{enumerate}
Then there exists an explicit family $\{\lambda_n\}$ of robust 
obstructions. 
\end{theorem}

Here we say that $\{\lambda_n\}$ is an explicit family of robust obstructions
if each $\lambda_n$ is short and easy to verify.  Short means 
$\bitlength{\lambda_n}$ is $O(\poly(n))$. Easy to verify means 
whether $\lambda_n$ is a robust 
obstruction can be verified in $O(\poly(n))$ time.

The $\poly(n)$ bound here and  in OH  is meant 
to be independent of $m$, as long as $m<< 2^n$;
i.e., it should hold even when $m=2^{\polylog(n)}$. In other words, the 
family $\{\lambda_n\}$ should  continue to remain an explicit  robust 
obstruction family, as we vary $m$ over all values $\le 2^{\polylog(n)}$,
and perhaps even values $\le 2^{o(n)}$, but will cease to be an obstruction
family for some large enough $m=2^{\Omega(n)}$. This is an 
important  uniformity condition. 

\proof 
OH basically says that  there exists a short robust obstruction $\lambda_n$
for every $n$. By Theorem~\ref{tverifyobsperm}, it is easy to verify. \qed

\subsection{Why should robust obstructions exist?}  \label{srobustex}
The main question now is: why should OH hold? That is, why should (short)
robust obstructions exist? 

As we already mentioned, the results in \cite{GCT1,GCT2} indicate that
strong obstructions should exist for every $n$, assuming $m=\poly(n)$. 
We shall give a heuristic argument for existence 
of robust obstructions  assuming that
strong  obstructions exist. This will crucially depend on the following 
SH for $m^d_\lambda$, which is essentially SH for the Kronecker problem
(i.e. specialization of Hypothesis~\ref{hshplethysmintro} to the Kronecker problem),  good experimental evidence for which is provided in \cite{rosas2}. 

\begin{hypo} {\bf (SH:)}  \label{hshperm}
\noindent (a): The  saturation index of 
$\tilde m^d_\lambda(k)$ is 
bounded by a polynomial in $m$. (Observe that the rank of $G$ is $\poly(m)$
and the height of $\lambda$ is at most $n^2+1$).
\noindent (b): The quasi-polynomial
$\tilde m^d_\lambda(n)$ is strictly saturated, i.e. the
saturation index is zero,  for almost all $\lambda$ (and $d$).
\end{hypo} 

If $V_\lambda(G)$ is a strong obstruction,
 $s_d^\lambda$ is nonzero but $m^d_\lambda$ is zero. Thus,
assumming PH1, 
there are three possibilities:

\begin{enumerate} 
\item $Q^d_\lambda$ is empty, and $P_{d}^\lambda$ is 
nonempty and contains an integer point.
\item  Both $Q^d\lambda$ and $P_{d}^\lambda$ are nonempty, 
the affine span of $Q^d_\lambda$ does not contain an integer point and 
$P_{d}^\lambda$ contains an integer point.
\item  Both $Q^d_\lambda$ and $P_{d}^\lambda$ are nonempty.
The affine span of $Q^d_\lambda$ contains an integer point, but 
$Q^d_\lambda$ does not. And $P_d^\lambda$ contains an integer point.
\end{enumerate}

In the first two cases, $\lambda$ is a robust obstruction. 
As per SH (Hypothesis~\ref{hshperm}),  for almost
all $\lambda$, 
the Ehrhart quasipolynomial of $Q^d_\lambda$ is  saturated: this means
(cf. the proof of Theorem~\ref{tindexquasi}),
if the affine span of $Q^d_\lambda$ contains an integer point then 
$Q^d_\lambda$ also contains an integer point. And hence, with a high 
probability, the third case should not occur.
In other words, strong obstructions can be  expected to be robust with
a high probability.

Let us call  a strong obstruction  $\lambda$ 
{\em fragile} if it is not robust; this means 
the affine span of $Q^d_\lambda$ contains an integer point, but $Q^d_\lambda$
does not.  By SH (Hypothesis~\ref{hshperm}), if $\lambda$ is fragile, then
for some $k=\poly(m)$, 
$Q_{k \lambda}^{k d}$ contains an intger point, and hence, $k \lambda$ 
is not obstruction. Thus  fragile obstructions are close to 
not being   obstructions,
and furthermore,  are expected to be rare, as argued above.
This is why we are focussing on robust obstructions.

It may be remarked that the only SH needed in the argument above 
is the one (Hypothesis~\ref{hshperm})
for the structural constant $m^d_\lambda$. 
This is a special case of the SH for the subgroup restriction problem 
(cf. Section~\ref{sssubgroup}) specialized to  the embedding
$G_g \hookrightarrow G$. 
In particular, we do not need SH for the structural constant $s_d^\lambda$;
i.e., for the more difficult decision problem in geometric invariant
theory (cf. Problem~\ref{pintrogit} and Section~\ref{sspgit}).

\subsection{On discovery of robust obstructions} 
It may be conjectured that not just the verification 
(cf. Theorem~\ref{tverifyobsperm})
but also the discovery of robust obstructions is easy for the 
problem under consideration. In this section we shall give 
an argument in support of this conjecture for geometric (robust) obstructions
(which may be conjectured to exist in the  problem under
consideration). 
For this we need to reformulate the notions of strong and robust obstructions
(Definition~\ref{drobustperm}) as follows.

Let  $T_\Z$ be the set of pairs $(d,\lambda)$ such that $s_{d}^\lambda$ is
nonzero and  $S_\Z$ the
set of pairs $(d,\lambda)$ such that $m^{d}_\lambda$ is
nonzero.

\begin{prop} \label{pfinitegenobs}
Assuming PH1 above (Hypotheses~\ref{hph1det}  and \ref{phph1Z}),
$T_\Z$ and $S_\Z$ are finitely generated semigroups
with respect to addition.
\end{prop} 
These  semi-groups 
 are analogues of the Littlewood-Richardson
semigroup (Section~\ref{slittlecone})  in this setting.

\proof  The proof is similar to that for 
 the Littlewood-Richardson semigroup \cite{zelevinsky}.

For  given $d$ and $\lambda$, 
the polytope 
$P_{d}^\lambda$ in PH1 for $s_{d}^\lambda$ (Hypothesis~\ref{phph1Z})  has a 
specification of the form 
\begin{equation}  \label{eqtconesatobs}
A x \le b
\end{equation} 
where $A$ depends only the variety $Z=\Delta_{W'}[f']$,
but not on $d$ or $\lambda$,
and $b$ depends homogeneously and linearly on $d$ and $\lambda$.
Let $P$ be the polytope defined by the inequalities (\ref{eqtconesatobs})
where both $d$ and $\lambda$ are treated as variables. Then
$P$ is a polyhedral cone (through the origin) in the ambient space 
containing $P$ with the coordinates $x,d$ and $\lambda$.
Let $P_\Z$ be the set of integer points in $P$.
It is a finitely generated 
semigroup since $P$ is a polyhedral cone. Let $T_\R$ be the 
orthogonal projection of $P$ on the 
hyperplane corresponding to the coordinates $d$ and $\lambda$.
Now $T_\Z$ is simply the projection of $P_\Z$.
Hence it is a finitely generated semigroup as well.

The proof for $S_\Z$ is similar, with $S_\R$ defined similarly. \qed 

The polyhedral cones $T_\R$ and $S_\R$ here
are analogues of the Littlewood-Richardson cone (Section~\ref{slittlecone})
in this setting. Note that $(d,\lambda) \in T_\R$ iff 
$P_d^\lambda$ is nonempty; similarly for $S_\R$.

A Weyl module $V_\lambda(G)$ is a strong  obstruction for the
pair $(f,g)$ of degree $d$ 
iff $(d,\lambda)$  occurs in $T_\Z$ but not in $S_\Z$. 
It is a robust obstruction iff it occurs in $T_R$ but not in $S_\Z$. 
It is a geometric obstruction iff it occurs in $T_\R$ but not in $S_\R$.
It is a modular obstruction iff it occurs in $T_\R$ and also in $S_\R$
but not in $S_\Z$.

Assuming PH1 (Hypothesis~\ref{phph1Z}),
whether $(d,\lambda)$  belongs to $T_\R$ can be determined 
in polynomial time by linear programming, since $(d,\lambda) \in T_\R$ 
iff $P_d^\lambda$ is nonempty.  Similarly, assuming PH1 
(Hypothesis~\ref{hph1det}),
whether $(d,\lambda) \in S_\R$ can be determined in polynomial time.

The following is a stronger complement to PH1.

\begin{hypo} {\bf (PH1*)} \label{hdisc}

Whether $T_\R \setminus S_\R$ is nonempty 
can be determined in polynomial time; i.e., 
$\poly(n, \bitlength{m})$ time. If so, the algorithm can
also output $(d,\lambda) \in T_\R \setminus S_\R$ of polynomial bit length.
\end{hypo} 

\begin{prop} Assuming PH1*, 
 given $n$ and $m$, 
the problem of deciding if a geometric  obstruction exists for the pair $(f,g)$,
and finding one if one exists, belongs to the complexity class $P$; i.e.,
it  can be done
in $\poly(n,\bitlength{m})$ time.
\end{prop} 
This immediately follows from Hypothesis~\ref{hdisc} since $(d,\lambda)$ is
a geometric  obstruction iff $(d,\lambda) \in T_\R \setminus S_\R$.

Hypothesis~\ref{hdisc}  is supported by the following: 

\begin{prop} 
Assuming PH1 (Hypotheses~\ref{hph1det}  and \ref{phph1Z}),
Hypothesis~\ref{hdisc} holds if 
$T_\R$ and $S_\R$ have polynomially many explicitly given
constraints with the specification of polynomial  bit length;
here polynomial means $\poly(n,\bitlength{m})$. 
\end{prop} 

The proposition  holds even if
the polytope $S_\R$ has  exponentially many
constraints, as long as it is given by a separation oracle that works in
polynomial time.

\proof
It suffices to check if $S_\R$ satisfies each constraint of $T_\R$. 
This can be done in polynomial time using the linear programming
algorithm in \cite{lovasz}. Specifically, let $l(y)\ge 0$ be a 
constraint of $T_\R$.
 Then we just need to minimize $l(y)$ on $S_\R$ and check if 
the minimum exceeds zero.
\qed 

But this method does not work when the number of constraints of 
$T_\R$ is exponential, as  expected in the context
of the lower bound problems under consideration.
In fact,  no
generic black-box-type   algorithm, like the 
one in \cite{lovasz} based on just a membership or separation oracle for
$T_\R$,  can be used to prove (4)  when the number 
of constraints of $T_\R$ is exponential. 

Fortunately, this is not a serious problem.
A basic principle in combinatorial optimization, as illustrated in
\cite{lovasz},
is that a complexity theoretic property
that holds for polytopes with polynomially many constraints will
also hold for polytopes with exponentially many constraints, provided
these constraints are sufficiently well-behaved.
For example, Edmond's perfect matching polytope for nonbipartite 
graphs has complexity-theoretic 
properties similar to the perfect matching polytope 
for bipartite graphs, though it can have exponentially many constraints.
We have already remarked that $T_\R$ and $S_\R$ are analogues
of the Littlewood-Richardson cones.
The facets of  the Littlewood-Richardson cone
have a very nice  explicit description 
\cite{kly,zelevinsky}. The cones $T_\R, S_\R$ here
are expected to have similar 
nice explicit description. This is 
why Hypothesis~\ref{hdisc} can be 
expected to hold even if the number of constraints 
of $T_\R$ is exponential, just as it holds even when $S_\R$ has
exponentially many constraints.
But a polynomial-time algorithm as in Hypothesis~\ref{hdisc}  would 
have to depend crucially
on the specific nature of the facets (constraints)
of $T_\R$ in the spirit of 
the linear-programming-based algorithm for the
construction of a maximum-weight perfect matching 
in nonbipartite graphs \cite{edmonds}, where too the number of constraints is
exponential but the algorithm still works because of the structure theorems
based on the specific nature of the constraints.

\section{Arithmetic form of the $P$ vs $NP$ problem in characteristic zero} 
\label{spvsnp}
We turn now to the arithemetic form of the $P$ vs. $NP$ 
problem in characteristic zero. The arguments are essentially verbatim 
translations of 
those for the arithmetic form of the $P^{\#P}$ vs. $NC$ problem in
the preceding section. Hence we shall be brief. 

In the preceding section $h(X)$ was $\perm(X)$ and $g(Y)$ was $\det(Y)$. 
Now $h(X)$ and $g(Y)$ would be explicit 
 (co)-NP-complete and $P$-complete
functions $E(X)$ and $H(Y)$ constructed in \cite{GCT1}. 
They can be thought of as  points in suitable $W=\sym^k(X)$ and 
$V=\sym^l[Y]$,  $k=O(n^2), l=O(m^2)$, 
 with the natural action of $GL(X)$ and $G=GL(Y)$,  where 
$n$ denotes the number of input parameters and $m$ denotes the 
circuit size parameter in the lower bound problem.  
These functions are extremely special
like the determinant and the permanent in the sense that 
they  are ``almost''  characterized by their stabilizers as explained
in \cite{GCT1}--and this is  enough for
our purposes. 

We again have a natural embedding $\phi: P(W) \rightarrow P(V)$,
which lets us define $f=\phi(h)$.
The class variety for $NP$ is defined to be $\Delta_V[f] \subseteq P(V)$,
the projective closure of the orbit $G f$.
The class variety for $P$ is
$\tilde \Delta_V[g] \subseteq P(V)$, 
which is defined to be the projective closure of $G [g]$, where $[g]$ denotes
the set of points in $P(V)$ that are stabilized by $G_g \subseteq G$, the
stabilizer of $g$. An explicit description of $G_g$
is given in \cite{GCT1}; cf.
Section 7 therein.  
To show $P\not = NP$ in characteristic zero,
it suffices  to show that 
$\Delta_V[f]$ is not a subvariety of $\tilde \Delta_V[g]$  for all
large enough $n$, if $m=\poly(n)$ (cf. Conjecture 7.4. in \cite{GCT1}).
For this, in turn, it suffices
to show existence of strong obstructions,
defined very much as in Section~\ref{sdetvsperm},
for all $n$, assumming $m=\poly(n)$.

We can then formulate 
PH1  for the  new 
$h(X)$ and $g(Y)$ just as in 
Hypotheses~\ref{hph1det}  and \ref{phph1Z}, and the notion of 
a robust obstruction as in Definition~\ref{drobustperm}. We then have:

\begin{theorem} (Verification of obstructions) 

Analogues of Theorems~\ref{tverifyobsperm}  and \ref{texplicitperm} 
holds for $h(X)=E(X)$ and $g(Y)=H(Y)$.
\end{theorem} 

Furthermore, even discovery of robust obstructions can be conjectured 
to be easy (poly-time)--this would follow from
the obvious analogue of Hypothesis~\ref{hdisc} here.

Heuristic argument for existence of robust obstructions is
very similar to the one in Section~\ref{srobustex}. It needs SH for the 
special case of the subgroup restriction problem for the 
embedding $G_g \hookrightarrow G$. 
The group $G_g$, as described in \cite{GCT1}, is 
a product of some copies of the algebraic torus
and the symmetric group. The subgroup restriction problem in this
case is akin to but harder than  the plethysm problem.

\end{document}